\documentclass[submission, Phys]{SciPost}
\usepackage{hyperref}
\def\doi{http://dx.doi.org/}
\usepackage{amsmath,braket,mathdots,mathtools,amssymb,xcolor,calligra,color}
\usepackage{amsthm} % JJ
\usepackage{bbm}  
\usepackage{fancyhdr}
\usepackage{float}
\usepackage{authblk}
\usepackage{color}
\usepackage{tikz}
\usetikzlibrary{spy}
\usetikzlibrary{patterns}
\usetikzlibrary{calc,arrows,positioning}
\usetikzlibrary{arrows,shadows}
\usetikzlibrary{decorations.pathmorphing}	% For Feynman 
\usetikzlibrary{decorations.markings}
\usepackage{pgfplots}

\usepackage{pgfplotstable}
\usepackage{filecontents}
\usepackage[utf8]{inputenc}

\newcommand{\be}{\begin{equation}}
\newcommand{\ee}{\end{equation}}
\newcommand{\bea}{\begin{eqnarray}}
\newcommand{\eea}{\end{eqnarray}}

\def\nn{\nonumber\\}

\def\fr#1{(\ref{#1})}

\def\fn{f_{\vec{\nu}}}

\def\bell{\ell}
\def\orho#1{{{{\cal O}\big(\rho_\beta^{#1}\big)}}}
\let\OLDthebibliography\thebibliography
\renewcommand\thebibliography[1]{
  \OLDthebibliography{#1}
  \setlength{\parskip}{0pt}
  \setlength{\itemsep}{0pt plus 0.3ex}
}

\newcommand{\sign}{\,\text{sgn}\,}
\newcommand{\Tr}{\,\text{tr}\,}

\newcommand{\li}{\,\text{Li}_2\,}
\newcommand{\si}{\,\text{Si}\,}

\begin{document}
%%%%%%%%%%%%%%%%%%%%%%%%%%%%%%%%%%%%%%%
\begin{center}
{\Large\bf Finite temperature and quench dynamics in the Transverse Field Ising Model from form factor expansions}
\end{center}
%%%%%%%%%%%%%%%%%%%%%%%%%%%%%%%%%%%%%%%
\begin{center}
Etienne Granet\textsuperscript{1$\star$}, Maurizio Fagotti\textsuperscript{2} and Fabian H. L. Essler\textsuperscript{1},
\end{center}
\begin{center}
{\bf 1} The Rudolf Peierls Centre for Theoretical Physics, Oxford
University, Oxford OX1 3PU, UK\\
{\bf 2} LPTMS, CNRS, Université Paris Sud, Université Paris-Saclay,
91405 Orsay, France \\
${}^\star$ {\small \sf etienne.granet@physics.ox.ac.uk}
\end{center}
\date{\today}

\section*{Abstract}
{\bf We consider the problems of calculating the dynamical order parameter
two-point function at finite temperatures and the one-point function
after a quantum quench in the transverse field Ising chain. Both of
these can be expressed in terms of form factor sums in the basis of
physical excitations of the model. We develop a general framework for
carrying out these sums based on a decomposition of form factors into
partial fractions, which leads to a factorization of the multiple sums
and permits them to be evaluated asymptotically. This naturally leads
to systematic low density expansions. At late times these expansions
can be summed to all orders by means of a determinant
representation. Our method has a natural generalization to semi-local
operators in interacting integrable models. 
}

\vspace{10pt}
\noindent\rule{\textwidth}{1pt}
\tableofcontents\thispagestyle{fancy}
\noindent\rule{\textwidth}{1pt}
\vspace{10pt}

\renewcommand\Affilfont{\fontsize{9}{10.8}\itshape}
%\renewcommand{\thesubsection}{\arabic{subsection}}

%\tableofcontents
%%%%%%%%%%%%%%%%%%%%%%%%%%
\section{Introduction}
%%%%%%%%%%%%%%%%%%%%%%%%%%

As a consequence of the existence of extensive numbers of conservation laws
with local densities the dynamical properties of quantum integrable
models at finite energy densities are both rich and unusual. The two
main settings of interest are finite temperature equilibrium response
and time evolution after quantum quenches. In the first setting the
aim is to determine two-point functions of the form
\be
\chi_{\cal AB}(x,t)=\frac{1}{Z(\beta)}{\rm Tr}\left[e^{-\beta H}{\cal
    A}(x,t){\cal B}(0,0)\right],
\label{2point}
\ee
where $Z(\beta)={\rm Tr}(e^{-\beta H})$ and ${\cal A}(x,t)$ is a
Heisenberg picture operator, while in the quench setting
one is interested in equal time expectation values
\be
\langle\Psi| {\cal O}(x,t)|\Psi\rangle,
\label{1point}
\ee
where $|\Psi\rangle$ is an initial state that is a linear
superposition of an exponentially (in system size) large number of
energy eigenstates.

%%%%%%%%%%%%%%%%%%%%%%%%%%%%%%%%%%%%%%%%%%%%
\subsection{Finite temperature dynamics}
%%%%%%%%%%%%%%%%%%%%%%%%%%%%%%%%%%%%%%%%%%%%
Early work on determining \fr{2point} focussed on the spin-1/2 XY
chain in a magnetic field, which can be mapped to a non-interacting
model of free fermions \cite{LSM}. 
In \cite{Perk80,Perk84} it was shown that two-point functions
fulfil systems of nonlinear differential equations, which in the
transverse-field Ising limit can be efficiently solved numerically
\cite{Perk09}. The dynamics at the Ising critical point was obtained
in \cite{MPS83a,MPS83b}.
The long time and distance asymptotics of two-point
functions in the XX limit was obtained from the solution of a
Riemann-Hilbert problem in \cite{IIKS93} arising from a Fredholm
determinant representation \cite{Colomo93,vladbook}.A
Fredholm determinant representation was also derived for the Ising
field theory \cite{leclair1996}. A semiclassical
approach to the low temperature regime in interacting integrable
models was pioneered in \cite{young} and has proved very useful
\cite{damle98,damle05,zarand09} due to its relative
simplicity. It is however limited in that it applies only to very low
temperatures and cannot be easily extended. Perhaps the most direct
approach to evaluating \fr{2point} or \fr{1point} is by introducing
spectral representations, e.g. 
\be
\chi_{\cal AB}(\ell,t)=\frac{1}{Z(\beta)}\sum_{n,m}e^{-\beta E_n}\langle n|{\cal
    A}(0,0)|m\rangle\langle m|{\cal B}(0,0)|n\rangle\ e^{it(E_n-E_m)-i\ell(P_n-P_m)},
\label{Lehmann}
\ee
where $|n\rangle$ are normalized eigenstates of energy $E_n$ and momentum
$P_n$. Early investigations of \fr{Lehmann} focussed on integrable
quantum field theories in the infinite volume
\cite{muss,Saleur00,CF02,rmk,AKT,Reyes06}, where the spectral
representations need to be regularized. This problem was solved in
\cite{PT08,takacs,EK08} and a systematic low temperature expansion of
dynamical two point functions in Fourier space was obtained
\cite{EK08,EK:finiteT,PT10,Steinberg}. For some correlators this expansion
exhibits divergences close to the zero temperature mass shell and
needs to be summed to all orders
%
%At zero-temperature it was understood recently how to sum all the form
%factors to obtain the large-distance and late-time asymptotics
%\cite{kozlowski1,kozlowski2,kozlowski3,kozlowskimaillet}.
%
%However, carrying out this summation exactly at non-zero temperature
is an open problem.
A similar approach was formulated for the case of
the Ising field theory in \cite{Doyon05} and used to obtain the
late-time asymptotic behaviour of the order parameter two-point
function \cite{doyongamsa}.

In order to go beyond the low temperature regime in interacting
integrable models it is useful to work in the micro-canonical ensemble
and employ typicality ideas. This provides a more efficient spectral
representation of the form 
\be
\chi_{\cal AB}(\ell,t)=\sum_m \langle E_\beta|{\cal A}(0,0)|m\rangle
\langle m|{\cal B}(0,0)|E_\beta\rangle\ e^{it(E_\beta-E_m)-i\ell(P_\beta-P_m)},
\label{Lehmann2}
\ee
where $E_\beta$ is a typical energy eigenstate at the energy density
corresponding to inverse temperature $\beta$ \cite{vladbook,PC14}. The
representation \fr{Lehmann2} can be analyzed numerically for finite
systems\cite{PC14}. Moreover, in particular limiting cases it appears
to be very efficient in that only a small number of states need to be
summed over \cite{DNP18}. In the zero temperature case
it has proved possible to formulate, and evaluate asymptotically, a
form factor expansion in the thermodynamic limit
\cite{kozlowski1,kozlowski2,kozlowski3,kozlowskimaillet}.
Very recently an axiomatic approach aimed at extending these ideas to
formulate form factors between states at finite energy densities in
the infinite volume limit was proposed \cite{CCP19} and used to
formulate a spectral representation. Using this representation 
to obtain explicit results for dynamical two-point functions remains
an open problem.

An alternative approach to finite temperature dynamics is based on the
Quantum Transfer Matrix approach \cite{Klumper93,DDV95}. The latter is
highly efficient for determining static properties
\cite{Damerau07,Boos08,Trippe10,Dugave13,Dugave14} and can be extended to
dynamical correlation functions \cite{GKKKS17}. Very recently this
method has been successfully applied to the XX model
\cite{GKS18,GKSS19,GKS19} and state-of-the-art results have been
obtained. The generalization to determine dynamical two-point
functions in interacting integrable models is an open problem.

The late time asymptotics of certain finite temperature two-point
functions can also be accessed by applying generalized hydrodynamics
\cite{CADY16,BCDF16} to the linear response regime, see \cite{Doyon18,Sarang19}.
%%%%%%%%%%%%%%%%%%%%%%%%%%%%%%%%%%%%%%%%%%%%
\subsection{Quench dynamics}
%%%%%%%%%%%%%%%%%%%%%%%%%%%%%%%%%%%%%%%%%%%%
Early work on quench dynamics again focussed on models that can be
analyzed by means of free fermion techniques
\cite{BMD70,BM71,SPS:04,RSMS09,RSMS10,CEF1,CEF2,CEF3,FE13}. Notably,
in \cite{CEF1,CEF2} exact results for the late time behaviour of one 
and two point functions of the order parameter in the transverse field
Ising model (TFIM) after quantum quenches were obtained. One way of going
beyond free theories is to employ a spectral representation
\be
\langle\Psi|{\cal O}(x,t)|\Psi\rangle=\sum_{n,m}
\langle\Psi|n\rangle\langle m|\Psi\rangle
\langle n|{\cal O}(0,0)|m\rangle\ e^{i(E_n-E_m)t-i(P_n-P_m)x}\ .
\ee
This was used to obtain the late time behaviour for small quenches in the
TFIM \cite{CEF1,CEF2,SE12,EEF12,KCC14,quenchLL} and the sine-Gordon model
\cite{Bertini14,CCS17,Horvath18}. The small quench regime is also accessible by
semiclassical methods \cite{IR11,RI11,BRI12,Evangelisti13,Kormos16},
which have the advantage of being significantly simpler to
implement. A much more efficient spectral representation is provided
by the Quench Action Approach \cite{CE13}. For translationally
invariant initial states this allows one to express expectation values
of local operators after a quantum quench from an initial state
$|\Psi\rangle$ as
\begin{equation}
  \lim _{L\rightarrow \infty }\left\langle \mathcal{O}\left( t\right)
  \right\rangle= \lim _{L\rightarrow \infty }\left(\frac {\left\langle
    \Psi \left| \mathcal{O}\left( t\right) \right| \Phi _{s}\right\rangle }
                 {2\left\langle \Psi  |\Phi _{s}\right\rangle }+\frac
                 {\left\langle \Phi _{s}\left| \mathcal{O}\left( t\right)
                   \right| \Psi \right\rangle } {2\left\langle \Phi
                   _{s}|\Psi \right\rangle }\right),
                 \label{QuenchA}
\end{equation}
where $|\Phi_s\rangle$ is a \emph{representative state} fixed by 
two requirements: first, it is a simultaneous eigenstate of the Hamiltonian and
of the (quasi)local conservation laws $I^{(n)}$ of the theory under
consideration, and, second, it correctly reproduces the expectation values
\be
\lim_{L\to\infty}\frac{\langle\Psi|I^{(n)}|\Psi\rangle}{L}=
\lim_{L\to\infty}\frac{\langle\Phi_s|I^{(n)}|\Phi_s\rangle}{L}.
\label{QA}
\ee
Expression \fr{QA} affords a more efficient spectral
representation involving only a single sum over energy eigenstates as
\be
\label{quenchfirst}
\langle\Psi | \mathcal{O}\left( t\right) |\Phi_{s}\rangle=
\sum_n\langle\Psi|n\rangle\langle n\left| \mathcal{O}\left( 0\right) \right|
\Phi_{s}\rangle\ e^{it(E_n-E_s)}\ .
\ee
The behaviour in the steady state reached in the limit $t\to\infty$ is
given by the expectation value in the representative state and this
has been analyzed in a number of cases
\cite{Wouters14,Brockmann14,Pozsgay14,Mestyan15,denardis14,piroli16}. The
time dependence is significantly more difficult to obtain. So far results are
restricted to a particular one-point function for small quenches in
the sine-Gordon model \cite{Bertini14} and density correlations at late times after a
quench in the repulsive Lieb-Linger model\cite{DNPC15}. 
%%%%%%%%%%%%%%%%%%%%%%%%%%%%%%%%%%%%%%%%%%%%
\subsection{Local vs semi-local operators}
%%%%%%%%%%%%%%%%%%%%%%%%%%%%%%%%%%%%%%%%%%%%
Locality properties of the operator of interest have important
implications in both finite temperature and quench contexts. For
quantum quenches this was emphasized in \cite{RSMS09,RSMS10} and
clarified through explicit calculations in Refs
\cite{CEF1,CEF2,CE13,Bertini14}. A precise definition of the mutual
locality index $\omega(A,B)$ of two operators exists in the context of
relativistic integrable quantum field theory, see e.g. \cite{lukyanov}; specifically, the product of operators $A(x,\tau)B(0,0)$ as a function of $(x,\tau)$ has the property
\be
{\cal A}_C\big[A(x,\tau)B(0,0)]=
e^{2\pi i\omega(A,B)}A(x,\tau)B(0,0),
\ee
where ${\cal A}_C$ denotes the analytic continuation along a
counter-clockwise contour $C$ around zero. Let us for simplicity
consider the case of a diagonal scattering theory with only a single
``elementary'' particle excitation created by the field $\Psi(x)$.
A convenient basis of energy eigenstates is given in terms of
scattering states of elementary excitations
\be
|\theta_1,\dots,\theta_n\rangle\ ,
\ee
where $\theta_j$ are rapidity variables related to the energy and
momentum of a single-particle excitation by
$\varepsilon(\theta)=M\cosh(\theta)$\ , $p(\theta)=\frac{M}{v}\sinh(\theta)$. 
Spectral representations of correlation functions (in the infinite
volume) involve form factors like
\be
\langle
\theta_1,\dots,\theta_N|A(0,0)|\theta_1',\dots,\theta'_M\rangle\ .
\ee
As we will see below the case $M=N$ is of particular
interest. \emph{Local operators} have vanishing mutual locality index
with $\Psi(x)$. As a consequence of kinematic poles \cite{smirnov} the
form factors become singular when rapidities in the set $\{\theta_j\}$
  approach those in $\{\theta'_j\}$. In the case $N=M$ the structure
  of singularities is \cite{takacs}
\bea
\label{localsing}
\langle\theta_1+\epsilon_1,\dots,\theta_N+\epsilon_N|A(0,0)|\theta_1,\dots,\theta_N\rangle
&=&\sum_{i_N=1}^N\dots\sum_{i_1=1}^Na_{i_1\dots
  i_N}(\theta_1,\dots,\theta_N)\frac{\epsilon_{i_1}\dots\epsilon_{i_N}}{\epsilon_1\dots\epsilon_N}\nn
&&+\dots
\eea
In contrast, for \emph{semi-local} operators $B$ with
$\omega(B,\Psi)=1$ one has instead
\bea
\langle\theta_1+\epsilon_1,\dots,\theta_N+\epsilon_N|A(0,0)|\theta_1,\dots,\theta_N\rangle
&=&\frac{2^N\langle 0|A(0,0)|0\rangle}{\epsilon_1\dots\epsilon_N}+\dots
\label{singularities}
\eea
This shows that form factors of such semi-local operators are much
more singular than the ones for local operators. Form factors in integrable lattice
models have analogous structures of singularities. 
%%%%%%%%%%%%%%%%%%%%%%%%%%%%%%%%%%%%%%%%%%%%
\subsection{One and two-point functions of semi-local operators}
%%%%%%%%%%%%%%%%%%%%%%%%%%%%%%%%%%%%%%%%%%%%
The nature of singularities for semi-local operators
\fr{singularities} has been exploited previously to obtain results for
1-point functions after \emph{small} quantum quenches
\cite{CE13,Bertini14}. The aim of this work is to extend this approach to general quantum quenches as well as to dynamical
two-point functions at finite temperatures. We focus on the case of
the order parameter in the TFIM because the form factors are
particularly simple in this case. This allows us to exhibit in
considerable detail which states in the respective spectral
representations contribute to the late time asymptotics of one and
two-point functions. These considerations can be generalized to interacting
integrable models, as will be shown in a following publication.

%%%%%%%%%%%%%%%%%%%%%%%%%%%%%%%%%%%%%%%%%%%%
\subsection{Outline and summary of the main results}
%%%%%%%%%%%%%%%%%%%%%%%%%%%%%%%%%%%%%%%%%%%%
We conclude our introduction with an outline of the following sections
and a brief summary of our key results.
\begin{itemize}
\item In Section \ref{sectfim} we briefly summarize a number of well known
results on the TFIM and then define in detail the two problems we
study in this paper, namely dynamical correlation functions at finite
temperature and time evolution of the order parameter after a quantum
quench. Although these two problems are of a very different physical
nature, we explain how they can both be formulated in terms of
sums over form factors and thus be addressed with similar techniques.

\item In Section \ref{pfd} we develop a novel framework for
organizing and (analytically) carrying out the sums over form factors
in both problems. It is based on a partial fraction decomposition of
the form factors, which organizes the sums according to the degree of the
poles the various terms exhibit, and naturally leads to an expansion
of the correlation functions in terms of the density of particles
$D=\int \rho(x)dx$ of the thermal/non-equilibrium stationary state of
interest, where $\rho$ is its particle density. We present in detail how this calculation works at order
${\cal O}(D^2)$ in the case of finite temperature equilibrium dynamics.
In order to make the expansion uniform in space and time it is
necessary to sum certain contributions to all orders in $D$. In this
way we obtain explicit expressions for the dynamical spin-spin
correlation function in an arbitrary macro-state $|\phi\rangle$, in particular thermal states. For a transverse magnetic field $h<1$ the results reads
\begin{align}
\langle \phi|\sigma^x_{\ell}(t)\sigma^x_{0}(0)|\phi\rangle
&\approx C\exp\left(-2 \int_{-\pi}^\pi \rho(x)(1+2\pi\rho(x))  |
t\varepsilon'(x)-\bell|dx\right)\ ,\nn
C&=\xi \exp\left(-2\int _{-\pi }^{\pi}\int _{-\pi }^{\pi}\frac {\rho (y) \rho '(x) } {\tan \left( \frac {x-y} {2}\right) }dxdy\right)\ .
\end{align}
The various quantities $\varepsilon,\xi$ entering this expression are defined below in Section \ref{sectfim}. This result is exact at order ${\cal O}(D^2)$, which means in
particular that higher orders in the expansion will contribute
additive terms in the exponents that involve third and higher powers
of the particle density $\rho(p)$. The expansion can be pursued to
higher orders in $D$ within the framework developed in Section \ref{pfd}.

We then turn to the time evolution of the order parameter after a
quantum quench. By combining our framework for carrying out form
factor sums with the quench action approach to quantum quenches
\cite{CE13} we obtain a systematic expansion of the order parameter
one-point function in powers of the particle density.

A key insight derived from our approach is that the late time
behaviour in both problems arises from processes that involve an
\emph{arbitrary} number of particle-hole excitations over respectively
the thermal and non-equilibrium steady state, but each of them is
``small'' in a sense that we make precise below. We argue that this is
a general feature of form factor expansions involving semi-local
operators, and represents a qualitative difference to the case of
local operators.

\item In Section \ref{qqd} we return to the quantum quench problem
and show how to determine the \emph{exact} exponent that characterizes the
exponential decay of the order parameter at late times. This
calculation is based on approximations valid at late times that allow 
the spectral sum to be cast in the form of a determinant. This
representation is similar to one obtained for the impenetrable Bose
gas in Ref. \cite{korepinslavnov}. The late time asymptotics can be
extracted from the determinant representation and leads to the result
\begin{equation}
\langle\sigma_\ell^x \rangle=C\exp\left(\frac{\vert
  t\vert}{\pi}\int_0^\pi \vert\varepsilon'(x)\vert
\log(1-4\pi\rho(x))dx\right)+\dots\ ,
\end{equation}
where the constant $C$ is known up to order ${\cal O}(D^2)$. This
exponent is in agreement with the exact expression of the decay time
obtained in a very different way in Ref. \cite{CEF2}, while our result
for $C$ is new.

\item In Section \ref{dynamicatemp} we generalize the approach of
Section \ref{qqd} to the case of the dynamical spin-spin correlation
function in an arbitrary macro state $|\phi\rangle$ described by a
density $\rho(p)$. We obtain the following expression of the late time
asymptotics 
\begin{align}
\langle \phi|\sigma^x_{\ell}(t)\sigma^x_{0}(0)|\phi\rangle &=C\exp\left(\frac{1}{2\pi}\int_{-\pi}^\pi|t\varepsilon'(x)-\bell|
\log(1-4\pi\rho(x))dx\right)+\dots\ .
\label{phiphi}
\end{align}
Here the exponent represents an exact result, while the constant $C$
is again only known to order ${\cal O}(D^2)$. This result is to the
best of our knowledge new. We compare \fr{phiphi} to numerically exact
results obtained using the representation of the finite temperature
correlator as a Pfaffian and find perfect agreement.
\end{itemize}

%%%%%%%%%%%%%%%%%%%%%%%%%%%%%%%%%%%%%%%%%%%%%%%%%%
%\section{Notations and statement of the problems}
\section{ Transverse Field Ising Model \label{sectfim}}
%%%%%%%%%%%%%%%%%%%%%%%%%%%%%%%%%%%%%%%%%%%%%%%%%%
%\subsection {The TFIM Hamiltonian}
The TFIM Hamiltonian on a ring with $ L $ sites reads
\begin{equation}
H(h)=-J\sum _{j=1}^{L}\left( \sigma _{j}^{x}\sigma _{j+1}^{x}+h\sigma _{j}^{z}\right)\ ,
\label{HTFIM}
\end{equation}
where $ \sigma _{j}^{\alpha }$ acts like the corresponding Pauli matrix at sites $ j
$ and like the identity elsewhere. We assume $ J,h >0 $ and consider periodic boundary
conditions. We refer the reader to Appendix A of \cite{CEF1} for
details about the diagonalization of this Hamiltonian. We simply
recall here that it can be expressed in terms of free fermions $
\alpha _{k}$ as  
\begin{equation}
H(h)=\sum _{k}\varepsilon(k) \bigg(\alpha _{k}^{\dagger}\alpha _{k}-\frac{1}{2}\bigg)\,, \qquad \varepsilon(k) =2J\sqrt {1+h^{2}-2h\cos k}\ ,
\end{equation}
and that the Hilbert space is divided into a Neveu Schwartz (NS) sector with states of the form
\begin{equation}
  |q_{1}\ldots q_{2n}\rangle =\alpha ^\dagger_{q_{1}}\ldots \alpha ^\dagger_{q_{2n}}|0\rangle _{\rm NS}\,, \qquad q_{i}=\frac {2\pi } {L}\left( n_{i}+\frac {1} {2}\right) \,, \qquad n_{i}=-\frac {L} {2},\ldots ,\frac {L} {2}-1\ ,
\end{equation}
and a Ramond (R) sector

\begin{equation}
  |p_{1}\ldots p_{2m+1}\rangle =\alpha ^\dagger_{p_{1}}\ldots \alpha ^\dagger_{p_{2m+1}}|0\rangle _{\rm R}\,, \qquad p_{i}=\frac {2\pi n_{i}} {L} \qquad n_{i}=-\frac {L} {2},\ldots ,\frac {L} {2}-1\ ,
\end{equation}
where the energy $ E\left( \left\{ q_{i}\right\} \right) $ and momentum $ P\left( \left\{ q_{i}\right\} \right) $ of such states are given by
\begin{equation}
E\left( \left\{ q_{i}\right\} \right)=\sum _{i=1}^{2n}\varepsilon \left( q_{i}\right) \,, \qquad P\left( \left\{ q_{i}\right\} \right)=\sum _{i=1}^{2n} q_{i}\ ,
\end{equation}
with an identical relation for the Ramond sector.

We will be interested in two problems involving the summation of form
factors of the order parameter over the full Hilbert space, which are given
by \cite {Bugrij,BL03,Gehlen,iorgov11} 

\begin{equation}
\label {ff}
\begin{aligned}
&{}_{\rm NS}\langle q_1,...,q_{2n}\vert\sigma_\ell^x\vert p_1...p_m\rangle_{\rm R}=e^{-i\ell(\sum_{j=1}^{2n}q_j-\sum_{l=1}^mp_m)}i^{\lfloor n+m/2\rfloor}(4J^2h)^{(m-2n)^2/4} \sqrt{\xi\xi_L}\\\times&\prod_{j=1}^{2n}\left(\frac{e^{\eta_{q_j}}}{L\varepsilon(q_j)}\right)^{1/2}
\prod _ { l= 1 } ^ {m}\left( \frac{ e ^ { - \eta_ {p_l}}}{ L\varepsilon(p_l)} \right)^{1 / 2 }\
 \prod_{ j < j'}^ { 2n}\frac{ \sin \frac{ q_ j - q _ {j ^\prime} }{ 2 }}{ \varepsilon_ { q_ j q_ {j ^\prime} }}\prod _ {l < l ^\prime } ^ { m}\frac{ \sin \frac{p_ l - p _ {l'} }{ 2 }}{ \varepsilon _ { p_ l p _ {l ^\prime} }}\prod _ { j = 1 } ^ { 2 n}\prod _ { l = 1} ^ {m}\frac{\varepsilon _ { q_  jp _l }}{\sin \frac{ q _ j - p _l }{ 2 }}   \ .
\end{aligned}
\end{equation}
Here $\xi=|1-h^2|^{1/4}$, $m$ is even (odd) for $h <1 $ ($h>1$), and
\be
\varepsilon_{ab}=\frac{\varepsilon(a)+\varepsilon(b)}{2}\ .
\label{veab}
\ee
The terms $\xi_L$ and $e^{\eta_k}$ do not depend on the momenta (except
$k$) and for large $L$ approach $1$ with exponential accuracy 
\begin{equation} 
\xi_L\approx 1\,,\quad e^{\eta_k}\approx 1\, ;
\end{equation}
thus, they will be set to $1$ in the following.

%%%%%%%%%%%%%%%%%%%%%%%%%%%%%%%%%%%%%%%%%%%%%%%%%%%%%%%%%%%%%%%%%%%%%
\subsection {Quenches in the quench action framework \label{quench}}
%%%%%%%%%%%%%%%%%%%%%%%%%%%%%%%%%%%%%%%%%%%%%%%%%%%%%%%%%%%%%%%%%%%%%

We consider the following quantum quench setup \cite{CEF1,CEF2}: at
time $t=0$ we prepare the system in the ground state of the TFIM
\fr{HTFIM} at a magnetic field $h_0<1$
\be
|\Psi\rangle=|0;h_0\rangle_{\rm NS}\ .
\ee
At times $t>0$ we evolve the system with Hamiltonian $H(h)$ with
$h_0\neq h<1$. As $|\Psi\rangle$ is not an eigenstate of $H(h)$ this
results in interesting dynamics. The order parameter one-point
function at time $t>0$ is given by \fr{QuenchA}, where the
representative state $|\Phi_s\rangle$ is characterized by the root
density \cite{CE13}
%Quantum quenches consist in studying the time evolution of local observables such as the order parameter $ \sigma ^x $ when at time $ t =0 $ the initial state is prepared in the ground state of the Hamiltonian at another magnetic field $ h _0 $; we refer the reader to \cite {CEF1,CEF2,CEF3} for an extensive study of them in the TFIM case. \\
%
%Among the different approaches to study quantum quenches a particularly appealling one is the quantum quench action \cite{CE13}. In this framework the full time evolution of a local observable $ \left\langle O\left( t\right) \right\rangle $ is given by the scalar product of the state of the system $ |\psi \rangle $ and a representative 'saddle point' eigenstate $ |\phi _{s}\rangle $
%\begin{equation}
%  \lim _{L\rightarrow \infty }\left\langle O\left( t\right) \right\rangle= \lim _{L\rightarrow \infty }\left(\frac {\left\langle \psi \left| O\left( t\right) \right| \phi _{s}\right\rangle } {2\left\langle \psi  |\phi _{s}\right\rangle }+\frac {\left\langle \phi _{s}\left| O\left( t\right) \right| \psi \right\rangle } {2\left\langle \phi _{s}|\psi \right\rangle }\right)\ .
%\end{equation}
%In the TFIM case the root density of the saddle point eigenstate in formula \eqref{quenchfirst} is known to be \cite {CE13}
\begin{align}
  \rho \left( k\right) &=\frac {1-\cos \Delta _{k}} {4\pi }\ ,\nn
  \cos \Delta _{k}&=\frac {hh_{0}-\left( h+h_{0}\right) \cos k+1} {\sqrt {1+h^{2}-2h\cos k}\sqrt {1+h_{0}^{2}-2h_{0}\cos k}}\ .
\end{align}
For later convenience we define the density of particles in the
representative state
\be
\rho_Q=\int_{-\pi}^\pi  \rho(x)dx\ ,
\ee
where the index $Q$ stands for 'quench'.
In a large finite volume $L$ we may choose \cite{CE13}
\be
|\Phi_s\rangle=\vert q_1,-q_1,...,q_N,-q_N\rangle_{\rm NS}\ ,
\ee
where the momenta $q_i$ are distributed according to the root density $\rho(k)$.
The time-evolved initial state is given by

\begin{equation}
  |\psi \left( t\right) \rangle =\prod _{p>0}\frac {1+ie^{-2it\varepsilon (p) }K(p) \alpha _{-p}^{\dagger}\alpha _{p}^{\dagger}} {\sqrt {1+K^{2}(p) }}|0\rangle \ ,
\end{equation}
with $ K(p) =\tan (\Delta _{p} / 2)$.

Equation \fr{QuenchA} thus provides the following representation for the
order parameter one-point function 
\begin{equation}
\label{bigsum}
\begin{aligned}
\left\langle \sigma^{x}_{\ell}\left( t\right) \right\rangle ={\rm Re}\Bigg[&
\sum _{M=0}^{\infty }\frac{(-1)^{M}}{i^{M-N}M!}\sum_{\substack{0<p_1,...,p_M\\\in \rm R}}{}_{\rm R}\langle
p_1,-p_1,...,p_M,-p_M\vert\sigma_{\bell}^x\vert q_1,-q_1,...,q_N,-q_N\rangle_{\rm NS}\\
&\qquad\qquad\qquad\qquad\qquad\times\ \prod_{j=1}^N\frac{e^{-2it\varepsilon(q_j)}}{K(q_j)}\prod_{j=1}^M K(p_j)e^{2it\varepsilon(p_j)}\Bigg]\ ,
\end{aligned}
\end{equation}
that is a sum of form factors over states that are expressed in terms of pairs of momenta. 
%%%%%%%%%%%%%%%%%%%%%%%%%%%%%%%%%%%%%%%%%%%%%%%%%%%%%%%%%%%%%%%%%%%%%%%%%%%%%
\subsection {Dynamical correlation functions at finite temperature}
\label{temp}
%%%%%%%%%%%%%%%%%%%%%%%%%%%%%%%%%%%%%%%%%%%%%%%%%%%%%%%%%%%%%%%%%%%%%%%%%%%%%
%Dynamical two-point  function of the order parameter at finite inverse temperature $ \beta $ is  defined as
%\begin{equation}
%  \chi^{xx}\left( l,t\right) =\frac {\tr\left( \sigma _{j+l}^{x}\left( t\right) \sigma _{j}^{x}\left( 0\right) e^{-\beta H}\right) } {\tr\left( e^{-\beta H}\right) }\ .
%\end{equation}
%The rationale we will be using is that in the scaling limit the trace over the full Hilbert space can be replaced by an expectation value between one representative state $ |q_{1},\ldots ,q_{N}\rangle _{NS}$  that dominates at this temperature, ie. whose root density solves the TBA equations. 

In the TFIM, the density of momenta $q$ of the representative state $|E_\beta\rangle$ in \eqref{Lehmann2} is
\begin{equation}
\label {densitytemp}
  \rho (q) =\frac {1} {2\pi }\frac {1} {1+e^{\beta \varepsilon (q) }}\ .
\end{equation}
For later convenience we define the corresponding density
\be
\rho_\beta=\int_{-\pi}^\pi  \rho(x)dx\ .
\ee

In practice a representative state is constructed from $\rho(q)$ as
follows. We first construct the particle counting function $z(q)$ by
integrating the root density
\be
z(q)=\int_{-\pi}^q \rho(y)dy\ .
\ee
We then solve the equations
\be
z(q^{(0)}_j)=\frac{2\pi j}{L}\ ,\quad j=1,\dots,N\ ,
\ee
where $N$ is fixed by the requirement that
$|q^{(0)}_j|\leq\pi$. Finally we set 
\be
q_j=\frac{2\pi}{L}\left(\Big\lfloor\frac{L}{2\pi}q_j^{(0)}-\frac{1}{2}\Big\rfloor+\frac{1}{2}\right)\ ,\quad
j=1,\dots,N.
\ee

Inserting a resolution of the identity between the two spin operators
in \eqref{Lehmann2} leads to the following spectral representation
\begin{equation}
\label {armtap}
   \chi^{xx}\left(\bell,t\right) =\sum _{M=0}^{+\infty }\frac {1}
       {M!}\sum _{\substack{p_{1} ,  \ldots, p_{M}\\ \in \rm R}}\vert
       \langle p_1,...,p_M\vert\sigma_l^x\vert
       q_1,...,q_{N}\rangle\vert ^2 e^{it(E\left( \left\{ q\right\}
         \right) -E\left( \left\{ p\right\} \right) )+ i\bell(P\left(
         \left\{ p\right\} \right) -P\left( \left\{ q\right\} \right)
         )}\ .
\end{equation}
The terms in the sum depend on the regime of the TFIM: in the ordered
phase, $ h <1 $, $M$ has the same (even/odd) parity as $N$, whereas in the
disordered phase, $ h >1 $, it has opposite parity. Moreover, in contrast
to the quench case, \fr{armtap} involves modulus squares of form factors, and
the intermediate states do not have a structure where momenta only
appear in pairs $\{-p_i,p_i\}$.

%%%%%%%%%%%%%%%%%%%%%%%%%%%%%%%%%%
\section{Systematic approach to form factor expansions for
  semi-local operators \label{pfd}} 
%%%%%%%%%%%%%%%%%%%%%%%%%%%%%%%%%%
In this section we present a general framework for carrying out 
the form factor sums \eqref {bigsum} and \eqref{armtap} analytically
at late times $Jt\gg 1$. It is based on decomposing 
the form factors \eqref {ff} into partial fractions so that the sums
over the $p$'s decouple and can be evaluated exactly. The key
observation is then that an oscillatory sum with a pole of order $d$ 
like $\sum _{n}\frac {e^{int}} {(n+1/2)^{d}}$ grows as $t^{d-1}$, so
that the leading poles give the leading time behaviour, and the terms
in the partial fraction decomposition can be organized according to
the total number of poles. This naturally leads to an expansion in
 the number of particles per unit site -- $N/L$ -- in the representative
state, which has already been proven very efficient for simpler
quantities such as the free energy in Bethe ansatz solvable
interacting models \cite{granetjacobsensaleur,ristivojevic}. 

In Sections \ref{hm1} and \ref{different} we consider the
application of this framework in the context of the finite temperature
case \eqref {armtap}, due to the more canonical sum over form factors
that it involves. Since the form factors differ for $h<1$, $h>1$ and $h=1$,
we will treat these cases separately.

We will be interested in large time or space asymptotics of
correlation functions, generically defined as requiring the phase
$it(E\left( \left\{ q\right\} \right) -E\left( \left\{ p\right\}
\right) )+ i\ell(P\left( \left\{ p\right\} \right) -P\left( \left\{
q\right\} \right) )$ in \eqref{armtap} to be large. This is in
particular the case of the large time and distance asymptotics 
at fixed
\begin{equation}
\alpha =\frac{t}{\ell}\ ,
\end{equation}
on which we will focus. However, the static correlations case $t=0$ and large $\ell$ is also covered by our calculations; we refer the reader to Section \ref{static} for details on this case. For later convenience we introduce the following notations
\begin{align}
\overline {\varepsilon }(x) &=\varepsilon (x)
-\frac {x} {\alpha }\ , \nn
v_{\rm max}&=\underset{|x|\leq\pi}{\max} \,\varepsilon'(x)\ ,
\label{vmax}
\end{align}
where $v_{\rm max}$ is the maximal group velocity of the elementary
fermion excitations in the TFIM. According to whether there exists an
$x_0$ such that $\overline {\varepsilon }'\left( x_0\right)=0$
('time-like region', $tv_{\rm max}>\bell$ for $t,\bell\geq 0$) or not
('space-like region', $tv_{\rm max}<\bell$ for $t,\bell\geq
0$), the next-to-leading terms in our expansions differ. We will
in the following compute these terms in the space-like region.
Their calculation in the time-like region is more involved and
  will be reported elsewhere.

In Section \ref{qu} we briefly present the application of our framework to
the dynamics after quantum quenches.

\subsection{\texorpdfstring{    Case $h < 1$: identical number of
    particles}{Lg} \label{hm1}}
In this subsection we treat the case $h < 1$, for which the sum
\eqref{armtap} includes intermediate states with the same number $N$ of 
particles as the representative state. We will show in Section
\ref{different} that the contributions of intermediate states with different
particle numbers $M\neq N$ is always subleading in time. We exploit this result right away and  focus on states with $M=N$ in this subsection.
%An important distinction has to be made according to whether there exists a saddle point $\overline {\varepsilon }^{'}\left( x_{0}\right) =0$, case that we will denote  'time-like regime', or whether $\overline {\varepsilon }\left( x\right) $ is a monotonous function $\forall x,\overline {\varepsilon }'\left( x\right) \neq 0$, case that we will denote  'space-like regime'. Indeed, in the space-like regime any oscillatory integral over a bounded function decays exponentially fast with time and can be fully neglected; whereas in the time-like region these integrals will decay as $\frac {1} {\sqrt {t}}$ according to the stationary phase approximation, and will thus contribute to 'relevant' sub leading corrections. The existence of a saddle point also directly affects the values of the sums over $p$'s below in \eqref {} since their leading contribution vanishes at them. For these reasons, we will treat separately the space-like and time-like regimes. 

%\subsubsection{Generalities}
\subsubsection{Partial fraction decomposition of form factors}
We recall that the partial fraction decomposition of a ratio of two
polynomials $\frac {P(X) } {\prod _{i}\left(X-x_{i}\right)
  ^{a_{i}}}$ with distinct $x_i$'s is the writing   
\begin{equation}
\label{pfdgen}
  \frac {P(X) } {\prod _{i}\left( X-x_{i}\right) ^{a_{i}}}=P_{0}(X) +\sum _{i}\sum _{\nu =1}^{a_{i}}\frac {B_{i,\nu}} {\left( X-x_{i}\right) ^{\nu }}\ ,
\end{equation}
with $P_{0}(X) $ a polynomial of degree $\text{deg}(P)-\sum _{i}a_{i}$ and $B_{i,\nu }$ independent of $X$, given by $B_{i,\nu }=\frac {1} {\left( a_{i}-\nu\right) !}\left( \frac {d} {dX}\right) ^{a_{i}-\nu }(P(X) \left( X-x_{i}\right) ^{a_{i}})\vert_{X=x_i}$.

The squared form factor appearing in \eqref{armtap} can be written as
\begin{equation}
\label{ff2}
\vert {}_{\rm NS}\langle q_1,...,q_{N}\vert\sigma_l^x\vert p_1,...,p_N\rangle_{\rm R}\vert ^2=\frac{\xi}{L^{2N}}F_{\{p_i\}}^{\{q_i\}}
\end{equation}
with\footnote{In $F_{U}^{V}$ the sets $U$ and $V$ can contain
  arbitrary momenta and are not meant to be each in the sectors $R$
  and $NS$ (this freedom will be indeed useful in Section
  \ref{recsect} below). For this reason we impose in the denominator
  that the momenta are different $u\neq v$, which is automatically
  satisfied if they are in different sectors, but not otherwise. } 

\begin{equation}
\label{fuv}
F_{U}^{V}=
\frac {\bigg|\displaystyle{\prod _{u\neq u'\in U}}\sin \left( \frac
  {u-u'} {2}\right)\prod _{v\neq v'\in V}\sin \left( \frac {v-v'}
  {2}\right)\bigg|} {\displaystyle\prod _{u\neq v\in U,V}\sin^2 \left(
  \frac {u-v} {2}\right)}
\frac {\displaystyle\prod _{u,v\in U,V}\varepsilon
  _{uv}^{2}} {\displaystyle\prod _{u,u'\in U}\varepsilon _{uu'} \prod _{v,v'\in
    V}\varepsilon _{vv'}}\ . 
\end{equation}

\begin{figure}[H]
\begin{center}
\begin{tikzpicture}[scale=1.3]
\draw[->] (0,0.85) -- (0,0.15);
\draw[->] (1,0.85) -- (1,0.15);
\draw[->] (2,0.85) -- (2,0.15);
\draw[->] (4,0.85) -- (4,0.15);
\filldraw[black] (0,0) circle (2pt);
\filldraw[black] (1,0) circle (2pt);
\filldraw[black] (2,0) circle (2pt);
\filldraw[black] (3,0) circle (2pt);
\filldraw[black] (4,0) circle (2pt);
\filldraw[red] (0,1) circle (2pt) node[above,black] {\tiny $\nu_1=2$};
\filldraw[red] (1,1) circle (2pt)node[above,black] {\tiny $\nu_2=2$};
\filldraw[red] (2,1) circle (2pt)node[above,black] {\tiny $\nu_3=1$};
\filldraw[red] (3,1) circle (2pt)node[above,black] {\tiny $\nu_4=0$};
\filldraw[red] (4,1) circle (2pt)node[above,black] {\tiny $\nu_5=2$};
\end{tikzpicture}
\hspace{1.5cm}
\begin{tikzpicture}[scale=1.3]
\draw[->] (0,0.85) -- (0,0.15);
\draw[->] (0.9,0.9) -- (0.1,0.1);
\draw[->] (2.1,0.9) -- (2.9,0.1);
\draw[->] (2.9,0.9) -- (2.1,0.1);
\draw[->] (4,0.85) -- (4,0.15);
\filldraw[black] (0,0) circle (2pt);
\filldraw[black] (1,0) circle (2pt);
\filldraw[black] (2,0) circle (2pt);
\filldraw[black] (3,0) circle (2pt);
\filldraw[black] (4,0) circle (2pt);
\filldraw[red] (0,1) circle (2pt) node[above,black] {\tiny $\nu_1=1$};
\filldraw[red] (1,1) circle (2pt)node[above,black] {\tiny $\nu_2=1$};
\filldraw[red] (2,1) circle (2pt)node[above,black] {\tiny $\nu_3=2$};
\filldraw[red] (3,1) circle (2pt)node[above,black] {\tiny $\nu_4=1$};
\filldraw[red] (4,1) circle (2pt)node[above,black] {\tiny $\nu_5=2$};
\end{tikzpicture}
\end{center}
\caption{Sketch of two examples of a function $f$ from $p$'s in red to $q$'s in black.}
\label{fig:exf}
\end{figure}
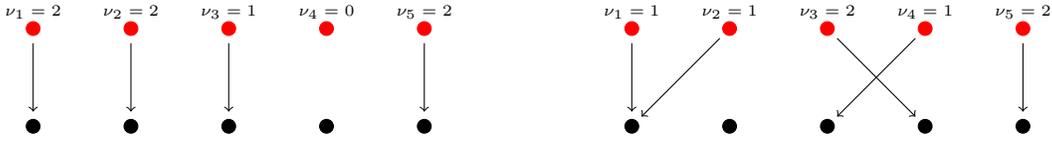

The $\varepsilon(p_1)$ factors are not polynomial in $p_1$ but are
nevertheless bounded and without zeros. Seen as a function of $p_1$,
the square of the form factor can thus be written $\sum _{i}\frac {A_{i}} {\sin ^{2}\left( \frac
  {p_{1}-q_{i}} {2}\right) }+\frac {B_{i}} {\sin \left( \frac
  {p_{1}-q_i} {2}\right) }+C$ with $A_{i},B_{i}$ independent of $p_1$
and $C$ a bounded function of $p_1$. Repeating the operation for the
other momenta, one can write 

\begin{equation}
\label{decomp}
\vert {}_{\rm NS}\langle q_1,...,q_{N}\vert\sigma_l^x\vert
p_1,...,p_N\rangle_{\rm R}\vert ^2= \frac {\xi } {L^{2N}}
\sum_{\nu_1,\dots,\nu_N=0}^2
\sum_{\{f_{\vec{\nu}}\}}\frac {\mathcal{A}\left( \left\{ q \right\}
  ,\{p\},\left\{ \nu \right\},
f_{\vec{\nu}}\right) } {\displaystyle\prod _{j=1}^N\sin ^{\nu_{j}}\left( \frac {p_{j}-q_{f_{\vec{\nu}}\left( j\right)} } {2}\right) }\ ,
\end{equation}
where the second sum is over a complete set of functions $f_{\vec{\nu}} :\{i\in \left\{ 1,\ldots ,N\right\}|\nu_i\neq 0\} \mapsto
\left\{ 1,\ldots ,N\right\} $, and where
$\mathcal{A}\left( \left\{ q \right\} ,\{p\},\left\{ \nu
\right\},f_{\vec{\nu}}\right)$ is a bounded function of $p_j$ if $\nu_j=0$, and
independent of $p_j$ otherwise. In Fig.~\ref{fig:exf} we show examples
of such functions $f_{\vec{\nu}}$. The important feature of \fr{decomp} is that each
$p$ appears at most once (however, the $q$'s may appear several
times).

\subsubsection{\texorpdfstring{Carrying the sum over the momenta $p_i$}{Lg}}

Let us briefly anticipate the method that we will use to carry out the sum over $p$ in \eqref {armtap}. If
there is a $\nu_i=0$ then the sum over $p_i$ is an oscillatory Riemann
sum of a bounded function, hence it decays to  zero with time. Thus the
leading behaviour  is obtained for  $\nu_i>0$ for all $i$. Then (and
only in this case) the coefficients $\mathcal{A}\left( \left\{ q
\right\} ,\{p\},\left\{ \nu \right\},f_{\vec{\nu}}\right)$ are independent of the
$p$'s and the maps will no longer depend on $\{\nu_j\}$.
These coefficients will be denoted by $A\left( \left\{ q \right\} ,\left\{ \nu 
\right\},f\right)$ and are obtained as
\begin{equation}
\label{aaa}
 A\left( \left\{ q \right\} ,\left\{ \nu \right\},f\right)=\left[\left(\prod _{j}\left( 2\frac {d} {dp_{j}}\right) ^{2-\nu _{j}}\sin ^{2}\left( \frac {p_{j}-q_{f(j)}} {2} \right) \right)F_{\{p_i\}_{i=1,...,N}}^{\{q_i\}_{i=1,...,N}} \right]\Bigg\vert_{\{p_{j}=q_{f\left( j\right) }\}}\ .
\end{equation}
This follows from the partial fraction decomposition, with a factor
$2$ because of the $1/2$ inside the sinus. 

From here on we will only consider terms in the partial fraction
decomposition such that $\nu_i>0$ for all $i$, so that \fr{aaa}
applies. We denote the corresponding contribution to the spectral
representation \eqref{armtap} of $\chi^{xx}(\bell,t)$ by $S$
\begin{align}
S=\frac{\xi}{N!L^{2N}}\sum_{\substack{p_1,...,p_N\\\in
    \rm{R}}}&\Bigg\{
\bigg[\sum_{\nu_1,\dots,\nu_N=1}^2
\sum_{\{f\}}\frac {A\left( \left\{ q \right\} ,\left\{ \nu \right\},f\right) } {\displaystyle\prod _{j=1}^N\sin ^{\nu_{j}}\left( \tfrac {p_{j}-q_{f\left( j\right)} } {2}\right) }\bigg]\nn
&\qquad\times\ e^{it(E\left( \left\{ q\right\} \right) -E\left( \left\{ p\right\} \right) )+ i\bell(P\left( \left\{ p\right\} \right) -P\left( \left\{ q\right\} \right) )}\Bigg\},
\end{align}
where the third sum is over a complete set of functions $f:\{1,...,N\}\mapsto\{1,...,N\}$.

In this form one can perform the sums over $p_j$ using the following
relations proven in Appendix \ref{appen} 

\begin{align}
\chi_1(q)&\equiv\sum _{p\in \rm{R}}\frac {e^{-it\big(\overline{\varepsilon }(p)-\overline{\varepsilon }(q)\big)
}} {L\sin \left( \frac {p-q} {2}\right) }=-i \sign
(t\overline{\varepsilon }'(q) )+\mathcal{O}(L^0t^{-1/2})\ ,\\
\chi_2(q)&\equiv\sum _{p\in \rm{R}}\frac {e^{-it\big( \overline{\varepsilon } \left(
    p\right) - \overline{\varepsilon } \left(q\right) \big)}}{L^2\sin^2  \left( \frac {p-q} {2}\right)}
=1-\frac{2\left| t \overline{\varepsilon } '(q)  \right| }{L}+\mathcal{O}(L^{-1}t^{-1/2})\ ,
\label{sumsval}
\end{align}
with $q\in\rm{NS}$, to obtain
\begin{equation}
S= \frac {\xi } {N!}\sum_{\nu_1,...,\nu_N=1}^2
\sum_{\{f\}} A\left( \left\{ q \right\} ,\left\{ \nu \right\},f\right)\prod_{i=1}^N\frac{\chi _{\nu_i}(q_{f(i)})}{L ^{2-\nu_i}}e ^{it(\overline{\varepsilon}(q_{i })-\overline{\varepsilon}(q_{f(i) }))}\ .
\end{equation}

Equations \eqref{sumsval} are valid only when
$\overline{\varepsilon}'(q)\neq 0$. If there is a point where
$\overline{\varepsilon}'(q)=0$, i.e. if we are in the time-like region,
 corrections in time to
\eqref{sumsval} have to be taken into account, and they are expected
to modify significantly the subleading corrections in the correlation
function. We leave this matter of discussion for future work. 

\subsubsection {\texorpdfstring{Constraints on the functions $f$}{Lg}}

The set of functions $f$ over which we need to sum is actually quite
constrained. First, since $F^{\{q_i\}_{i=1,\dots
    N}}_{\{p_i\}_{i=1,\dots N}}=0$ whenever $p_{i}=p_{j}$ we must have  
$f\left( i\right) \neq f\left( j\right) $ whenever
$\nu_{i}=\nu_{j}=2$.  

We also need $f\left( i\right) \neq f\left( j\right) $ if $\nu_i=1$
and $\nu_j=2$. Indeed, if $f(i)=f(j)$ then in \eqref{aaa}
there is a $\sin^2 \tfrac{p_i-p_j}{2}$ factor in the numerator of
$F^{\{q_n\}_{n=1,\dots  ,N}}_{\{p_n\}_{n=1,\dots
      ,N}}$, but as there is only one derivative with respect to $p_i$
and none with respect to $p_j$ this factor will make the coefficient
$A(\{q\},\{\nu\},f)$ vanish upon taking
  $p_i=q_{f(i)}=q_{f(j)}=p_j$.

More generally, if $f$ takes $k$ times the same value at points with $\nu_i=1$, then all the $k(k-1)/2$ terms $\sin^2 \tfrac{p_i-p_j}{2}$ contribute to a zero of order $k(k-1)$; since the number of derivatives is equal to $k$, we must have $k=2$. 

These arguments show that the sum  over $f$ can be
replaced by a sum over three disjoint subsets $I_0, I_1,
I_2\subset\{1,\ldots,N\}$, where $I_k$ is the set of points with
$\nu=1$ attained $k$ times by $f$. The remaining points
$\{1,...,N\}-(I_0\cup I_1\cup I_2)$ all have $\nu=2$. There is a
combinatorial factor $\frac {N!} {2^{\left| I_{2}\right| }}$
corresponding to the number of such functions with this precise
ouput. It follows that  $A\left( \left\{ q \right\} ,\left\{ \nu
\right\},f\right)$ depends only on the sets $I_{0},I_1,I_2$ and we have 
\begin{equation}
\begin{aligned}
\label{mainsum}
 S=\xi \sum _{\substack{I_{0}, I_1, I_2 \subset\{1,\ldots,N\}\\\left| I_{0}\right| =\left| I_{2}\right| , \text{ all disjoint}}}A (I_{0}, I_1, I_2)2 ^{-\left| I_{2}\right| }&e ^{it(\sum _{i\in I_0}\overline {\varepsilon }\left( q_{i}\right) -\sum _{i\in I_2}\overline {\varepsilon }\left( q_{i}\right) )}\\
 \times&\prod _{i\in I_{1}}\frac {\chi_1\left( q_{i}\right) } {L} \prod _{i\in I_{2}}\frac {\chi_1^2\left( q_{i}\right) } {L^2} \prod _{i\notin I_{0, 1,2}}\chi_2\left( q_{i}\right)\ .
\end{aligned}
\end{equation}

The expression for the coefficients $A (I_{0},I_1, I_2)$ can be
simplified as follows. We observe that, whenever we have $\nu_i=2$ in
\fr{aaa}, the various factors depending on $p_i$ and $q_{f(i)}$ precisely
compensate one another. Hence we can work with a reduced form factor
involving only momenta in $I_0,I_1,I_2$
\begin{equation}
\label{Agen}
A\left(I_0,I_1,I_2\right)=\left[\left(\prod _{j=1}^{n+2m}\left( 2\frac
  {d} {dp_{j}}\right) \sin ^{2}\left( \frac {p_{j}-q_{f(j)}} {2}
  \right) \right) F_{\{p_i\}_{i=1,...,n+2m}}^{\{q_i\}_{i\in I_0\cup
      I_1\cup I_2}} \right]\Bigg\vert_{\{p_{j}=q_{f\left( j\right)
  }\}}\ , 
\end{equation}
for any function $f$ such that $\{f(i)\}_{i=1,...,n}=I_1$,
$\{f(i)\}_{i=n+1,...,n+m}=I_2$, 

\noindent and $\{f(i)\}_{i=n+m+1,...,n+2m}=I_2$. The set $I_2$ does
appear twice by construction, and $I_0$ does not appear at all. The
decomposition of $\{1,...,n+2m\}$ into $\{1,...,n\}$, $\{n+1,...,n+m\}$,
$\{n+m+1,...,n+2m\}$ is arbitrary, and it needs only to involve one set
with $n$ elements and two sets with $m$ elements.\\ 

The sum of all the terms in \eqref{mainsum} with
$|I_1|=n,|I_0|=|I_2|=m$ will be denoted by $S_{n,2m}$. We can factorize
$S_{0,0}$ and, using the explicit expression for $\chi_1(q)$ and
$\chi_2(q)$, write

\begin{equation}
\begin{aligned}
\label{mainsum2}
S_{n,2m} =\frac{(-i)^nS_{0,0}}{(-2)^mL^{n+2m}}
\sum _{\substack{I_{0,1,2} \subset\{1,\ldots,N\}\\
\left| I_{0}\right| =\left| I_{2}\right|=m\\
 |I_1|=n \\ \text{all disjoint}}}&A (I_{0}, I_1, I_2)
\frac{\displaystyle\prod_{i\in
    I_1}\sign(t\overline{\varepsilon}'(q_i))}{\displaystyle\prod_{i\in
    I_{0,1,2}}
\big(1-\tfrac{2|t\overline{\varepsilon}'(q_i)|}{L}\big)}\nn
&\qquad\times e ^{it(\sum _{i\in
    I_0}\overline {\varepsilon }\left( q_{i}\right) -\sum _{i\in
    I_2}\overline {\varepsilon }\left( q_{i}\right) )}\ .
\end{aligned}
\end{equation}
If $n$ and $m$ stay finite in the limit $L\to\infty$ we have
$\prod_{i\in
  I_{0,1,2}}\big(1-\tfrac{2|t\overline{\varepsilon}'(q_i)|}{L}\big)=1+\mathcal{O}(L^{-1})$
and hence 
\begin{equation}
\begin{aligned}
\label{snm}
S_{n,2m}=&\frac{(-i)^nS_{0,0}}{(-2)^mL^{n+2m}}\sum_{\substack{q_1^0<...<q_m^0\\q_1^1<...<q_n^1\\q_1^2<...<q_m^2\\\text{all distinct}}}A(\{q^0\},\{q^1\},\{q^2\})e^{it\sum_{i=1}^m\overline {\varepsilon }(q_i^0)-\overline {\varepsilon }(q^2_i) }\prod_{i=1}^n \sign(t\overline {\varepsilon }'(q_i^1))\\
&+\mathcal{O}(L^{-1})\ .
\end{aligned}
\end{equation}

Since the momenta selected by the sets $I_{0,1,2}$ are drawn from the
momenta $\{q_j|j=1,\dots,N\}$ of the representative state with density
$\rho$, the term $S_{n,2m}$ is of order $(N/L)^{n+2m}$ times
$S_{0,0}$. Hence this expansion naturally leads to an expansion in $N/L$.

\subsubsection{\texorpdfstring{Example: correlation function at
   ${\cal O}(\rho_\beta)$ uniformly in $t$ at large $t$}{Lg}} 
Let us give some examples. With $I_0=I_1=I_2=\{\}$ we have $A(I_0,I_1,I_2)=1$ hence the term
\begin{equation}
\begin{aligned}
\label{s00}
  S_{0,0}=\xi \prod _{i=1}^{N}\left(1-\frac {2\left| t
    \overline{\varepsilon } '\left( q_i\right) \right| } {L}\right)=
  &\xi\exp\left(-2\int _{-\pi }^{\pi }\left| t\overline{\varepsilon}
  '(x) \right| \rho ( x) dx\right)+\mathcal{O}(L^{-1})\ , 
\end{aligned}
\end{equation}
where, for a function $f(x)$, we used 
\begin{equation}
\underset{L\to\infty}{\lim}\prod_{k=0}^{L-1}\left(1+\frac{f(k/L)}{L}\right)=\exp \int_0^1f(x)dx\ .
\end{equation}
This term is the  correlation function at order $1$ in $\rho$ uniformly in $t$ at large $t$.\\
\begin{figure}[H]
\begin{center}
\begin{tikzpicture}[scale=1.3]
\draw[->] (0,0.85) -- (0,0.15);
\draw[->] (1,0.85) -- (1,0.15);
\draw[->] (2,0.85) -- (2,0.15);
\draw[->] (3,0.85) -- (3,0.15);
\draw[->] (4,0.85) -- (4,0.15);
\filldraw[black] (0,0) circle (2pt);
\filldraw[black] (1,0) circle (2pt);
\filldraw[black] (2,0) circle (2pt);
\filldraw[black] (3,0) circle (2pt);
\filldraw[black] (4,0) circle (2pt);
\filldraw[red] (0,1) circle (2pt) node[above,black] {\tiny $\nu_1=2$};
\filldraw[red] (1,1) circle (2pt)node[above,black] {\tiny $\nu_2=2$};
\filldraw[red] (2,1) circle (2pt)node[above,black] {\tiny $\nu_3=2$};
\filldraw[red] (3,1) circle (2pt)node[above,black] {\tiny $\nu_4=2$};
\filldraw[red] (4,1) circle (2pt)node[above,black] {\tiny $\nu_5=2$};
\end{tikzpicture}
\end{center}
\label{exf00}
\caption{Sketch of configurations contributing to $S_{0,0}$.}
\end{figure}
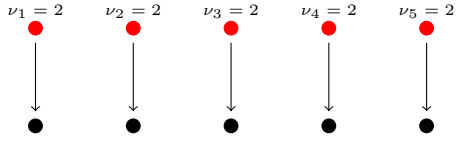

\subsubsection{\texorpdfstring{Example: correlation function at
    ${\cal O}(\rho_\beta^2)$ }{Lg} \label{rho2fixt}}
With $I_1=\{i,j\}$ and $I_0=I_2=\{\}$ we have 
\begin{equation}
A(I_0,I_1,I_2)=\frac {2} {\sin ^{2}\left( \frac {q_{i}-q_{j}} {2}\right) }+ \frac {8\varepsilon '\left( q_{i}\right)  \varepsilon'\left( q_{j}\right) } {\left( \varepsilon \left( q_{i}\right) +\varepsilon \left( q_{j}\right) \right) ^{2}}\ .
\end{equation}
This term leads to
\begin{equation}\label{vanish}
 S_{2,0}= -\frac {S_{0,0}} {L^{2}}\sum _{i< j}\left( \frac {2} {\sin ^{2}\left( \frac {q_{i}-q_{j}} {2}\right) }+ \frac {8\varepsilon '\left( q_{i}\right)  \varepsilon'\left( q_{j}\right) } {\left( \varepsilon \left( q_{i}\right) +\varepsilon \left( q_{j}\right) \right) ^{2}}\right)\sign( \overline{\varepsilon}'(q_j) \overline{\varepsilon}'(q_i))\ .
\end{equation}
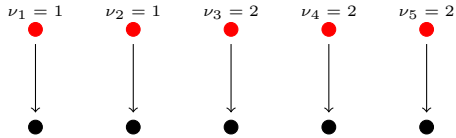
\begin{figure}[H]
\begin{center}
\begin{tikzpicture}[scale=1.3]
\draw[->] (0,0.85) -- (0,0.15);
\draw[->] (1,0.85) -- (1,0.15);
\draw[->] (2,0.85) -- (2,0.15);
\draw[->] (3,0.85) -- (3,0.15);
\draw[->] (4,0.85) -- (4,0.15);
\filldraw[black] (0,0) circle (2pt);
\filldraw[black] (1,0) circle (2pt);
\filldraw[black] (2,0) circle (2pt);
\filldraw[black] (3,0) circle (2pt);
\filldraw[black] (4,0) circle (2pt);
\filldraw[red] (0,1) circle (2pt) node[above,black] {\tiny $\nu_1=1$};
\filldraw[red] (1,1) circle (2pt)node[above,black] {\tiny $\nu_2=1$};
\filldraw[red] (2,1) circle (2pt)node[above,black] {\tiny $\nu_3=2$};
\filldraw[red] (3,1) circle (2pt)node[above,black] {\tiny $\nu_4=2$};
\filldraw[red] (4,1) circle (2pt)node[above,black] {\tiny $\nu_5=2$};
\end{tikzpicture}
\end{center}
\label{exf20}
\caption{Sketch of configurations contributing to $S_{2,0}$.}
\end{figure}

For the choice $I_1=\{\}$ and $I_0=\{i\},I_2=\{j\}$ we have 
\begin{equation}
A(I_0,I_1,I_2)=-\frac {2} {\sin ^{2}\left( \frac {q_{i}-q_{j}} {2}\right) }\ ,
\end{equation}
and
\begin{equation}
S_{0,2}=  \frac {S_{0,0}} {L^{2}}\sum _{i\neq j} \frac {e^{it( \overline{\varepsilon}(q_j)- \overline{\varepsilon}(q_i))}} {\sin ^{2}\left( \frac {q_{i}-q_{j}} {2}\right) }\sign( \overline{\varepsilon}'(q_j) \overline{\varepsilon}'(q_i))\ .
  \end{equation}
  \begin{figure}[H]
\begin{center}
\begin{tikzpicture}[scale=1.3]
\draw[->] (0,0.85) -- (0,0.15);
\draw[->] (1,0.9) -- (0.1,0.1);
\draw[->] (2,0.85) -- (2,0.15);
\draw[->] (3,0.85) -- (3,0.15);
\draw[->] (4,0.85) -- (4,0.15);
\filldraw[black] (0,0) circle (2pt);
\filldraw[black] (1,0) circle (2pt);
\filldraw[black] (2,0) circle (2pt);
\filldraw[black] (3,0) circle (2pt);
\filldraw[black] (4,0) circle (2pt);
\filldraw[red] (0,1) circle (2pt) node[above,black] {\tiny $\nu_1=1$};
\filldraw[red] (1,1) circle (2pt)node[above,black] {\tiny $\nu_2=1$};
\filldraw[red] (2,1) circle (2pt)node[above,black] {\tiny $\nu_3=2$};
\filldraw[red] (3,1) circle (2pt)node[above,black] {\tiny $\nu_4=2$};
\filldraw[red] (4,1) circle (2pt)node[above,black] {\tiny $\nu_5=2$};
\end{tikzpicture}
\end{center}
\label{exf02}
\caption{Sketch of configurations contributing to $S_{0,2}$.}
\end{figure}
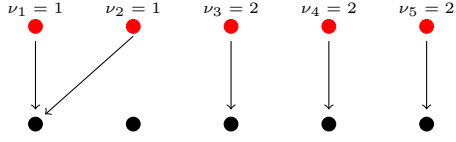
Although they are individually both divergent in $L$ in the scaling limit $L\to\infty$, their sum is not divergent and is, see Appendix \ref{appen},
\begin{equation}
S_{2,0}+S_{0,2}=-S_{0,0}\left(4\pi \int _{-\pi }^{\pi }\left| t\overline{\varepsilon} '(x) \right| \rho^2 ( x) dx+c\right)\ ,
\end{equation}
with the following value in the space-like regime where $\sign(\overline{\varepsilon}'(x))$ is constant
\begin{equation}
\begin{aligned}
\label{valc}
c=&2\int _{-\pi }^{\pi}\int _{-\pi }^{\pi}\frac {\rho ( y) \rho '( x) } {\tan \left( \frac {x-y} {2}\right) }dxdy\ .
\end{aligned}
\end{equation}
%\begin{equation}
%\begin{aligned}
%\label{valc}
%c=&2\int _{-\pi }^{\pi}\int _{-\pi }^{\pi}\frac {\rho ( y) \rho '( x) } {\tan \left( \frac {x-y} {2}\right) }\sign(\overline{\varepsilon } '( x) \overline{\varepsilon } '( y) )dxdy\\
%&\qquad\qquad+ 4\int _{-\pi }^{\pi}\int _{-\pi }^{\pi}\frac {\varepsilon '(x) \varepsilon '( y) \sign(\overline{\varepsilon } '( x) \overline{\varepsilon } '( y) )} {( \varepsilon ( x) +\varepsilon( y) ) ^{2}}\rho ( x) \rho ( y) dxdy\ .
%\end{aligned}
%\end{equation}
Contrarily to the previous case, this order $\rho^2$ of the correlation function at fixed large $t$ cannot be uniform in $t$, since it diverges for $t\to\infty$. In fact, the summation of other simple-pole contributions will lead to an exponentiation of this $\rho^2$ term and hence a correction of the exponent in the exponential decay.

We remark that in the very different context of the master equation approach for zero-temperature ground state correlations of a local operator in the XXZ spin chain, chains of double poles arising in some cycle integrals were observed to yield sub-leading exponential behaviours as well\cite{Kitanine}, which could suggest some yet not clear structural commonalities.

\subsubsection{\texorpdfstring{Recursive structure of $A(I_0,I_1,I_2)$}{Lg} \label{recsect}}
In the general case the amplitudes $A(I_0,I_1,I_2)$ are obtained from
\eqref{Agen}, but the sums over momenta associated with the index sets
$I_0,I_2$ in \eqref{snm} cannot be carried out as simply as in the
cases treated above. 
In fact, as we noted earlier, the derivatives corresponding to $I_2,I_0$
in \eqref{Agen}  have to be applied on the double zero
$\sin^2\frac{p_i-p_j}{2}$ in the numerator to give a non-vanishing
result, so that one actually has  

\begin{equation}
\label{rec}
A(I_0,I_1,I_2)=(-2)^{|I_2|}\bigg[\bigg(\prod_{i\in
    I_1}\left(2\frac{d}{dp_i}\right)\sin^2\frac{p_i-q_i}{2}\bigg)F_{\{p_i\}_{i\in
      I_1}\cup \{q_i\}_{i\in  I_2}}^{\{q_i\}_{i\in I_1\cup
      I_0}}\bigg]\Bigg|_{p_j=q_j,j\in I_1}\ . 
\end{equation}

The form factor $F_{\{p_i\}_{i\in I_1}\cup \{q_i\}_{i\in  I_2}}^{\{q_i\}_{i\in I_1\cup I_0}}$ can itself be decomposed into partial fractions. One obtains
\begin{equation}
\label{a2}
A(I_0,I_1,I_2)=(-2)^{|I_2|}\sum_{\substack{\nu_i\in\{0,1,2\}\\ i\in I_1\cup I_2\\ \nu_i=1\text{ if }i\in I_1}}\sum_{\substack{f:\{i\in I_1\cup I_2| \nu_i>0\}\\\mapsto I_1\cup I_0\\ f(i)=i \text{ if }i\in I_1}}\frac{A(\{q_i\}_{i\in I_1\cup I_0},\{\nu\},f)}{\prod_{j\in I_2}\sin^{\nu_j}\frac{q_j-q_{f(j)}}{2}}\ .
\end{equation}
We observe that the sum over $\{q^2\}$ in \eqref{snm} will play a
similar role to the sum over $p$'s in \eqref{armtap}, with however the
important difference that they are drawn from the original $q$'s of
the representative state and are not arbitrary momenta as is the case
for the $p$'s in \eqref{armtap}.

\subsubsection{\texorpdfstring{Partial fraction in $A(I_0,I_1,I_2)$ leading in density}{Lg}}
Relation \eqref{a2} reveals a recursive structure in the calculation
of $A(I_0,I_1,I_2)$. However, we will not develop this
recursion further here, but will rather focus on the leading partial
fractions of \eqref{a2} obtained with a set $\{\nu\}$ such that
$\nu_i=2$ for $i\in I_2$ and $\nu_i=1$ for $i\in I_1$, and functions
$f: I_1\cup I_2\mapsto I_1\cup I_0$  that map $I_2$ to $I_0$ in a
one-to-one fashion and fulfil $f(i)=i$ for $i\in I_1$.

 \begin{figure}[H]
\begin{center}
\begin{tikzpicture}[scale=1.3]
\draw[->] (0,0.85) -- (0,0.15);
\draw[->] (1,0.9) -- (0.1,0.1);
\draw[->] (2,0.85) -- (2,0.15);
\draw[->] (3,0.9) -- (3.9,0.1);
\draw[->] (4,0.85) -- (4,0.15);
\draw[->] (5,0.85) -- (5,0.15);
\draw[->] (6,0.85) -- (6,0.15);
\draw[->] (7,0.85) -- (7,0.15);
\filldraw[blue] (0,0) circle (2pt) node[below left, black]{\tiny $\nu'_1=2$};
\filldraw[black] (1,0) circle (2pt);
\filldraw[black] (2,0) circle (2pt);
\filldraw[black] (3,0) circle (2pt);
\filldraw[blue] (4,0) circle (2pt) node[below right, black]{\tiny $\nu'_2=2$};
\filldraw[black] (5,0) circle (2pt);
\filldraw[black] (6,0) circle (2pt);
\filldraw[blue] (7,0) circle (2pt) node[below right, black]{\tiny $\nu'_3=1$};
\filldraw[red] (0,1) circle (2pt) node[above,black] {\tiny $\nu_1=1$};
\filldraw[red] (1,1) circle (2pt)node[above,black] {\tiny $\nu_2=1$};
\filldraw[red] (2,1) circle (2pt)node[above,black] {\tiny $\nu_3=2$};
\filldraw[red] (3,1) circle (2pt)node[above,black] {\tiny $\nu_4=1$};
\filldraw[red] (4,1) circle (2pt)node[above,black] {\tiny $\nu_5=1$};
\filldraw[red] (5,1) circle (2pt)node[above,black] {\tiny $\nu_6=2$};
\filldraw[red] (6,1) circle (2pt)node[above,black] {\tiny $\nu_7=2$};
\filldraw[red] (7,1) circle (2pt)node[above,black] {\tiny $\nu_8=1$};
\filldraw[black] (1,-1) circle (2pt);
\filldraw[black] (3,-1) circle (2pt);
\filldraw[black] (7,-1) circle (2pt);
\draw[->] (0.1,-0.1) -- (0.9,-0.9);
\draw[->] (3.9,-0.1) -- (3.1,-0.9);
\draw[->] (7,-0.15) -- (7,-0.85);
\end{tikzpicture}
\end{center}
\label{exfrec}
\caption{Sketch of the two functions $f$ after one step of
  recursion. In red are indicated the $p$'s, in blue the $q$'s that
  play the role of the $p$'s after the first step of the recursion,
  that are those with index in $I_1$ or $I_2$. The 'new' $q$'s on the
  last row are those with index in $I_1$ or $I_0$.} 
\end{figure}
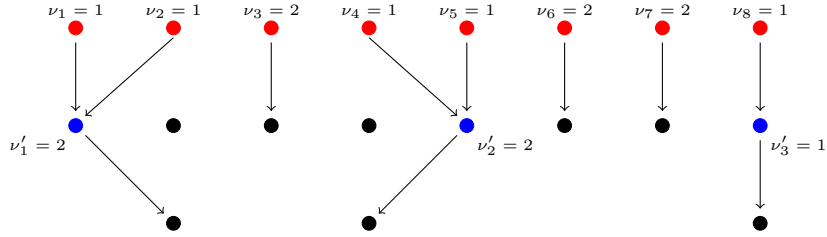

The coefficient then reads
\begin{equation}
A(\{q_i\}_{i\in I_1\cup I_0},\{\nu\},f)=\bigg[\bigg(\prod_{i\in I_1}\left(2\frac{d}{dp_i}\right)\sin^2\frac{p_i-q_{i}}{2}\bigg)F_{\{p_i\}_{i\in I_1}}^{\{q_i\}_{i\in I_1}}\bigg]\Bigg|_{p_i=q_{i},i\in I_1}\ .
\end{equation}
Let us select one $i\in I_1$ and introduce a reduced set
$I_1'=I_1-\{i\}$. Performing the derivative with respect to $p_i$ gives
\begin{equation}
\begin{aligned}
A(\{q_i\}_{i\in I_1\cup I_0},\{\nu\},&f)=\left(\prod_{k\in I_1'}\left(2\frac{d}{dp_k}\right)\sin^2\frac{p_k-q_{k}}{2}\right)\\
&2\left(\sum_{k\in I_1'}\tfrac{1}{\tan \frac{q_{i}-p_k}{2}}-\tfrac{\varepsilon'(q_{i})}{\varepsilon_{q_{i}p_k}}-\sum_{k\in I_1'}\tfrac{1}{\tan \frac{q_{i}-q_k}{2}}-\tfrac{\varepsilon'(q_{i})}{\varepsilon_{q_{i}q_k}}\right)F_{\{p_k\}_{k\in I_1'}}^{\{q_k\}_{k\in I_1'}}\Bigg|_{p_k=q_{k},k\in I_1'}\ .
\end{aligned}
\end{equation}
We observe that the first factor in the second line
%extra term in the middle
has to be differentiated precisely one more time for
the result not to vanish (the fact that $f$ is one-to-one on $I_1$ to
$I_1$ is essential for this), so that
%Moreover the extra term vanishes if it is differentiated twice more.
%Hence 
\begin{equation}
\begin{aligned}
A(\{q_i\}_{i\in I_1\cup I_0},\{\nu\},f)=\sum_{j\in I_1'}&\left(\frac{2}{\sin^2\frac{q_{i}-q_{j}}{2}}+2\frac{\varepsilon'(q_{i})\varepsilon'(q_{j})}{\varepsilon^2_{q_{i}q_{j}}}\right)\\
&\times\left[\left(\prod_{k\in I_1''}\left(2\frac{d}{dp_k}\right)\sin^2\frac{p_k-q_{k}}{2}\right)F_{\{p_k\}_{k\in I_1''}}^{\{q_k\}_{k\in I''_1}}\right]\Bigg|_{p_k=q_{k},k\in I_1''}\ ,
\end{aligned}
\end{equation}
with $I_1''=I_1-\{i,j\}$. Applying the same reasoning to the remaining momenta yields
\begin{equation}
\label{phia}
A(\{q_i\}_{i\in I_1\cup I_0},\{\nu\},f)=\sum_{P\text{ pairings of }I_1}\prod_{(i,j)\in P}\left(\frac{2}{\sin^2\frac{q_{i}-q_{j}}{2}}+8\frac{\varepsilon'(q_{i})\varepsilon'(q_{j})}{(\varepsilon(q_i)+\varepsilon(q_j))^2}\right)\equiv \phi(I_1)\ .
\end{equation}

\subsubsection {\texorpdfstring{Result: correlation function at
    ${\cal O}(\rho_\beta^2)$ uniformly in $t$ at large $t$}{Lg}}
In analogy with our notations for the first level of the recursive
structure we denote by $S_{n,2m|0,0}$ all contributions to \eqref{snm}
that arise by specifying $\nu_i=2$ for $i\in I_2$ in \eqref{a2}. We
observe that in \eqref{rec} the form factor vanishes if there are
coinciding momenta in $I_0$ and in $I_2$, or momenta that occur both in $I_1$
and $I_0$ or $I_2$. Hence one only has to impose that momenta in $I_1$
are distinct among themselves, and that momenta in $I_0$ are distinct
from those of $I_2$. This gives
\begin{equation}
\begin{aligned}
&S_{n,2m|0,0}=\frac{(-i)^nS_{0,0}}{n!(m!)^2L^{n+2m}}\times\\
&\sum_{\substack{q_1^1,...,q_n^1\\\text{all distinct}}}\sum_{\substack{q_1^0,...,q_m^0\\q_1^2,...,q_m^2\\q^0\text{ distinct from }q^2}}\phi(\{q^1\})\sum_{\substack{f:\{q^2\}\to\{q^0\}\\\text{one-to-one}}}\frac{e^{it\sum_{i=1}^m\overline {\varepsilon }(q_i^0)-\overline {\varepsilon }(q^2_i) }}{\prod_{j=1}^m\sin^2 \tfrac{q_j^2-q^0_{f(j)}}{2}}\prod_{i=1}^n \sign(t\overline {\varepsilon }'(q_i^1))\ .
\end{aligned}
\end{equation}
The two sets of sums factorize. We have
\begin{equation}
\begin{aligned}
\frac{1}{(m!)^2L^{2m}}&\sum_{\substack{q_1^0,...,q_m^0\\q_1^2,...,q_m^2\\q^0\text{
      distinct from
    }q^2}}\sum_{\substack{f:\{q^2\}\to\{q^0\}\\\text{one-to-one}}}\frac{e^{it\sum_{i=1}^m\overline
    {\varepsilon }(q_i^0)-\overline {\varepsilon }(q^2_i)
}}{\prod_{j=1}^m\sin^2
  \tfrac{q^2_j-q^0_{f(j)}}{2}}=\frac{1}{m!}\left(\frac{1}{L^2}\sum_{q_i\neq
  q_j}\frac{e^{it(\overline {\varepsilon }(q_i)-\overline {\varepsilon
    }(q_j) )}}{\sin^2 \tfrac{q_i-q_j}{2}}\right)^m\\ 
%&=\frac{1}{m!}\left(L \frac{4\pi^2}{3}\int_{-\pi}^\pi\rho(x)^3dx-4\pi
%\int_{-\pi}^\pi \rho(x)^2  |
%t\overline{\varepsilon}'(x)|dx+{\cal O}\big(L^0t^{-\frac{1}{2}}\big)\right)^m.
\end{aligned}
\end{equation}
As for the terms in the sum over $\{q^1\}$, they vanish for $n$ odd,
while for even $n=2p$ they are
\begin{equation}
\begin{aligned}
&\frac{(-i)^n}{n!L^n}\sum_{\substack{q_1^1,...,q_n^1\\\text{all distinct}}}\phi(\{q^1\})\prod_{i=1}^n \sign(t\overline {\varepsilon }'(q_i^1))\\
&=\frac{(-1)^p}{(2p)!}\frac{(2p)!}{p!2^p}\left(\frac{1}{L^2}\sum_{q_i\neq q_j}\sign(\overline {\varepsilon }'(q_i)\overline {\varepsilon }'(q_j))\left(\frac{2}{\sin^2\frac{q_{i}-q_{j}}{2}}+8\frac{\varepsilon'(q_{i})\varepsilon'(q_{j})}{(\varepsilon(q_i)+\varepsilon(q_j))^2}\right)\right)^p\,.
%&=\frac{(-1)^p}{p!}\left(L
%  \frac{4\pi^2}{3}\int_{-\pi}^\pi\rho(x)^3dx+c+{\cal
%      O}\big(L^{-1}\big)\right)^p\ ,
\end{aligned}
\end{equation}
It follows
that the infinite volume limit of the sum of all
$S_{n,2m|0,0}$ reads
\begin{equation}
\begin{aligned}
\sum_{n,m\geq 0}S_{n,2m|0,0}&=S_{0,0}\exp\left(\frac{1}{L^2}\sum_{q_i\neq
  q_j}  \Sigma_{ij} \sign(\overline {\varepsilon }'(q_i)\overline {\varepsilon }'(q_j))\right)\\
    \Sigma_{ij}&\equiv\frac{e^{it(\overline {\varepsilon }(q_i)-\overline {\varepsilon
    }(q_j) )}-1}{\sin^2 \tfrac{q_i-q_j}{2}}-4\frac{\varepsilon'(q_{i})\varepsilon'(q_{j})}{(\varepsilon(q_i)+\varepsilon(q_j))^2}\ .
\end{aligned}
\end{equation}
We note that it
involves the same sums as in $S_{0,2}$ and $S_{2,0}$ above. We obtain
\begin{equation}
\sum_{n,m\geq 0}S_{n,2m|0,0}=S_{0,0}\exp\left(-c-4\pi \int_{-\pi}^\pi \rho(x)^2  | t\overline{\varepsilon}'(x)|dx+{\cal O}\big(t^{-\frac{1}{2}}\big)\right)\ ,
\end{equation}
where $S_{0,0}$ has been computed in \eqref{s00}. We have thus
obtained the order ${\cal O}(\rho_\beta^2)$ contribution to the
correlation function uniformly in $t$
\begin{equation}
\label{correlchixxord}
\chi^{xx}(t,\bell)\approx C\exp\left(-2 \int_{-\pi}^\pi \rho(x)(1+2\pi\rho(x))  | t\varepsilon'(x)-\bell|dx\right)\ ,
\end{equation}
with
\begin{equation}
C=\xi \exp\left(-2\int _{-\pi }^{\pi}\int _{-\pi }^{\pi}\frac {\rho (y) \rho '(x) } {\tan \left( \frac {x-y} {2}\right) }dxdy\right)\ .
\end{equation}
In the time-like region, the exponent would be the same, but the constant would differ.

This result should be compared to the semiclassical approach of Sachdev
and Young \cite{young,sachdevbook}, which gives
\be
\chi^{xx}_{\rm SY}(t,\bell)\approx \xi\exp\left(- \int_{-\pi}^\pi
\frac{dk}{\pi}e^{-\beta\varepsilon(k)}| t\varepsilon'(k)-\bell|\right)\ . 
\ee
As expected our result reduces to the semiclassical one in the limit $\beta J\gg 1$.\\

\subsection {\texorpdfstring{Case $h > 1$: different numbers of
    particles \label{different}}{Lg}} 
We now turn to the case $h > 1$. Here the sum \eqref {armtap}
involves only intermediate states with numbers of particles that are
different from that of the representative state. Hence we must study
form factor sums with $M\neq N$. We will compute the prefactors at
order $\orho{1}$ in the space-like region, and because of saddle points
effects only at order $\orho{0}$ in the time-like region.

\subsubsection{General structure}

We start by considering the general structure of contributions with
$N\neq M$. This discussion applies also to the $h<1$ case and in
particular shows that there the dominant contributions arise from $N=M$.
As in the case $M=N$, the form factor 

\noindent $\vert {}_{\rm NS}\langle
q_1,...,q_{N}\vert\sigma_l^x\vert p_1,...,p_M\rangle_{\rm R}\vert ^2$ can be
decomposed into partial fractions 

\begin{equation}
\label{pfdhp1}
  \vert {}_{\rm NS}\langle q_1,...,q_{N}\vert\sigma_l^x\vert
  p_1,...,p_M\rangle_{\rm R}\vert ^2= \frac {\xi (2J\sqrt{h})^{(M-N)^2}}
  {L^{N+M}}\sum _{\nu_1,\dots,\nu_M=0}^2\sum_{\{f_{\vec{\nu}}\}}\frac {A\left( \left\{ q
    \right\} ,\{p\},\left\{ \nu \right\},f_{\vec{\nu}}\right) } {\prod _{j=1}^M\sin
    ^{\nu_{j}}\left( \frac {p_{j}-q_{f_{\vec{\nu}}\left( j\right)} } {2}\right)
  }\ , 
\end{equation}
with $f :\{i\in \left\{ 1,\ldots ,M\right\}|\nu_i\neq 0\} \mapsto
\left\{ 1,\ldots ,N\right\} $ any function, and $A\left( \left\{ q
\right\},\{p\} ,\left\{ \nu \right\},f_{\vec{\nu}}\right)$ is a
bounded function of $p_j$ if $\nu_j=0$, and independent of $p_j$ otherwise.

The important difference from the case $M=N$ is that we cannot always
neglect the contributions with $\nu_j=0$. Indeed, let us denote by $k$
the number of $\nu_j=0$. The corresponding contributions
give rise to
%will contribute to
$k$ oscillatory bounded integrals that will each decay with
time. On the other hand each of the $M-k$ sums over the other momenta $p_i$
will generate an oscillating factor
$e^{-it\overline{\varepsilon}(q_{f_{\vec{\nu}}(j)})}$ according to \eqref{sumsval}, while
$N$ factors $e^{it\overline{\varepsilon}(q_k)}$ are already
present. The resulting oscillatory sums may have singularities, but
according to \eqref{sumsval} summing these singularities does not
consume any oscillatory factor, it only lowers the number of
singularities. Hence in the end we will be left with $|N-M+k|$ oscillatory
bounded integrals to perform.
%, and these extra oscillatory terms will yield $|N-M+k|$ of them.
In total, there are thus $k+|N-M+k|$ of such integrals. Hence if $M\leq N$
the case $k=0$ is still dominant at late times, but if $M>N$
then all the cases $0\leq k\leq M-N$ are a priori of the same
order. In both cases these leading terms involve $|M-N|$ oscillatory
bounded integrals, so that we can conclude that the terms $M\neq N$ are
exponentially smaller than the case $M=N$ in the space-like regime,
and typically around\footnote{Clearly, there are specific degenerate
cases where the stationary phase approximation will give a
different factor than $\tfrac{1}{\sqrt{t}}$. Due to the possible
time-dependence of the integrand apart from the oscillatory term
this factor may also be marginally corrected by logarithms.}
$t^{-|M-N|/2}$ smaller in the time-like regime. 

It follows that for $h>1$ the dominant terms in \eqref{armtap} are
obtained for $M=N\pm 1$, which we now consider in turn.

\subsubsection{\texorpdfstring{Case $M=N-1$}{Lg}}

For $M=N-1$ the dominant contribution in density in \eqref{pfdhp1} is
obtained with $\nu_i=2$ for all $i=1,...,N-1$. The function 
$f:\{1,...,N-1\}\mapsto \{1,...,N\}$ has to be injective so there is
one $q_i$ not attained by $f$, leaving $(N-1)!$ possible equivalent
choices for $f$ once $q_i$ is chosen. Then one has
$A\left( \left\{ q\right\},\{p\} ,\left\{ \nu
\right\},f_{\vec{\nu}}\right)=\tfrac{1}{\varepsilon(q_i)}$. Using \eqref{sumsval},
the corresponding contribution to \eqref{armtap} is
\begin{equation}
S_{0,0}^-=\frac{2J\xi
  \sqrt{h}}{L}\sum_{i=1}^N\frac{e^{it\overline{\varepsilon}(q_i)}}{\varepsilon(q_i)}\prod_{j\neq
  i}\chi_2(q_j)\ . 
\end{equation}
This term is of order $\rho$, so in the time-like region it vanishes
at the order of our computation. In the space-like region we have 
\begin{equation}
S_{0,0}^-=2J\xi \sqrt{h}\left(\int_{-\pi}^\pi\frac{e^{it\overline{\varepsilon}(x)}}{\varepsilon(x)}\rho(x)dx\right)\exp\left(-2\int _{-\pi }^{\pi }\left| t\overline{\varepsilon} '(x) \right| \rho(x) dx\right)\ .
\end{equation}
There, $\overline{\varepsilon}(x)$ is monotonous and
$e^{it\overline{\varepsilon}(x)}$ is periodic (because the distance $\ell$ is an
integer) so the first integral decays with time faster than any
power-law and cannot be simplified further. 
\begin{figure}[H]
\begin{center}
\begin{tikzpicture}[scale=1.3]
\draw[->] (0,0.85) -- (0,0.15);
\draw[->] (1,0.85) -- (1,0.15);
\draw[->] (2.1,0.9) -- (2.9,0.1);
\draw[->] (3.1,0.9) -- (3.9,0.1);
\filldraw[black] (0,0) circle (2pt);
\filldraw[black] (1,0) circle (2pt);
\filldraw[black] (2,0) circle (2pt);
\filldraw[black] (3,0) circle (2pt);
\filldraw[black] (4,0) circle (2pt);
\filldraw[red] (0,1) circle (2pt) node[above,black] {\tiny $\nu_1=2$};
\filldraw[red] (1,1) circle (2pt)node[above,black] {\tiny $\nu_2=2$};
\filldraw[red] (2,1) circle (2pt)node[above,black] {\tiny $\nu_3=2$};
\filldraw[red] (3,1) circle (2pt)node[above,black] {\tiny $\nu_4=2$};
\end{tikzpicture}
\end{center}
\label{exfnm1}
\caption{Sketch of the leading configurations contributing to $M=N-1$.}
\end{figure}
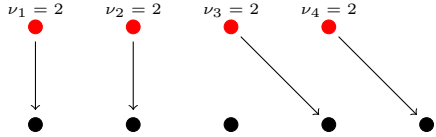
%However in the time-like region it is dominated by saddle points $SP=\{x, \overline{\varepsilon}'(x)=0\}$ and we have
%\begin{equation}
%S_{0,0}^-=\frac{2J\xi \sqrt{2\pi h}}{\sqrt{t}}\left(\sum_{s\in SP}\frac{e^{\mp i\pi/4}\rho(s)}{\varepsilon(s)\sqrt{|\overline{\varepsilon}''(s)|}}e^{-it \overline{\varepsilon}(s)}\right)\exp\left(-2t\int _{-\pi }^{\pi }\left| \overline{\varepsilon} '\left( x\right) \right| \rho \left( x\right) dx\right)
%\end{equation}
%where $\pm$ is the sign of $\overline{\varepsilon}''(s)$.

\subsubsection{\texorpdfstring{Case $M=N+1$ and $k=1$}{Lg}}
For $M=N+1$ with one $\nu_i=0$, the dominant contribution in
\eqref{pfdhp1} is obtained with the remaining $\nu_j=2$ for $j\neq
i$. Thus $f:\{1,...,N+1\}-\{i\}\mapsto \{1,...,N\}$ has to be one-to-one,
leaving $N!$ equivalent choices. Then one has
$A\left( \left\{ q\right\},\{p\} ,\left\{ \nu
\right\},\fn\right)=\tfrac{1}{\varepsilon(p_i)}$. Using \eqref{sumsval},
the corresponding contribution to \eqref{armtap} is
\begin{equation}
S_{0,0}^+=\frac{2J\xi \sqrt{h}}{(N+1)L}\sum_{i=1}^{N+1}\sum_{p_i}\frac{e^{-it\overline{\varepsilon}(p_i)}}{\varepsilon(p_i)}\prod_{j=1}^N\chi_2(q_j)\ .
\end{equation}
%Since the prefactor is of order $\rho^0$, we have to take into account saddle point effects in $\chi_2$ to get the full order $\rho$ result. The corrections to \eqref{sumsval} are
%\begin{equation}
%\chi_2(q)=1-\frac{2t|\overline{\varepsilon}'(q)|}{L}+\frac{2\sqrt{2t|\overline{\varepsilon}''(q)|}}{\pi L}\zeta_2^\pm(\sqrt{\tfrac{2t}{|\overline{\varepsilon}''(q)|}}\overline{\varepsilon}'(q))
%\end{equation}
%with $\zeta_2^\pm(u)$ given in \eqref{}, see appendix \ref{appen}. Then
%\begin{equation}
%S_{0,0}^+=2J\xi \sqrt{h}\left(\int_{-\pi}^\pi\frac{e^{it\overline{\varepsilon}(x)}}{2\pi\varepsilon(x)}dx\right)\exp\left(-2t\int _{-\pi }^{\pi }\left| \overline{\varepsilon} '\left( x\right) \right| \rho \left( x\right) dx+\frac{2}{\pi}\int_{-\pi}^\pi \sqrt{2t|\overline{\varepsilon}''(x)|}\zeta_2^\pm(\sqrt{\tfrac{2t}{|\overline{\varepsilon}''(x)|}}\overline{\varepsilon}'(x))\rho(x)dx\right)
%\end{equation}
%with $\pm$ the sign of $\overline{\varepsilon}''(x)$. In the space-like region we always have $\overline{\varepsilon}'(x)\neq 0$, so that $\zeta_2^\pm$ can be replaced by its expansion \eqref{}, revealing that this extra term is $\tfrac{1}{\sqrt{t}}$ times an oscillatory integral that decays faster than any power-law. Hence
In the infinite volume limit we obtain the following result in the
  space-like regime
\begin{equation}
S_{0,0}^+=2J\xi \sqrt{h}\left(\int_{-\pi}^\pi\frac{e^{-it\overline{\varepsilon}(x)}}{2\pi\varepsilon(x)}dx\right)\exp\left(-2\int _{-\pi }^{\pi }\left| t\overline{\varepsilon} '(x) \right| \rho (x) dx\right)\ ,
\end{equation}
where the prefactor is accurate to first order in the density $\rho_\beta$.
%In the time-like region however, \eqref{} cannot be used everywhere, and close to a saddle point $s\in SP$ one has to rescale $x=s+\tfrac{u}{\sqrt{2t|\overline{\varepsilon}''(s)|}}$ to obtain
%\begin{equation}
%\int_{-\pi}^\pi \sqrt{2t|\overline{\varepsilon}''(x)|}\zeta_2^\pm(\sqrt{\tfrac{2t}{|\overline{\varepsilon}''(x)|}}\overline{\varepsilon}'(x))\rho(x)dx=\sum_{s\in SP}\rho(s)\int_{-\infty}^\infty \zeta_2^\pm(u)du
%\end{equation}
%With $\int_{-\infty}^\infty \zeta_2^\pm(u)du=\pm \pi i$ one obtains
In the time-like region the prefactor is only accurate to ${\cal
  O}(\rho_\beta^0)$, because saddle point effects arise at order
${\cal O}(\rho_\beta)$. A saddle point approximation gives in this
regime
\begin{equation}
\begin{aligned}
S_{0,0}^+=\frac{2J\xi \sqrt{h}}{\sqrt{2\pi |t|}}\left(\sum_{s\in SP}\frac{e^{\mp i\pi/4}}{\varepsilon(s)\sqrt{|\overline{\varepsilon}''(s)|}}e^{-it \overline{\varepsilon}(s)}\right)\exp\left(-2\int _{-\pi }^{\pi }\left|t \overline{\varepsilon} '(x) \right| \rho(x) dx\right)\ ,
\end{aligned}
\end{equation}
where $SP=\{s,\overline{\varepsilon}'(s)=0\}$ denotes the set of
  saddle points and $\pm$ the sign of $t\overline{\varepsilon}'(s)$.
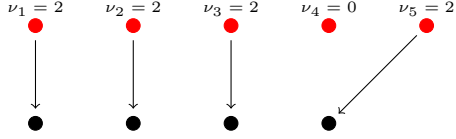
\begin{figure}[H]
\begin{center}
\begin{tikzpicture}[scale=1.3]
\draw[->] (0,0.85) -- (0,0.15);
\draw[->] (1,0.85) -- (1,0.15);
\draw[->] (2,0.85) -- (2,0.15);
\draw[->] (3.9,0.9) -- (3.1,0.1);
\filldraw[black] (0,0) circle (2pt);
\filldraw[black] (1,0) circle (2pt);
\filldraw[black] (2,0) circle (2pt);
\filldraw[black] (3,0) circle (2pt);
\filldraw[red] (0,1) circle (2pt) node[above,black] {\tiny $\nu_1=2$};
\filldraw[red] (1,1) circle (2pt)node[above,black] {\tiny $\nu_2=2$};
\filldraw[red] (2,1) circle (2pt)node[above,black] {\tiny $\nu_3=2$};
\filldraw[red] (3,1) circle (2pt)node[above,black] {\tiny $\nu_4=0$};
\filldraw[red] (4,1) circle (2pt)node[above,black] {\tiny $\nu_5=2$};
\end{tikzpicture}
\end{center}
\label{exfnp1k1}
\caption{Sketch of the leading configurations contributing to $M=N+1$ and $k=1$.}
\end{figure}

\subsubsection{\texorpdfstring{Case $M=N+1$ and $k=0$}{Lg}}
For $M=N+1$ and all $\nu_i\neq 0$, the dominant contribution in
\eqref{pfdhp1} is obtained by $\{\nu\},\fn$ such that
$\nu_i=\nu_j=1$ with $\fn(i)=\fn(j)$, and the remaining $\nu_l=2$ for
$l\neq i,j$. Once $q_k=q_{\fn(i)}=q_{\fn(j)}$ are chosen, there are ${
  N+1\choose 2} N!$ possibilities for $\fn$. All of these lead to
$A\left( \left\{ q \right\},\{p\},\left\{ \nu
\right\},\fn\right)=\tfrac{-2}{\varepsilon(q_k)}$. Using
\eqref{sumsval}, the corresponding contribution to \eqref{armtap} becomes

\begin{equation}
S_{0,2}^+=-\frac{2J\xi \sqrt{h}}{L}\sum_{k=1}^{N}\frac{e^{-it\overline{\varepsilon}(q_k)}}{\varepsilon(q_k)}\chi_1^2(q_k)\prod_{j\neq k}^N\chi_2(q_j)\ .
\end{equation}
%The first sum is of order $\rho$, so saddle point effects can be ignored in $\chi_2$ at order $\rho$. Hence
Taking the infinite volume limit we obtain
\begin{equation}
S_{0,2}^+=-2J\xi \sqrt{h}\left(\int_{-\pi}^\pi\frac{e^{-it\overline{\varepsilon}(x)}}{\varepsilon(x)}\chi_1^2(x)\rho(x)dx\right)\exp\left(-2\int _{-\pi }^{\pi }\left| t\overline{\varepsilon} '(x) \right| \rho(x) dx\right)\ .
\end{equation}
In the space-like region $\chi_1(x)=i\sign(\overline{\varepsilon}'(x))$ and one has with the prefactor at order $\rho^1$
\begin{equation}
S_{0,2}^+=2J\xi \sqrt{h}\left(\int_{-\pi}^\pi\frac{e^{-it\overline{\varepsilon}(x)}}{\varepsilon(x)}\rho(x)dx\right)\exp\left(-2\int _{-\pi }^{\pi }\left|t \overline{\varepsilon} '(x) \right| \rho(x) dx\right)\ .
\end{equation}
In time-like region, this term vanishes at order $\rho^0$.
\begin{figure}[H]
\begin{center}
\begin{tikzpicture}[scale=1.3]
\draw[->] (0,0.85) -- (0,0.15);
\draw[->] (1,0.85) -- (1,0.15);
\draw[->] (2,0.85) -- (2,0.15);
\draw[->] (2.9,0.9) -- (2.1,0.1);
\draw[->] (3.9,0.9) -- (3.1,0.1);
\filldraw[black] (0,0) circle (2pt);
\filldraw[black] (1,0) circle (2pt);
\filldraw[black] (2,0) circle (2pt);
\filldraw[black] (3,0) circle (2pt);
\filldraw[red] (0,1) circle (2pt) node[above,black] {\tiny $\nu_1=2$};
\filldraw[red] (1,1) circle (2pt)node[above,black] {\tiny $\nu_2=2$};
\filldraw[red] (2,1) circle (2pt)node[above,black] {\tiny $\nu_3=1$};
\filldraw[red] (3,1) circle (2pt)node[above,black] {\tiny $\nu_4=1$};
\filldraw[red] (4,1) circle (2pt)node[above,black] {\tiny $\nu_5=2$};
\end{tikzpicture}
\end{center}
%\label{exfnp1k0}
\caption{Sketch of the leading configurations contributing to $M=N+1$ and $k=0$.}
\end{figure}
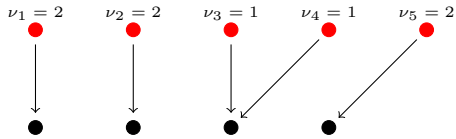

\subsubsection{\texorpdfstring{Result: correlation functions for $h>1$
at leading order in $\rho_\beta$}{Lg}}
Putting everything together, in the space-like regime and at leading
order in density we obtain the following result 
\begin{equation}
\label{chixxdisor}
%\chi^{xx}(\bell,t)\approx 2J\xi\sqrt{h}
%\int_{-\pi}^\pi\left(\frac{e^{-it\overline{\varepsilon}(x)}}{2\pi}+2\cos(t\overline{\va%repsilon}(x))\rho(x)
%\right) \frac{dx}{\varepsilon(x)}\exp\left(-2\int _{-\pi }^{\pi
%}\left|t \overline{\varepsilon} '\left( x\right) \right| \rho \left(
%x\right) dx\right)\, ,
\chi^{xx}(\ell,t)\approx 2J\xi\sqrt{h}
\left[\int_{-\pi}^\pi dx\frac{e^{-it\overline{\varepsilon}(x)}}{2\pi\varepsilon(x)}
\big(1+4\pi\rho(x)\big)\right]
\exp\left(-2\int _{-\pi }^{\pi
}\left|t \overline{\varepsilon} '(x) \right| \rho \left(
x\right) dx\right)\, , 
\end{equation}
In the time-like regime we have instead% with the prefactor at order $\rho^0$
\begin{equation}
\label{chixxdisor2}
\chi^{xx}(\bell,t)\approx \frac{2J\xi \sqrt{h}}{\sqrt{2\pi |t|}}\left(\sum_{s\in SP}\frac{e^{\mp i\pi/4}}{\varepsilon(s)\sqrt{|\overline{\varepsilon}''(s)|}}e^{-it \overline{\varepsilon}(s)}\right)\exp\left(-2\int _{-\pi }^{\pi }\left|t \overline{\varepsilon} '(x) \right| \rho(x) dx\right)\, ,
\end{equation}

where the sum is over the saddle points $s$ of $\overline{\varepsilon}(x)$.

This result should be compared to the semiclassical approach of Sachdev
and Young \cite{young,sachdevbook}, which gives
\be
\chi^{xx}_{\rm SY}(t,\bell)\approx
2J\xi\sqrt{h}
\int_{-\pi}^\pi \frac{dx}{2\pi}\frac{e^{-it\overline{\varepsilon}(x)}}
{\varepsilon(x)}\ \exp\left(- \int_{-\pi}^\pi \frac{dk}{\pi}e^{-\beta\varepsilon(k)}| t\varepsilon'(k)-\bell|\right)\ .
\ee
As expected our result reduces to the semiclassical one in the limit $\beta J\gg 1$.

\subsection{Case $h=1$}
In the case $h=1$ the structure of the form factor \eqref{ff} is
modified compared to the case $h\neq 1$: since  $\varepsilon(0)=0$, there are additional
poles. The nature of these
poles is moreover different from those appearing in the partial
fraction decomposition of the previous sections, since they involve
pairs of momenta: for $\varepsilon_{pp'}$ to vanish we must have
$p=p'=0$.  

We note that at zero temperature the model becomes critical at $h=1$
and correlation functions should exhibit power-law decays. These
%aspects should be discussed in a subsequent publication. 
issues are beyond the scope of this work.

\subsection {\texorpdfstring{Quantum quench case \label{qu}}{Lg}}
We now turn to the time evolution of the order parameter
  one-point function after a quench of the transverse field within the
ordered phase. Our aim is to evaluate the spectral representation
\fr{bigsum} obtained in the framework of the quench action approach.

\subsubsection{Generalities}
The sum over form factors appearing in the context of quantum quench
dynamics \eqref{bigsum} differs from the finite temperature case
\eqref{armtap} notably because now in both states of the form factors the momenta come in pairs $p_i,-p_i$. One can then write  

\begin{equation}
\label{rewrite}
\begin{aligned}
&_{\rm{R}}\langle p_1,-p_1,...,p_N,-p_N\vert\sigma_\ell^x\vert
q_1,-q_1,...,q_N,-q_N\rangle_{\rm{NS}}
=\frac{(-4)^N\sqrt{\xi}}{L^{2N}}\prod_{j=1}^N\sin q_j  \sin p_j\\
&\qquad\qquad\qquad\qquad\times\
\prod_{i,j=1}^N\frac{\varepsilon^4_{q_jp_i}}{\varepsilon^2_{q_jq_i}\varepsilon^2_{p_jp_i}}
\frac{\displaystyle\prod_{i\neq j=1}^N(\cos q_i-\cos q_j)(\cos p_i-\cos p_j)}{\displaystyle\prod_{i,j=1}^N(\cos q_i-\cos p_j)^2}\ .
\end{aligned}
\end{equation}

Focusing again on $M=N$ in \eqref{bigsum}, we have
\begin{equation}
\label{bigsum2}
\begin{aligned}
\left\langle \sigma^{x}_{\ell}\left( t\right) \right\rangle ={\rm Re}\Bigg[
\frac{\sqrt{\xi}}{N!L^{2N}}&\sum_{\substack{0<p_1,...,p_N\\\in\rm{R}}}\prod_{j=1}^N4\frac{f(p_j)}{f(q_j)}\sin p_j\sin q_je^{2it(\varepsilon_{p_j}-\varepsilon_{q_j})}\\
&\times\ \prod_{i,j}\frac{\varepsilon^4_{q_jp_i}}{\varepsilon^2_{q_jq_i}\varepsilon^2_{p_jp_i}}\frac{\prod_{i\neq j}(\cos q_i-\cos q_j)(\cos p_i-\cos p_j)}{\prod_{i,j}(\cos q_i-\cos p_j)^2}\Bigg]\ ,
\end{aligned}
\end{equation}
with
\begin{equation}
\label{f}
f(p)\equiv K(p)=\sqrt{\frac{2\pi\rho(p)}{1-(2\pi\rho(p))^2}}\ .
\end{equation}

\subsubsection{Differences from the finite temperature case}

We apply a partial fraction decomposition to the second line of
\eqref{bigsum2}, seen as a ratio of polynomials in the $\cos p_j$. 
The procedure is the same as in Section \ref{hm1}. More
precisely, we define  
\begin{equation}
\label{fuvcos}
 \widetilde{F}_{U}^{V}=\frac {\displaystyle\bigg|\prod _{u\neq u'\in
     U}\left( \cos u-\cos u'\right)\prod _{v\neq v'\in V}\left(\cos
   v-\cos v'\right)\bigg|} {\displaystyle\prod _{u\neq v\in U,V}(\cos
   u-\cos v)^2}\frac {\displaystyle\prod _{u,v\in U,V}
   \varepsilon_{uv}^{{4}}} {\displaystyle\prod_{u,u'\in U}\varepsilon_{uu'}^{{2}}\prod _{v,v'\in V}\varepsilon_{vv'}^{{2}}}\ ,
\end{equation}
which we decompose into partial fractions with $\cos u$ taken to be
the relevant variables. This gives
\begin{equation}
\begin{aligned}
\label{decompcos}
%_{\rm{R}}\langle p_1,-p_1,...,p_N,-p_N\vert\sigma_l^x\vert
%q_1,-q_1,...,q_N,-q_N\rangle_{\rm{NS}}= &\frac {(-4)^N\sqrt{\xi} } {L^{2N}}\prod_{j=1}^N \sin q_j \sin p_j\\
%&\times
\widetilde{F}^{\{q_i\}}_{\{p_i\}}
=\sum_{\nu_1,\dots,\nu_N=0}^2\sum _{\{\fn\}}\frac{\mathcal{A}\left( \left\{ q
  \right\} ,\{p\},\left\{ \nu\right\},\fn\right) } {\displaystyle\prod _{j=1}^N (\cos
  p_j-\cos q_{\fn(j)})^{\nu_j}}\ , 
\end{aligned}
\end{equation}
where the second sum is over any function $f :\{i\in \left\{ 1,\ldots ,N\right\}| \nu_i\neq 0\}
\mapsto \left\{ 1,\ldots ,N\right\} $, and where
$\mathcal{A}\left( \left\{ q \right\} ,\{p\},\left\{ \nu
\right\},\fn\right)$ is a bounded function of $p_j$ if $\nu_j=0$, and
independent of $p_j$ otherwise.

There are however some noteworthy differences from Section \ref{hm1} brought by the presence
of factors involving the function $f(p)$ and the fact that there is
still a $p_i$ dependence in the sum \eqref{bigsum2} outside the
partial fraction decomposition. In place of \eqref{sumsval} we now have
\begin{equation}
\label{use2}
\begin{aligned}
\frac{4}{L}\sum_{p>0,\in\rm{R}}\frac{\sin p \sin q' f(p)/f(q')}{\cos q-\cos p}e^{2it\varepsilon(p)}&=2i\sign(t\varepsilon'(q))\sin q'\frac{f(q)}{f(q')}e^{2it\varepsilon(q)}+\mathcal{O}(L^0 t^{-1/2})\\
\frac{4}{L^2}\sum_{p>0,\in\rm{R}}\frac{\sin p \sin q f(p)/f(q)}{(\cos q-\cos p)^2}e^{2it\varepsilon(p)}&=\left(1-\frac{4 t |\varepsilon'(q)|}{L}+\frac{2i\sign(\varepsilon'(q))}{L}\frac{f'(q)}{f(q)}\right)e^{2it\varepsilon(q)}\\
&\qquad\qquad\qquad\qquad\qquad+\mathcal{O}(L^{-1} t^{-1/2})\ ,
\end{aligned}
\end{equation}
as shown in Appendix \ref{appen}. The analog of equation \eqref{snm} is
  given by
\begin{equation}
\begin{aligned}
\label{snm2}
S_{n,2m}=\frac{(2i)^nS_{0,0}}{(-2)^mL^{n+2m}}\sum_{\substack{q_1^0<...<q_m^0\\q_1^1<...<q_n^1\\q_1^2<...<q_m^2\\\text{all distinct}}}&A(\{q^0\},\{q^1\},\{q^2\})\prod_{j=1}^m\left(4\sin q^0_j\sin q^2_j \frac{f(q^2_j)}{f(q^0_j)}\right)\\
&\times e^{2it\sum_{i=1}^m\varepsilon (q_i^2)-\varepsilon (q^0_i) }\prod_{i=1}^n \sin q_i^1\sign(\varepsilon '(q_i^1))\ ,
\end{aligned}
\end{equation}
with
\begin{equation}
S_{0,0}=\sqrt{\xi}\exp\left(-4\int _{0 }^{\pi }\left|t \varepsilon '(x) \right| \rho(x) dx\right)\ .
\end{equation}
Here we used that $f(p)$ \fr{f} fulfils
\begin{equation}
\label{int0}
\int_0^\pi \frac{f'(x)}{f(x)}\rho(x)dx=0\ .
\end{equation}
The possibility of differentiating $f(p)$ also modifies the derivation
of \eqref{phia}, which now takes the form
\begin{equation}
\begin{aligned}
\label{phia2}
&A(\{q_i\}_{i\in I_1\cup I_0},\{\tilde{\nu}\},\tilde{f})=\\
&\sum_{K\subset I_1}\prod_{k\in K}\frac{f'(q_k)}{f(q_k)}\sum_{P\text{ pairings of }I_1-K}\prod_{(i,j)\in P}\left(\frac{1}{(\cos q_i-\cos q_j)^2}+4\frac{\partial_{\cos k}\varepsilon(q_{i})\partial_{\cos k}\varepsilon(q_{j})}{(\varepsilon(q_i)+\varepsilon(q_j))^2}\right)\ .
\end{aligned}
\end{equation}
Because of \eqref{int0}, however, only $K=\{\}$ remains after the sum over $q_k$.

\subsubsection{\texorpdfstring{Result: ${\cal O}(\rho_Q^2)$ uniformly
    in $t$ at large $t$}{Lg}}

The final result of the above calculation for the time
evolution of the order parameter one-point function after a quench of the transverse
field within the ordered phase is
\begin{equation}
\label{cquench}
\left\langle \sigma^{x}_\ell\left( t\right) \right\rangle=C\exp\left(-4 \int_{0}^\pi \rho(x)(1+2\pi\rho(x))  | t\varepsilon'(x)|dx\right)\ ,
\end{equation}
where
\begin{equation}
\label{ccquench}
C=\sqrt{\xi} \exp\left(-4\int _{0}^{\pi}dx\int _{0 }^{\pi}dy\rho(y)\Big[
\frac {\rho '(x) \sin y} {\cos y-\cos x }-2\frac {|\varepsilon '(x) \varepsilon '(y)|} {\left( \varepsilon (x) +\varepsilon (y) \right) ^{2}}\rho(x)\Big]\right).
\end{equation}
This result holds at the second order in the density ${\cal O}(\rho_Q^2)$.
The reader may have noted that, since $\varepsilon'(0)=\varepsilon'(\pi)=0$, the quantum quench dynamics is a
'time-like region' case, so
the saddle point effects might modify this prefactor at order
$\rho^1$. It turns out, however, that $\rho(0)=\rho'(0)=\rho(\pi)=\rho'(\pi)=0$, so the saddle point corrections are higher order in time and do not affect the prefactor. 

\subsection{Comments on the form factor summation}
Having carried out form factor summations to obtain the leading
late-time asymptotics in the low-density regime in both the finite
temperature and the quantum quench contexts it is useful to take
stock and stress some features that we expect to be of a general
nature.
\subsubsection{Which states govern the late time dynamics?\label{govern}}

The leading late time behaviour follows from equations \eqref{sumsval}
that enter the sum over $p$'s of the partial fraction decomposition
\eqref{decomp}. These formulas are obtained by isolating the
singularity and dropping the integral of an oscillatory bounded
function, \emph{cf.} Appendix \ref{appen}.
The singular part involves momenta $p_j$ in \eqref {armtap} that
  are at distance $\mathcal{O}(L^{-1})$ of any of the $q_i$. The aim of this
section is to quantify more precisely the number of $p_j$ that have to
be summed in order to recover the late time dynamics.
In other words, we would like to know the smallest function
$\eta(L)$ such that  
\begin{equation}
\label  {sumeta}
\sum _{n=-L\eta(L)}^{L\eta(L)}\frac {e^{it ( \overline{\varepsilon } (q)-\overline{\varepsilon } \left( p_n\right) )}} {L^2\sin^2  \left( \frac {p_n-q} {2}\right) }=\left(1-\frac {2\left| t \overline{\varepsilon } '(q) \right| } {L}\right)+{\cal O}(L^{-1}t^{-1/2})+{\cal O}(L^{-2})\ ,
\end{equation}
with $p_n=q+\frac{2\pi}{L}(n+1/2)$. We first observe that if $ L\eta
\left( L\right)\to N_0 $ remains finite when we take $L\to\infty$, then  
\begin{equation}
\begin{aligned}
\sum _{n=-L\eta(L)}^{L\eta(L)}\frac {e^{it ( \overline{\varepsilon } (q)-\overline{\varepsilon } \left( p_n\right) )}} {L^2\sin^2  \left( \frac {p_n-q} {2}\right) }&\approx\sum _{n=-L\eta(L)}^{L\eta(L)}\frac {1} {L^2\sin^2  \left( \frac {p_n-q} {2}\right) }+\mathcal{O}(t^2/L^2) \\
 &\approx \sum _{n=-N_0}^{N_0}\frac {1} {\pi^2(n+1/2)^2 } <\sum _{n=-\infty}^{\infty}\frac {1} {\pi^2(n+1/2)^2 }=1\ .
\end{aligned}
\end{equation}
As our spectral representation involves $N\propto L$ sums of this kind
we obtain an infinite product over factors that are strictly smaller
than $1$ and obtain a vanishing answer. Hence retaining only a finite
number of $p_j$'s in our sum is clearly insufficient.
  
%After being multiplied for each of the $N\gg 1$ momenta $q$ in the
%scaling limit, the full sum over form factors with this restriction
%over the momenta $p$ thus vanishes. 

Conversely, if $ \eta (L)=\eta$ stays finite we do have \eqref {sumeta} at leading order in time because the terms we drop compared to having limits $\pm\infty$ contribute to an oscillatory integral of a bounded function
\begin{equation}
\int_{\eta}^\infty \frac{e^{itx}}{x^2}dx+\int_{-\infty}^{-\eta} \frac{e^{itx}}{x^2}dx\ ,
\end{equation}
 that vanishes at large times.
 
 What if now $ \eta (L)\to 0 $ with $ L \eta (L)\to \infty $? We have, by turning the sum into an integral with Euler-Maclaurin correction terms
\begin{equation}
\label{emll}
  \sum _{n>\eta \left( L\right) L}\frac {e^{i\frac {\theta } {L}\left( n+1 / 2\right) }} {\left( n+1 / 2\right) ^{2}}=\frac {1} {L}\int _{\eta }^{\infty }\frac {e^{i\theta x}} {x^{2}}dx+\mathcal{O}\left( \frac {1} {L^{n}\eta ^{n}},n\geq 3\right) \ ,
\end{equation}
and
\begin{equation}
\label{emll2}
  \int _{\eta }^{\infty }\frac {e^{i\theta x}} {x^{2}}dx=\frac{1}{\eta}E_2(-i\theta \eta)=\frac {i} {\eta ^{2}\theta }e^{i\eta \theta }+\mathcal{O}\left( \frac {1} {\theta ^{n+2}\eta ^{n+3}},n\geq 0\right) \ ,
\end{equation}
obtained from the expansion of the exponential integral
$E_2(x)=\int_1^\infty \tfrac{e^{-xu}}{u^2}du$. If we want this term to
become negligible compared to \eqref{sumsval} at late times in the
scaling limit, then we need $L\eta^{n+2}(L)\to\infty$ for all $n\geq 0$ in
order to ensure that both the Euler MacLaurin correction terms in
\eqref{emll} and \eqref{emll2} are negligible. Hence any power-law
$\eta(L)=L^{-\nu}$ with $\nu>0$ will not suffice. Stated differently, the
number $L\eta(L)$ has to be larger than any $L^\nu$ for $0<\nu<1$, but
any macroscopic fraction $\epsilon L$ with $\epsilon>0$ is sufficient. We will denote by \textit{mesoscopic} this number of states (in contrast with microscopic $\mathcal{O}(1)$ and macroscopic $\mathcal{O}(L)$).

Finally, it is clear from e.g. \eqref{s00} that in order to recover an
exponential decay in time, one should multiply a $\mathcal{O}(L)$ number of
terms like \eqref{sumsval}, and to recover the right exponent one
should take into account 'almost all' these terms at each momentum
$q$.

We conclude that if the sum over the $p_j$ in \eqref{armtap} is viewed
in terms of particle-hole excitations over the $q_i$, the leading late time
behaviour emerges from a \textit{mesoscopic} number (i.e. larger than any $L^\nu$ for $0<\nu<1$, but smaller than $\epsilon L$ for any $\epsilon>0$) of
particle-hole excitations \emph{around each $q_i$}. It represents an exponential number of states, but still sub-entropic, in the sense that it includes only states whose macroscopic state is the representative state itself. We expect this
to be a general feature of form factor expansions of semi-local
operators that holds also in interacting theories.

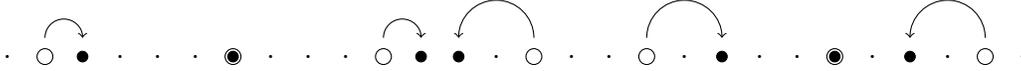
\begin{figure}[H]
\begin{center}
\begin{tikzpicture}[scale=1]
\draw[->]     (-0.5,0.25) arc (180: 0:0.25);
\draw[->]     (4,0.25) arc (180: 0:0.25);
\draw[->]     (6,0.25) arc (0: 180:0.5);
\draw[->]     (7.5,0.25) arc (180: 0:0.5);
\draw[->]     (12,0.25) arc (0: 180:0.5);
\node at (-1,0) {.};
\draw[black] (-0.5,0) circle (3pt);
\node at (0,0) {.};
\node at (0.5,0) {.};
\node at (1,0) {.};
\node at (1.5,0) {.};
\draw[black] (2,0) circle (3pt);
\node at (2.5,0) {.};
\node at (3,0) {.};
\node at (3.5,0) {.};
\draw[black] (4,0) circle (3pt);
\node at (4.5,0) {.};
\node at (5,0) {.};
\node at (5.5,0) {.};
\draw[black] (6,0) circle (3pt);
\node at (6.5,0) {.};
\node at (7,0) {.};
\draw[black] (7.5,0) circle (3pt);
\node at (8,0) {.};
\node at (8.5,0) {.};
\node at (9,0) {.};
\node at (9.5,0) {.};
\draw[black] (10,0) circle (3pt);
\node at (10.5,0) {.};
\node at (11,0) {.};
\node at (11.5,0) {.};
\draw[black] (12,0) circle (3pt);
\node at (12.5,0) {.};
\filldraw[black] (0,0) circle (2pt);
\filldraw[black] (2,0) circle (2pt);
\filldraw[black] (4.5,0) circle (2pt);
\filldraw[black] (5,0) circle (2pt);
\filldraw[black] (8.5,0) circle (2pt);
\filldraw[black] (10,0) circle (2pt);
\filldraw[black] (11,0) circle (2pt);
\end{tikzpicture}
\end{center}
\label{exfnp1k0}
\caption{Sketch of the states contributing to the leading asymptotics: position of the momenta of the representative state (empty circles), position of the momenta of the intermediate state (filled circles), and position of the holes (dots).}
\end{figure}

%We note that this observation is compatible with pre-existing work at zero temperature where the correlation function is expanded into a finite number of particle-hole excitations (ref needed). Indeed, at zero temperature the momenta $q$ are fully packed between a certain $-q_F$ and $q_F>0$. Since there are no holes inside the sea of particles, particle-hole excitations emerging from the bulk have to be macroscopic, thus negligible at late times. It follows that only particle-hole excitations emerging from the edges can contribute to the late time asymptotics, and the farther they are from the edges the larger they have to be, thus the smaller they contribute. Hence in this case the correlation function is naturally expanded in terms of a finite number of particle-hole excitations from the edges.

\subsubsection {Non-vanishing low-density limit of form factors}
The leading order term both in time and density \eqref{s00}, obtained
by keeping only the double poles in the partial fraction decomposition
\eqref{decomp}, has an interesting physical interpretation in terms of
\textit{low-density limit} of the form factor \eqref{ff}. This limit
consists in assuming that $\rho(x)$ is small everywhere, i.e. that
$L(q_{i+1}-q_i)\to\infty$ when $L\to\infty$ in the scaling
limit\footnote{Strictly speaking, this is a different limit than
merely $N/L\ll 1$, since this latter one does not require
$\rho(x)={\cal O}(N/L)$  everywhere. For example, a ground state
at zero temperature and high chemical potential would have $N/L\ll 1$
but not $\rho(0)\ll 1$.}. One can observe then that the square of the
form factor \eqref{ff2} vanishes in the scaling limit $L\to\infty$
unless each $p$ is at a distance of order $\mathcal{O}(L^{-1})$ from a $q$, and
because of the low-density assumption this $q$ becomes unique in the
limit $L\to\infty$. Thus one can write 
\begin{equation}
\label{lowd}
p_i=q_i+\frac{2\pi}{L}(n_i+1/2)\ ,
\end{equation}
with $n_i$ an integer. In the low-density limit we have

\begin{equation}
\label{simple}
\begin{aligned}
\underset{L\to\infty}{\lim}\,\frac{\sin \frac{q_j-q_{j'}}{2}\sin \frac{p_j-p_{j'}}{2}}{\sin \frac{q_j-p_{j'}}{2}\sin \frac{q_{j'}-p_{j}}{2}}=-1\,,\qquad \underset{L\to\infty}{\lim}\, \frac{\varepsilon_{q_j p_{j'}}\varepsilon_{q_{j'} p_{j}}}{\varepsilon_{q_j q_{j'}}\varepsilon_{p_j p_{j'}}}=1 \,,\quad \text{for }j<j'\ ,
\end{aligned}
\end{equation}
and so in the scaling limit
\begin{equation}
  \vert \langle q_1,...,q_{N}\vert\sigma_l^x\vert p_1,...,p_{N}\rangle\vert ^2 =\frac {\xi } {\pi^{2N}}\prod _{j=1}^{N}\frac {1} {\left( n_{j}+1 / 2\right) ^{2}}\ ,
\end{equation}
which constitutes the low-density approximation of the form factor in the case it is not vanishing in the scaling limit.

 Now, we trade the $1/N!$ in \eqref{armtap} for an ordering $p_1<...<p_N$ and choose an ordering $q_1<...<q_N$. In the low-density limit, we have $L(q_{i+1}-q_i)\to\infty$ in the large $L$ limit: Hence whenever the integers $n_i$ in \eqref{lowd} are ${\cal O}(L^0)$ the constraint on the $p's$ is automatically satisfied, and one can sum, in the large $L$ limit, on arbitrary integers. The fact that both the $1/N!$ and the constraint $p_1<...<p_N$ are removed in the low-density limit is a simplifying feature very specific to this regime.
 
% Such a constraint translates into a constraint on the $n_j$ in \eqref{lowd}, namely $n_{j+1}>n_j-\tfrac{L}{2\pi}(q_{j+1}-q_{j})\approx n_j-\tfrac{1}{2\pi \rho(q_j)}$. At leading order in $\rho$, this constraint becomes $-\tfrac{1}{4\pi \rho(q_j)}<n_j<\tfrac{1}{4\pi \rho(q_j)}$, so that \eqref{armtap} becomes
%
%\begin{equation}
%\chi^{xx}\left(\ell,t\right) =\sum _{n_{1}=-\lceil\tfrac{1}{ 4\pi
%    \rho(q_1)}\rceil}^{\lfloor\tfrac{1}{4\pi \rho(q_1)}\rfloor} ...
%\sum_{n_{N}=\lceil-\tfrac{1}{4\pi \rho(q_N)}\rceil}^{\lfloor\tfrac{1}{4\pi \rho(q_N)}\rfloor}\frac
%    {\xi } {\pi^{2N}}\prod _{j=1}^{N}\frac {e^{-2i\pi
%        t\sum_{j=1}^N\overline{\varepsilon}'(q_j)\frac{(n_j+1/2)}{L}}}
%    {\left( n_{j}+1 / 2\right) ^{2}}\ . 
%\end{equation}

Hence in the low density limit we have
\begin{equation}
   \chi^{xx}\left(\ell,t\right) =\sum _{n_{1}=-\infty}^{\infty} ...\sum _{n_{N}=-\infty}^{\infty}\frac {\xi } {\pi^{2N}}\prod _{j=1}^{N}\frac {e^{-2i\pi t\sum_{j=1}^N\overline{\varepsilon}'(q_j)\frac{(n_j+1/2)}{L}}} {\left( n_{j}+1 / 2\right) ^{2}}\ .
\end{equation}
This expression factorizes and can be computed along \eqref{s00} with formulas \eqref{sumuseful}.

\subsubsection{Static correlations \label{static}}
Results \eqref{correlchixxord}, \eqref{chixxdisor},
\eqref{chixxdisor2} and their analogues hereafter
\eqref{chitemperatureordered}, \eqref{chitemperaturedisordered} for
the finite temperature correlations are obtained in the regime $\ell
,t\to\infty$ at fixed $\alpha=t/\ell$. In their derivation, however, we have only used that the
phase $it(E\left( \left\{ q\right\} \right) -E\left( \left\{ p\right\}
\right) )+ i\ell(P\left( \left\{ p\right\} \right) -P\left( \left\{
q\right\} \right) )$ in \eqref{armtap} is large, therefore our calculations are also applicable to
static correlations $t=0$ at large $\ell$. It amounts to replacing
$t\overline{\varepsilon}(x)$ by $-\ell x$ in all the phases
$e^{-it\overline{\varepsilon}(x)}$. All the oscillatory integrals that
we neglected because of their large time behaviour $\sim t^{-1/2}$ will now decay at large distances at
least as $\ell^{-1}$ (and in general much faster). In particular
\eqref{sumsval} still hold, but with corrections
$\mathcal{O}(L^0 \ell^{-1})$ and $\mathcal{O}(L^{-1} \ell^{-1})$ respectively. 
Static
correlations are then obtained by replacing
$|t\overline{\varepsilon}'(x)|$ (i.e., $|t\varepsilon'(x)-\ell |$) by $|\ell|$ in the final results.

\section {Quantum quench dynamics \label{qqd} beyond low densities}
The framework presented in Section \ref{pfd} is general and permits us to compute the late time behaviour and subleading
corrections of form factor sums, order by order in $\rho_Q=N/L$. In
particular, it yields the expression of the observables of interest as
$Ce^{-t/\tau}$, where both $C$ and $\tau$ are the exact expressions at
a given order (here second order) in $\rho_Q$. The
computation of a generic order in the density is however  rather involved. In this section and in the following one we focus on the
exponent $\tau$ of the exponential decay, for which another more 
efficient but less general approach can be used to calculate it at all
orders in $\rho_Q$, i.e. writing the correlation function as $e^{-t/\tau}$ where $\tau$ includes all orders in $\rho_Q$.

This first section treats the quantum quench problem introduced in Section \ref {quench}.
%%%%%%%%%%%%%%%%%%%%%%%%%%%%%%%%%%
\subsection{Determinant representation}
%%%%%%%%%%%%%%%%%%%%%%%%%%%%%%%%%%
As shown in Section \ref{different}, the leading contribution in
\eqref {bigsum} is obtained for $N=M$, on which we will focus. The
starting observation is that the last term of \eqref{rewrite} can be
written as a Cauchy determinant. Indeed we have
\begin{equation}
\label{cauch}
\frac{\prod_{i\neq j}(\cos q_i-\cos q_j)(\cos p_i-\cos p_j)}{\prod_{i,j}(\cos q_i-\cos p_j)^2}=(\det C)^2=\det C^TC\,,\qquad C_{ij}=\frac{1}{\cos p_i-\cos q_j}\ .
\end{equation}
Let us define $\bar{M}=C^TC$ and $M^j_{ik}=C_{ij}C_{kj}$, so that
$\bar{M}_{ik}=\sum_jM_{ik}^j$. The determinant of $\bar{M}$ can be expanded as follows
\begin{equation}
\begin{aligned}
\det\bar{M}&=\sum_{\tau\in\mathfrak{S}_N}\sign(\tau)\bar{M}_{1\tau(1)}...\bar{M}_{N\tau(N)}\\
&=\sum_{j_1,...,j_N\in\{1,...,N\}}\sum_{\tau\in\mathfrak{S}_N}\sign(\tau)M^{j_1}_{1\tau(1)}...M^{j_N}_{N\tau(N)}\ .
\end{aligned}
\end{equation}
The term $M^{j_a}_{a\tau(a)}M^{j_b}_{b\tau(b)}$ is invariant under the replacement $\tau\to\tau\cdot (a\, b)$ if $j_a=j_b$, whereas the sign of the permutation changes. Hence the sum over $\tau$ vanishes unless all the $j$'s are distinct, i.e., if they are a permutation of $1,..., N$. In conclusion we find
\begin{equation}
\det\bar{M}=\sum_{\sigma\in\mathfrak{S}_N}\det M^\sigma\ ,
\end{equation}
with $M^\sigma_{ik}=M^{\sigma(i)}_{ik}$. This relation permits one to
eliminate the $1/N!$ factor in \eqref{bigsum2}, which then reads
\begin{equation}
\label{bigsum3}
\begin{aligned}
\left\langle \sigma^{x}_{\ell}\left( t\right) \right\rangle ={\rm Re}\Bigg[\frac{\sqrt{\xi}}{L^{2N}}\sum_{0<p_1,...,p_N}&\prod_{j=1}^N4\frac{f(p_j)}{f(q_j)}\sin p_j\sin q_je^{2it(\varepsilon(p_j)-\varepsilon(q_j))} \prod_{i,j}\frac{\varepsilon^4_{q_jp_i}}{\varepsilon^2_{q_jq_i}\varepsilon^2_{p_jp_i}}\det M\Bigg]\ ,
\end{aligned}
\end{equation}
where $f(p)$ is defined in \eqref{f} and $M$ is explicitly given by 
\begin{equation}
M_{ij}=\frac{1}{\cos q_i-\cos p_i}\frac{1}{\cos q_j-\cos p_i}\ .
\end{equation}

\subsubsection{Approximation \label{appro}}
In order to proceed we now drop one of the factors in \fr{bigsum3},
which we argue is justified at late times.
%Now enters the approximation that we make in order to focus on the
%leading time behaviour only. 
The factor involving the $\varepsilon_{k,k'}$ in \eqref{bigsum3} is a
function 
\be
g_{\{q_1,...,q_N\}}(p_1,...,p_N)=
\prod_{i,j}\frac{\varepsilon^4_{q_jp_i}}{\varepsilon^2_{q_jq_i}\varepsilon^2_{p_jp_i}}
\ee
such that for $\sigma\in\mathfrak{S}_N$
\begin{align}
\label{approx}
&\text{(i)}\ g_{\{q_1,...,q_N\}}(p_1,...,p_N)\text{ is regular, symmetric and has no poles in }p_1,...,p_N;\nn
&\text{(ii)}\ g_{\{q_1,...,q_N\}}(q_{\sigma(1)},...,q_{\sigma(N)})=1;\nn
&\text{(iii)}\ \forall k=1,...,N,\quad g_{\{q_1,...,q_N\}}(p_1,...,p_N)\vert_{\forall j=k+1,...,N,\,p_j=q_{\sigma(j)}}=g_{\{q_{\sigma(1)},...,q_{\sigma(k)}\}}(p_1,...,p_k);\nn
&\text{(iv)}\ \forall
i=1,...,N,\quad\partial_{p_i}g_{\{q_1,...,q_N\}}(p_1,...,p_N)\vert_{\forall
  j\   p_j=q_{\sigma(j)}}=0\ \ . 
\end{align}

The first two properties (i) and (ii) follow immediately from the
definition \fr{veab}. The third one (iii) means that if some $p$'s are
set to some $q$'s in a one-to-one fashion, then one recovers the same
function $g$ with the remaining $p$'s and $q$'s. As for property (iv),
it means that (ii) holds at order $(p_i-q_{\sigma(i)})^2$. 

We will now argue that by virtue of these properties setting $g$ to
$1$ does not affect the leading behaviour at late times $t$. 
First, property (i) ensures that $g$ does not modify the general
structure \eqref{decomp} by allowing e.g. for higher order poles, or
poles at other momenta. Property (ii) ensures that $g$ does not
modify the values $A (I_{0}, I_1, I_2)$ when $I_1=\{\}$, because in
these cases there is always a double zero to differentiate and $g$
is then evaluated at a permutation of the momenta. When $I_1\neq
\{\}$, the function $g$ does change $A (I_{0}, I_1, I_2)$, but, 
because of property (iv), it will not modify the pairing structure of
\eqref{phia}, and will only modify the 
%inner coefficient 
factors in \fr{phia} by an extra additive term. Finally, property
(iii) allows one to repeat these steps recursively in
\eqref{rec}. We now observe that the resulting partial fraction
decomposition will always boil down to evaluating 
sums of the form \eqref{qqq}. The contribution of $g$ is an additional term to \eqref{qqq}, and so the leading time
behaviour will never depend on $g$.  

Based on these arguments we now make the approximation of setting
$g=1$
\begin{equation}
\label {matrix}
\begin{aligned}
\langle \sigma^{x}_{\ell}\left( t\right) \rangle
&\approx{\rm Re}\frac{4^N\sqrt{\xi}}{L^{2N}}\sum_{\substack{0<p_1,...,p_N\\\in \rm{R}}}\underset{i,j}{\det} \Bigg\lvert \frac{e^{it(2\varepsilon(p_i)-\varepsilon(q_i)-\varepsilon(q_j))}\sin p_i\sin q_jf(p_i)/f(q_j)}{(\cos q_i-\cos p_i)(\cos q_j-\cos p_i)} \Bigg\rvert\\
    &=\sqrt{\xi}{\det}\big(A\big)\ ,
\end{aligned}
\end{equation}
where the matrix $A$ is given by
\begin{equation}
\label {matrixA}
\begin{aligned}
A_{ij}=\frac{4}{L^2}\sum_{0<p\in\rm{R}}\frac{e^{it(2\varepsilon(p)-\varepsilon(q_i)-\varepsilon(q_j))}\sin p\sin q_jf(p)/f(q_j)}{(\cos
  q_i-\cos p)(\cos q_j-\cos p)}\ ,\quad i,j=1,\dots,N\ .
\end{aligned}
\end{equation}

We note that the above analysis is very similar to the one employed by
Korepin and Slavnov in their work on the single-particle Green's
function in the impenetrable Bose gas \cite{korepinslavnov}.

%%%%%%%%%%%%%%%%%%%%%%%%%%%%%%%%%%%%%%%%%%%%%
\subsection {Asymptotic forms of the matrix elements \label{asympfme}}
%%%%%%%%%%%%%%%%%%%%%%%%%%%%%%%%%%%%%%%%%%%%%
In the next step we work out the large-$L$ asymptotics of the
  matrix elements $A_{ij}$.
%%%%%%%%%%%%%%%%%%%%%%%%%%%%%%%%%%%%%%%
\subsubsection {Diagonal matrix elements}
%%%%%%%%%%%%%%%%%%%%%%%%%%%%%%%%%%%%%%%
The diagonal matrix elements were already computed in \eqref{use2},
\emph{cf.} Appendix \ref{appen}. They read 
\begin{equation}
\begin {aligned}
A_{ii}&=1-\frac {\left| \theta _{i}\right| } {\pi }+\frac {2i\sign(t\varepsilon'(q_i))} {L}\frac {f'(q_i)} {f(q_i)}+\mathcal{O}\left( L^{-1}t^{-1/2}\right) \ ,
  \end {aligned}
\label{Aii}
\end{equation}
where we have defined
\begin{equation}
\label{thet}
\theta _{i}=\frac {4\pi t\varepsilon'\left( q_{i}\right) } {L}\ .
\end{equation}
%%%%%%%%%%%%%%%%%%%%%%%%%%%%%%%%%%%%%%%%%%%%%%%%%
\subsubsection {Off-diagonal matrix elements}
%%%%%%%%%%%%%%%%%%%%%%%%%%%%%%%%%%%%%%%%%%%%%%%%%
To compute the off-diagonal elements, we write 
\begin{equation}
\frac {1} {\left( \cos q_{i}-\cos
  p\right) \left( \cos q_{j}-\cos p\right) }=\frac {1} {\cos
  q_{i}-\cos q_{j}}\left(\frac {1} {\cos
  q_{j}-\cos p}-\frac {1} {\cos q_{i}-\cos
  p}\right)\ ,
\end{equation}
and use the first equation in \eqref{use2}, see also Appendix
\ref{appen}. We obtain 
  \begin{equation}
A_{ij}=\frac {2i\sign(t \varepsilon'(q_j))\sin q_{j}} {L(\cos  q_{i}
  -\cos q_{j}) }\Big[ e^{it\left( \varepsilon( q_{j}) -\varepsilon(
    q_{i}) \right) }-\frac {f(q_i)\sign(\varepsilon'(q_j)
    \varepsilon'(q_i))} {f(q_j) }e^{it\left( \varepsilon( q_{i})
    -\varepsilon( q_{j}) \right) }\Big]\ . 
\label{Aij}
\end{equation}
%%%%%%%%%%%%%%%%%%%%%%%%%%%%%%%%%%%%%%%%%%%%%%%%%
\subsubsection {Approximate determinant representation}
%%%%%%%%%%%%%%%%%%%%%%%%%%%%%%%%%%%%%%%%%%%%%%%%%
Combining \fr{Aii} and \fr{Aij} provides the following approximate determinant representation
\begin{equation}
\begin{aligned}
 \left\langle \sigma^{x}_{\ell}\left( t\right) \right\rangle &\approx{\rm Re}\sqrt {\xi }\det(I-\Xi)\ ,
\end{aligned}
\end{equation}
where
\begin{equation}
\begin {aligned}
\Xi_{ij}=&\delta_{ij}\Big[\frac{\vert\theta_i\vert}{\pi} -\frac
  {2i\sign(t\varepsilon'(q_i))} {L}\frac{f'(q_i)}{f(q_i)} \Big]\\
+&(1-\delta_{ij}) \frac {2i\sign(t \varepsilon'(q_j))\sin q_{j}} {L(\cos  q_{j} -\cos q_{i}) }\Big[ e^{it\left( \varepsilon( q_{j}) -\varepsilon( q_{i}) \right) }-\frac {f(q_i)\sign(\varepsilon'(q_j) \varepsilon'(q_i))} {f(q_j) }e^{it\left( \varepsilon( q_{i}) -\varepsilon( q_{j}) \right) }\Big]\\
+&\mathcal{O}\left( L^{-1}t^{-1/2}\right)\ .
\end {aligned}
\end{equation}
%%%%%%%%%%%%%%%%%%%%%%%%%%%%%%%%%%%%%%%%
\subsection {Evaluating the determinant}
%%%%%%%%%%%%%%%%%%%%%%%%%%%%%%%%%%%%%%%%
We now write $\det M=\exp\Tr\log M$ and use the expansion
\begin{equation}
\log(I-\Xi)=-\sum_{n\geq 1}\frac{\Xi^n}{n}\ .
\label{logxi}
\end{equation}
\subsubsection {First order}
The first order gives 
\begin{equation}
\begin{aligned}
\Tr\Xi&=4\vert t\vert\int_0^\pi \vert \varepsilon'(x)\vert \rho(x)dx+2i\int _{0}^{\pi }\frac{f'(x)}{f(x)}\rho (x)dx+\mathcal{O}(L^{0}t^{-1/2})\\
&=4\vert t\vert\int_0^\pi \vert \varepsilon'(x)\vert \rho(x)dx+\mathcal{O}(L^{0}t^{-1/2})\ .
\end{aligned}
\label{xi1}
\end{equation}
%%%%%%%%%%%%%%%%%%%%%%%%%%%%%%%%
\subsubsection {Second  order}
%%%%%%%%%%%%%%%%%%%%%%%%%%%%%%%%
Since $ \sum _{i}\Xi _{ii}^{2}=\mathcal{O}\left( L^{-1}\right) $ the second order reads 
\be
\Tr (\Xi^2)=\sum_{i\neq j}\Xi_{ij}\Xi_{ji}+{\cal
  O}(L^{-1})=S_1+S_2+{\cal  O}(L^{-1})\ ,
\ee
where
\be
S_1=\frac {8} {L^{2}}\sum_{i\neq j}\frac {\sin q_{i}\sin q_{j}}
      {\left( \cos q_{i}-\cos  q_{j}\right) ^{2}}\ ,\quad
S_2=-\frac {8}
      {L^{2}}\sum _{i\neq j}\frac {\sin q_{i}\sin q_{j}
        \tfrac{f(q_i)}{f(q_j)}} {\left( \cos q_{i}-\cos
        q_{j}\right) ^{2}} e^{2it\left( \varepsilon \left( q_{i}\right) -\varepsilon
          \left( q_{j}\right) \right) }\ .
\ee
These sums are computed in Appendix \ref{appen}. We obtain
\begin{equation}
\begin{aligned}
\Tr (\Xi^2)&=
4\vert t\vert\int_0^\pi dx\ \vert \varepsilon'(x)\vert 4\pi\rho(x)^2
+ 8\int dx dy\ \frac {\sin y} {\cos y-\cos x}\rho '(x)
\rho (y)+\mathcal{O}\left( t^{-1/2}\right)\ .
\end{aligned}
\label{xi2}
\end{equation}
Once exponentiated in the determinant, we recognize the terms obtained
earlier with the partial fraction decomposition approach
\eqref{cquench}, \fr{ccquench}, but without the contributions
  involving the $\varepsilon$ in the prefactor $C$. This difference is
  a direct consequence of our approximation $g=1$.
%\begin{equation}
%\begin{aligned}
%\tr (\Xi^2)&=
%4\vert t\vert\int_0^\pi dx\ \vert \epsilon'(x)\vert 4\pi\rho(x)^2
%+ 8\int dx dy\ \frac {\sin y} {\cos y-\cos x}\rho '\left( x\right)
%\rho \left( y\right)\nn
%&- 8\pi i\int \frac {\partial _{2}f\left( x,x\right) } {\sin
%  ^{2}x}\rho ^{2}\left( x\right) dx +O\left( t^{-1/2}\right).
%\end{aligned}
%\end{equation}

%%%%%%%%%%%%%%%%%%%%%%%%%%%%%%%%%%%%%%%%%%%%%%%%%%%%%%%%%%%%%%%
\subsubsection{Leading late-time contribution at all orders}
%%%%%%%%%%%%%%%%%%%%%%%%%%%%%%%%%%%%%%%%%%%%%%%%%%%%%%%%%%%%%%%
In order to compute the $ {\cal O}(t)$ term at higher orders in $\rho$, we first
notice that it can only arise from the second order poles in the matrix
entries, hence by pairs of momenta separated by $ o (L^0)$. In this
regime the matrix elements become 
\begin{equation}
\label{leadinghigher}
  \Xi_{ij}=\delta _{ij}\frac {\left| \theta _{i}\right| } {\pi }+\left( 1-\delta _{ij}\right) \frac {-i\sign(\theta)} {\pi }\frac{e^{i\theta _{i}T_{i}\Delta _{ij} / 2}-e^{-i\theta _{i}T_{i}\Delta _{ij} / 2}}{T_{i}\Delta _{ij}}\ ,
\end{equation}
with $ \Delta _{ij}=i-j$. 
We obtain in this regime 
\begin{equation}
\begin{aligned}
\Tr (\Xi^n)&=\sum_{i_1,i_2,...,i_n}\Xi_{i_1i_2}...\Xi_{i_ni_1}\\
&=-\sum_i\frac{1}{(i\sign(\theta_i)\pi)^nT_i^n}\sum_{\Delta_1,...,\Delta_{n-1}\neq
  0}\Big[\prod_{m=1}^{n-1}\frac{1-e^{-i\theta_iT_i\Delta_m}}{\Delta_m}\Big]
\frac{1-e^{i\theta_iT_i(\Delta_1+...+\Delta_{n-1})}}{\Delta_1+...+\Delta_{n-1}}\ .
\end{aligned}
\end{equation}
The sums over $\Delta_j$ can be carried out using that for $\Delta'\neq 0$
\begin{equation}
  \sum _{\Delta \neq 0,-\Delta'}\frac {1-e^{-i\theta \Delta }} {\Delta }\frac {1-e^{i\theta \left( \Delta +\Delta'\right) }} {\Delta +\Delta '}=2i(\pi-|\theta|) \sign(\theta)\frac {1-e^{i\theta \Delta '}} {\Delta '}\ ,
\label{433}
\end{equation}
which is obtained from
$\tfrac{1}{\Delta(\Delta+\Delta')}=\tfrac{1}{\Delta'}(\tfrac{1}{\Delta}-\tfrac{1}{\Delta+\Delta'})$
and relations \eqref{sumuseful} by carefully treating the cases
$\Delta=0,-\Delta'$. Using that $\theta=\mathcal{O}(L^{-1})$ to neglect the
$|\theta|$ term in \fr{433} we arrive at
\begin{equation}
\begin{aligned}
\Tr (\Xi^n)&=-\sum_i\frac{(2i\sign(\theta_i)\pi)^{n-2}}{(i\sign(\theta_i)\pi)^nT_i^n}\sum_{\Delta_1\neq 0}\frac{(1-e^{-i\theta_iT_i\Delta_1})(1-e^{i\theta_iT_i\Delta_1})}{\Delta_1^2}+\mathcal{O}(L^{-1})\\
&=\sum_i\frac{2^{n-1}}{\pi^2T_i^n}(2\li(1)-\li(e^{i\theta_i T_i})-\li(e^{-i\theta_i T_i}))+\mathcal{O}(L^{-1})\\
&=\sum_i\frac{2^{n-1}\vert\theta_i\vert}{\pi T_i^{n-1}}+\mathcal{O}(L^{-1})\\
&=4\vert t\vert\int_0^\pi \vert \varepsilon'(x)\vert (4\pi\rho(x))^{n-1}\rho(x)dx+\mathcal{O}(L^{-1})\ .
\label{xin}
\end{aligned}
\end{equation}
\subsubsection {Influence of the boundaries\label{boundaries}}
In the discussion above the momenta $p$ are constrained to be
positive. In order to use equations \eqref{sumsval} we therefore had to neglect
possible boundary effects for $q$'s close to zero. We now verify that
this does not influence the result. We have 
\begin{equation}
  \sum _{n=-\eta _{1}L}^{\eta _{2}L}\frac {e^{i\left( n+1 / 2\right) \frac {w} {L}t}} {\left( n+1 / 2\right) ^{2}}= \sum _{n=-\infty}^{\infty}\frac {e^{i\left( n+1 / 2\right) \frac {w} {L}t}} {\left( n+1 / 2\right) ^{2}}-\frac {1} {\eta _{2}L}E_{2}\left( -iw\eta _{2}t\right) -\frac {1} {\eta _{1}L}E_{2}\left( iw\eta _{1}t\right) \ ,
\end{equation}
with $ E_{2}(x) =\int _{1}^{\infty }\frac {e^{-xt}} {t^{2}}dt$ the exponential integral function. $ L\eta _{1,2}$ are the number of vacancies between $ 0 , \pi$ and $ q _i $. They are $ \eta _{1}=\frac {q_{i}} {2\pi }$ and $ \eta _{2}=\frac {\pi -q_{i}} {2\pi }$. Hence the correction to $ \Tr \Xi $ is
\begin{equation}
  -\frac {2} {\pi }\int _{0}^{\pi }\left(\frac {E_{2}\left( -2i\varepsilon '(x) xt\right) } {x}+\frac {E_{2}\left( 2i\varepsilon '(x) \left( \pi -x\right) t\right) } {\pi -x}\right)\rho(x) dx\ ,
\end{equation}
which goes to zero for large $ t $ because  the density vanishes quadratically at $ 0 $ and $ \pi $. 

%%%%%%%%%%%%%%%%%%%%%%%%%%%%
\subsection{Result: late-time asymptotics of the order parameter after
a quench}
%%%%%%%%%%%%%%%%%%%%%%%%%%%%
Substituting \fr{xi1}, \fr{xi2} and \fr{xin} into
\fr{logxi} we arrive at the following result for the late-time
asymptotics of the order parameter one-point function after a quench
within the ferromagnetic phase
\begin{equation}
\label {res}
\frac{\langle \Psi_N(t)\vert\sigma_\ell^x \vert\Psi_s(t)\rangle }{\langle
  \Psi_N(t)\vert\Psi_s(t)\rangle}=C\exp\left(\frac{\vert
  t\vert}{\pi}\int_0^\pi \vert\varepsilon'(x)\vert
\log(1-4\pi\rho(x))dx\right)\equiv Ce^{-t/\tau}\ ,
\end{equation}
with $C$ given in \eqref{ccquench} at order ${\cal O}(\rho_Q^2)$. The
decay rate reproduces the exact result obtained in
\cite{CEF1,CEF2}. However, the prefactor $C$ differs from the one 
conjectured in \cite{CEF1,CEF2}. We address this difference in Section 
\ref{numerical} below.  

%%%%%%%%%%%%%%%%%%%%%%%%%%%%%%%%%%%%%%%%%%%%%%%%%%
\subsection {Numerical Checks}
\label{numerical}
%%%%%%%%%%%%%%%%%%%%%%%%%%%%%%%%%%%%%%%%%%%%%%%%%%

We now present some numerical checks of equation \eqref{res}
for the time evolution of the order parameter. 
In the limit of large separations
$\ell\gg 2Jt$, using the Lieb-Robinson bound and the clustering properties of
the initial state $|\Psi\rangle$, the two-point function factorizes into the
product of two one-point functions that are identical by translational invariance
\be
\left\langle\Psi| \sigma _{\ell+1}^{x}\left( t\right) \sigma _{1}^{x}\left(
t\right)|\Psi \right\rangle=
%\left\langle\Psi| \sigma^{x}_{\ell+1}\left(t\right)|\Psi \right\rangle
\left\langle\Psi| \sigma^{x}_1\left(t\right)|\Psi
\right\rangle^2+\mathcal{O}(e^{-\gamma (\bell-2Jt)})\, ,
\ee
with $\gamma$ a constant of order $1$. 
We can then obtain the one-point function
$ \left\langle\Psi| \sigma^{x}_1\left(t\right)|\Psi \right\rangle $ as
the square root of the two-point function
$ \left\langle \sigma _{\ell+1}^{x}\left( t\right) \sigma _{1}^{x}\left(
t\right) \right\rangle $ in the limit 
$\ell\gg 2Jt$, which can be efficiently computed
numerically, as it can be expressed as the determinant of a block Toeplitz matrix even in the
thermodynamic limit \cite {CEF2}. 

\begin{figure}[H]
\begin{tikzpicture}[scale=0.9]
\begin{axis}[
    enlargelimits=false,
    xlabel = $Jt$,
    ymin=0.9651,
    ymax =0.9657,
    xmin =30 , 
     y tick label style={
        /pgf/number format/.cd,
            fixed,
            fixed zerofill,
            precision=4,
        /tikz/.cd
    }
]
\addplot+[
    only marks,
    mark=+,
    mark size=2.9pt,
    color=red]
table{tenew.dat};
\addplot [
    domain=0:150, 
    samples=100, 
    color=blue,
    ]
    {0.96549815};
\addplot [
    domain=0:150, 
    samples=100, 
    color=brown,
    dashed
    ]
     {0.96525};
     \node[] at (axis cs: 132,0.965215) {\small $\sqrt{C^x_{\rm FF}(\infty)}$};
\end{axis}
\end{tikzpicture}
%\begin{tikzpicture} [scale=1]
%\begin{axis}[
%    enlargelimits=false,
%    xlabel = $Jt$,
%    ymin=0.976,
%    ymax =0.981,
%    xmin =10 , 
%    y tick label style={
%        /pgf/number format/.cd,
%            fixed,
%            fixed zerofill,
%            precision=3,
%        /tikz/.cd
%    }
%]
%\addplot+[
%    only marks,
%    scatter,
%    mark=+,
%    mark size=2.9pt,
%    color=red]
%table{0v7to_0v3.dat};
%\addplot [
%    domain=0:150, 
%    samples=100, 
%    color=blue,
%    ]
%    {0.980649};
%\end{axis}
%\end{tikzpicture}
\begin{tikzpicture}[scale=0.9,spy using outlines={rectangle, magnification=1.6,connect spies}]
\begin{axis}[
    enlargelimits=false,
    xlabel = $h$,
    ymin =0.996, 
    ymax =1.05,
     y tick label style={
        /pgf/number format/.cd,
            fixed,
            fixed zerofill,
            precision=2,
        /tikz/.cd
    }
]
\addplot[
    only marks,
    mark=+,
    mark size=2.9pt,
    color=red]
table{prefactor.dat};

%\addplot[
%    color=blue]
%table{prefactor_th.dat};
\addplot[
    color=blue]
table{prefquench.dat};

\coordinate (spypoint) at (axis cs:0.5,1.0028);
\coordinate (spyviewer) at (axis cs:0.45,1.035);
\spy[width=5cm,height=2cm] on (spypoint) in node [fill=white] at (spyviewer);
\end{axis}
\end{tikzpicture}
\caption {Left: Numerical results for $\langle \sigma^{x}_1\left(
  t\right) \rangle e^{t / 
    \tau }$ with $\tau$ defined in \eqref{res} as a function of $t$
  for a quench from $h_{0}=0.1$ to $h=0.5$, corresponding to a
    density $\rho_Q=0.0218$. The prefactor at order ${\cal
      O}(\rho_Q^2)$ \eqref{ccquench} is shown in blue
  and is seen to be compatible with our numerical results. For
  comparison we also show $\sqrt{C^x_{\rm FF}(\infty)}=0.96525$. 
  Right: Numerically determined (red) and calculated leading order
  (blue) prefactor \eqref{ccquench} divided by $\sqrt {\xi}$, for
  quenches from $h_0=0.1$ to $h$ as a function of $h$. Because of the
  oscillations, the measured value is an average of six points between
  times $t=75$ and $100$.}  
\label {prefactorf}
\end{figure}
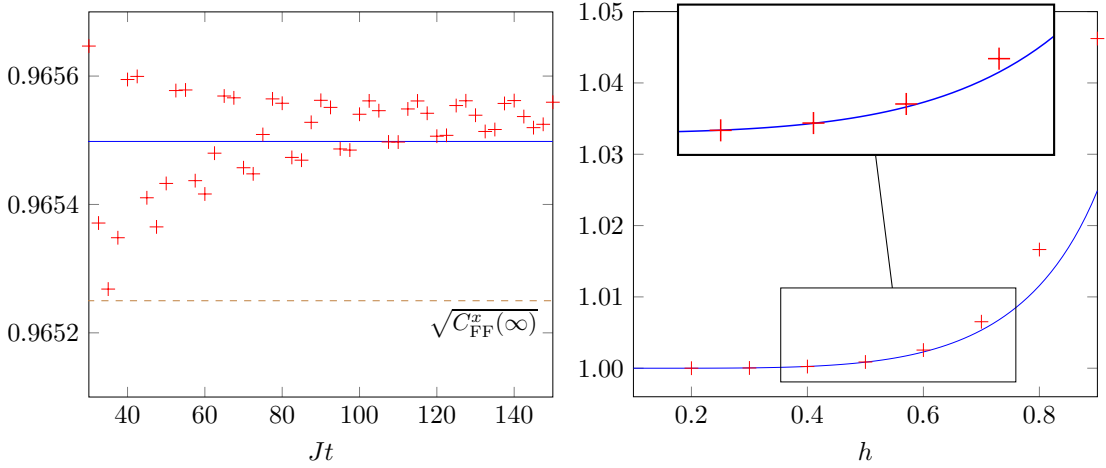

Numerical checks of the exponential decay in \eqref {res} have already 
been reported in \cite {CEF2}. Since our prediction \fr{ccquench} differs from the one conjectured
in \cite {CEF2}, we will  focus on the
prefactor ${\cal C}^x_{\rm FF}(\alpha)$ of the asymptotic behaviour of the two-point
function in the limit $\ell,t\to\infty$, $\alpha=t/\ell$ fixed
\begin{align}
\left\langle\Psi| \sigma _{\ell+1}^{x}\left( t\right) \sigma _{1}^{x}\left(
t\right)|\Psi \right\rangle &\simeq {\cal C}^x_{\rm FF}(\alpha)
\exp\Big[\ell \int_0^\pi \frac{\mathrm d k}{\pi}\log\left|\cos\Delta_k\right|
\theta_H\big(2\varepsilon'_h(k)t-\ell\big)\Big] 
\nn &\times
\exp\Big[
2t \int_0^\pi\frac{\mathrm d k}{\pi} \varepsilon'_h(k)
\log\left |\cos\Delta_k\right|\theta_H\big(\ell-2\varepsilon'_h(k)t\big)\Big].
\label{eq:predictionintr}
\end{align}
 In \cite {CEF2} it was assumed that the constant $C^x_{\rm FF}(\alpha)$ is 
 independent of $\alpha$.
Calculating the asymptotics of the correlator for $\alpha\to\infty$
then leads to \cite {CEF3}
\be
C^x_{\rm FF}(\infty)=\frac{1-h h_0+\sqrt{(1-h^2)(1-h_0^2)}}{2\sqrt{1-h 
    h_0}\sqrt[4]{1-h_0^2}}\ .
\ee
From \eqref{ccquench} it however follows $C\neq \sqrt{C^x_{\rm FF}(\infty)}$, suggesting in turn that
$C^x_{\rm FF}(\alpha)$ is in fact $\alpha$-dependent. This is indeed
supported by our numerical results, even though the difference
$|C-\sqrt{C^x_{\rm FF}(\infty)}|$ is tiny.   
In Figs~\ref {prefactorf} we show that our results \eqref {res} and
\eqref{ccquench} are in agreement with numerical calculations of the
order parameter one-point function.

\section{Dynamical correlation functions at arbitrary finite temperatures\label{dynamicatemp}}
%%%%%%%%%%%%%%%%%%%%%%%%%%%%%%%%%%%%%%%%%%%%%%%%%%%%%%%%%%%%%%%%%%%%%%%%%%%%%%%%%%
In this section we follow the same reasoning as in Section \ref{qqd}
but for dynamical correlation functions at finite temperature. 
%Namely, we obtain them as $C e^{-t/\tau}$ with $\tau$ including all
%orders in $\rho$. 
We again treat the two cases $h<1$ and $h>1$ separately.

\subsection {\texorpdfstring{Ordered phase $h<1$}{Lg}}
%\subsubsection {An approximate determinant representation}
In the ordered phase, the sum in \eqref {armtap} involves states with
the same number of particles as in the quench case \eqref {bigsum}. However,
it is not possible to express each term as a Cauchy determinant as 
in Section \ref{qqd}. This can nevertheless be overcome for the leading late time behaviour, where one can work with an
approximate version of the form factor \eqref{ff}.

We focus again on intermediate states with $M=N$ in \eqref{armtap}, which were argued in Section \ref{different} to give the leading late time
behaviour. As discussed in Section \ref{appro}, the dominant
contribution to the correlation function arises from the sums \eqref{type1}, and the term
proportional to $t$ on the right-hand side of \eqref{type1} has its
origin in the isolated singularity $\frac{1}{x^2}$ in $\frac{1}{\sin^2
  x}$. Hence, replacing terms of the form $\frac{1}{\sin^2 x}$
by $\frac{1}{x^2}$ in the form factor \eqref{ff} will not affect the
leading behaviour at late times. It follows that if we define
\begin{align}
\label {armtapap}
\tilde{\chi}^{xx}\left( \ell,t\right) &=\frac {1} {N!}\sum
_{\substack{p_{1} ,  \ldots,
    p_{N}\\\in\rm{R}}}\tilde{F}_{\{p_1,...,p_N\}}^{\{q_1,...,q_N\}}
e^{it(E\left( \left\{ q\right\} \right) -E\left( \left\{ p\right\}
  \right) )+ i\bell(P\left( \left\{ p\right\} \right) -P\left( \left\{
  q\right\} \right) )}\ ,\\ 
\label{ff2a}
\tilde{F}_{\{p_1,...,p_N\}}^{\{q_1,...,q_N\}}&=\frac {\xi } {L^{2N}}\frac {\prod _{i\neq j}\frac {q_{i}-q_{j}} {2}   \frac {p_{i}-p_{j}} {2} } {\prod _{i,j}\left( \frac {q_{i}-p_{j}} {2}\right)^2 }\ ,
\end{align}
then we have
\be
{\chi}^{xx}(\ell,t)=\widetilde{\chi}^{xx}(\ell,t)\kappa(\ell,t)\ ,
\ee
with $\kappa(\ell,t)$ a function that is subleading in $\ell,t$ with respect to ${\chi}^{xx}(\ell,t)$.

The form factor $\tilde{F}$ has the same Cauchy determinant structure
we encountered in the quench case \eqref{cauch}. Hence we can follow
through the same steps and obtain 
\begin{equation}
\tilde{\chi}^{xx}\left( \ell,t\right) =\xi\underset{i,j=1,...,N}{\det}\left[\frac{4}{L^2}\sum_{p\in\rm{R}}\frac{e^{i\tfrac{t}{2}(\overline{\varepsilon}(q_i)+\overline{\varepsilon}(q_j)-2\overline{\varepsilon}(p)})}{(q_j-p)(q_i-p)}\right]\ .
\end{equation}
The leading time behaviour of the coefficients of this matrix can be computed as in Section \ref{asympfme} with formulas \eqref{sumuseful}. One obtains
\begin{equation}
\tilde{\chi}^{xx}\left( \ell,t\right) =\xi\det(1-\Xi)\ ,
\end{equation}
with
\begin{equation}
\begin {aligned}
\Xi_{ij}=&\delta_{ij}\frac{|2t\overline{\varepsilon}'(q_i)|}{L}\\
+&(1-\delta_{ij}) \frac {2i\sign(t \overline{\varepsilon}'(q_j))} {L(q_j-q_i) }\Big[ e^{i\tfrac{t}{2}\left( \overline{\varepsilon}( q_{i}) -\overline{\varepsilon}( q_{j}) \right) }-\sign(\overline{\varepsilon}'(q_j) \overline{\varepsilon}'(q_i))e^{i\tfrac{t}{2}\left( \overline{\varepsilon}( q_{j}) -\overline{\varepsilon}( q_{i}) \right) }\Big]\\
+&\mathcal{O}\left( L^{-1}t^{-1/2}\right)\ .
\end {aligned}
\end{equation}
Since one is interested only in the late time dynamics, we can focus on terms $i,j$ such that $q_i-q_j=o(L^0)$ as in \eqref{leadinghigher}. Then one obtains the same formula as in \eqref{leadinghigher}
\begin{equation}
  \Xi_{ij}=\delta _{ij}\frac {\left| \theta _{i}\right| } {\pi }+\left( 1-\delta _{ij}\right) \frac {-i\sign(\theta)} {\pi }\frac{e^{i\theta _{i}T_{i}\Delta _{ij} / 2}-e^{-i\theta _{i}T_{i}\Delta _{ij} / 2}}{T_{i}\Delta _{ij}}\ ,
\end{equation}
with $\Delta _{ij}=i-j$, $T_{i}=\frac{1}{2\pi\rho(q_i)}$ and $\theta _{i}=\frac{2\pi t\overline{\varepsilon}'(q_i)}{L}$.

\subsubsection {Two-point dynamical correlation functions in the ordered phase}
 
Following through the same steps as in the quench case we arrive at
\begin{equation}
\label{chitemperatureordered}
\begin{aligned}
\chi^{xx}(\ell,t)&= 
%C \exp\left(\frac{1}{2\pi}\int_{-\pi}^\pi |t\overline{\varepsilon}'(x)| \log(1-4\pi\rho(x))dx\right)\\
%&=C 
C\exp\left(\frac{1}{2\pi}\int_{-\pi}^\pi|t\varepsilon'(x)-\bell|
\log(1-4\pi\rho(x))dx\right)\ , 
\end{aligned}
\end{equation}
where, using the results of Section \ref{pfd},
\begin{equation}
\label{crho}
C=\begin{cases}
\xi \exp\left(-2\int _{-\pi }^{\pi}\int _{-\pi }^{\pi}\frac {\rho
  (y) \rho '(x) } {\tan \left( \frac {x-y}
  {2}\right) }dxdy\right)\qquad\text{in the space like-region at order
}\rho_\beta^2\\ 
\xi \qquad\text{in the time like-region at order }\rho_\beta^0\ .\\
\end{cases}
\end{equation}
As far as we know the result \fr{chitemperatureordered} has not
  previously been obtained in the literature.
\subsubsection{Numerical checks}
%%%%%%%%%%%%%%%%%%%%%%%%%%%%%%%%%%

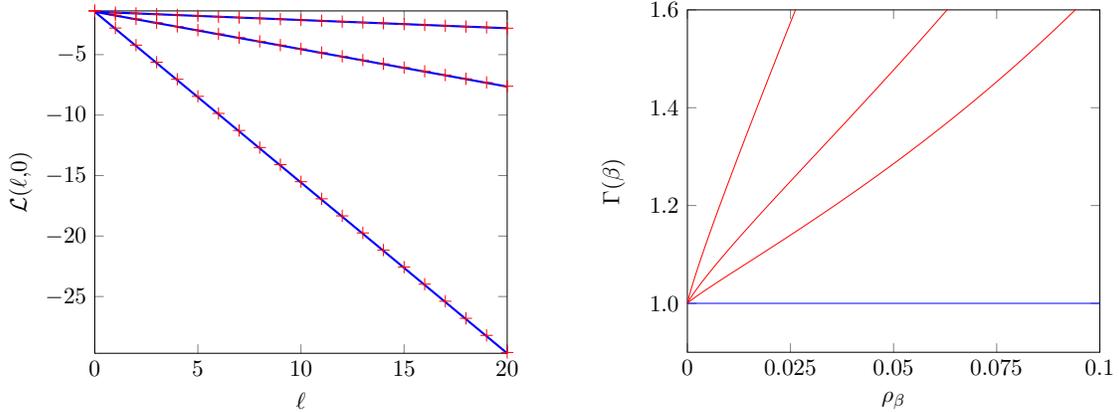
\begin{figure}[H]
\begin{center}
\begin{tikzpicture}[scale=0.8]
\begin{axis}[
    enlargelimits=false,
    xlabel = $\ell$,
    ylabel=${\cal L}(\ell{,}0)$,
     y tick label style={
        /pgf/number format/.cd,
            fixed,
            fixed zerofill,
            precision=0,
        /tikz/.cd
    }
]
\addplot[
    only marks,
    mark=+,
    mark size=2.9pt,
    color=red]
table[
           x expr=\thisrowno{0}, 
           y expr=ln(\thisrowno{1} )
         ]{temp_h0v5_b0v25.dat};

\addplot[
    line width=1pt,
    color=blue]
table[
           x expr=\thisrowno{0}, 
           y expr=ln(\thisrowno{1} )
         ]{temp_h0v5_b0v25_th.dat};

\addplot[
    only marks,
    mark=+,
    mark size=2.9pt,
    color=red]
table[
           x expr=\thisrowno{0}, 
           y expr=ln(\thisrowno{1} )
         ]{temp_h0v5_b2.dat};

\addplot[
    line width=1pt,
    color=blue]
table[
           x expr=\thisrowno{0}, 
           y expr=ln(\thisrowno{1} )
         ]{temp_h0v5_b2_th.dat};

\addplot[
    only marks,
    mark=+,
    mark size=2.9pt,
    color=red]
table[
           x expr=\thisrowno{0}, 
           y expr=ln(\thisrowno{1} )
         ]{temp_h0v5_b1.dat};

\addplot[
    line width=1pt,
    color=blue]
table[
           x expr=\thisrowno{0}, 
           y expr=ln(\thisrowno{1} )
         ]{temp_h0v5_b1_th.dat};

\end{axis}
\end{tikzpicture}
\qquad
\begin{tikzpicture}[scale=0.8]
\begin{axis}[
    enlargelimits=false,
    xlabel =$\rho_\beta$,
    ylabel =$\Gamma(\beta)$,
    ymin=0.9,
    ymax =1.6,
    xmin =0 , 
    xmax=0.1,
    xtick={0,0.025,0.05,0.075,0.1},
    xticklabels={0,0.025,0.05,0.075,0.1},
    y tick label style={
        /pgf/number format/.cd,
            fixed,
            fixed zerofill,
            precision=1,
        /tikz/.cd
    }
]
\addplot[red] table[
x index=1,
y index=2,
mark=none]
{tempconst_h0v25_T_r_okrho.dat};
\addplot[red] table[
x index=1,
y index=2,
mark=none]
{tempconst_h0v5_T_r_okrho.dat};
\addplot[red] table[
x index=1,
y index=2,
mark=none]
{tempconst_h0v75_T_r_okrho.dat};
 \addplot [
    domain=0:1, 
    samples=100, 
    color=blue,
    ]
    {1};
\end{axis}
\end{tikzpicture}
\end{center}
\caption {Left: 
%$\log\left\langle \sigma ^{x}_{\ell+1}\left( 0\right) \sigma
%  ^{x}_1\left( 0\right)\right\rangle $ 
${\cal L}(\ell,0)$ for  $h=0.5$ and $\beta=2,1,0.25$ (top to
  bottom). Numerical results for a $L=200$ site open chain
are shown as red crosses and equation \eqref{chitemperatureordered} by a
continuous blue line. Right: 
Numerically determined $\Gamma(\beta)$ as a function of $\rho_\beta$ for
$\ell=30$ and $h=3/4,1/2,1/4$ from top to bottom; in blue is the expected
value as $\rho_\beta\to 0$.}  
\label {spaceordered11}
\end {figure}

In order to check the accuracy of \eqref{chitemperatureordered} at
finite times we have carried out numerical simulations following
Ref.~\cite{derzkho}, where the finite temperature dynamical
two-point function for a finite open chain is computed exactly as a
Pfaffian of a known matrix. As long as the two points are sufficiently
far from the boundaries, then they take almost the same values as in an
infinite chain.  
In Fig.~\ref{spaceordered11} we compare \eqref{chitemperatureordered}
to numerical results in the space-like region by considering the
logarithm of the correlator
\be
{\cal L}(\ell,t)=\log\big(\langle\sigma^x_{\ell+1}(t)\sigma^x_1(0)\rangle\big).
\ee
For simplicity we take the extreme case $\alpha=0$, which corresponds
to setting $t=0$. We recall indeed that static correlations are also covered by our calculations, as explained in Section \ref{static}. In the left panel we plot
\eqref{chitemperatureordered} as a function of distance $\ell$ and mark by red crosses
 numerical results obtained for a chain of 
$L=200$ sites. We see that the asymptotic
result \fr{chitemperatureordered} is in excellent agreement for $\alpha=0$. In the right panel we test
the accuracy of the ${\cal O}(\rho_\beta^2)$ value of 
the prefactor \eqref{crho} by considering the quantity
\be
\Gamma(\beta)=\frac{\xi-\left\langle \sigma ^{x}_{\ell+1}\left( 0\right)
    \sigma ^{x}_1\left( 0\right)\right\rangle 
e^{\ell\int_{-\pi}^\pi \frac{dx}{2\pi}\log(1-4\pi\rho(x))}}{\xi-C}\ ,
\ee
where $C$ is given by \fr{crho}.

We see that $\Gamma(\beta)$ approaches $1$ for small values of $\rho_\beta$,
which means that the prefactor \eqref{crho} is indeed correct to order
$\orho{2}$. The linear increase in $\rho_\beta$ shows that the
correction to our result for $C$ is $\orho{3}$.

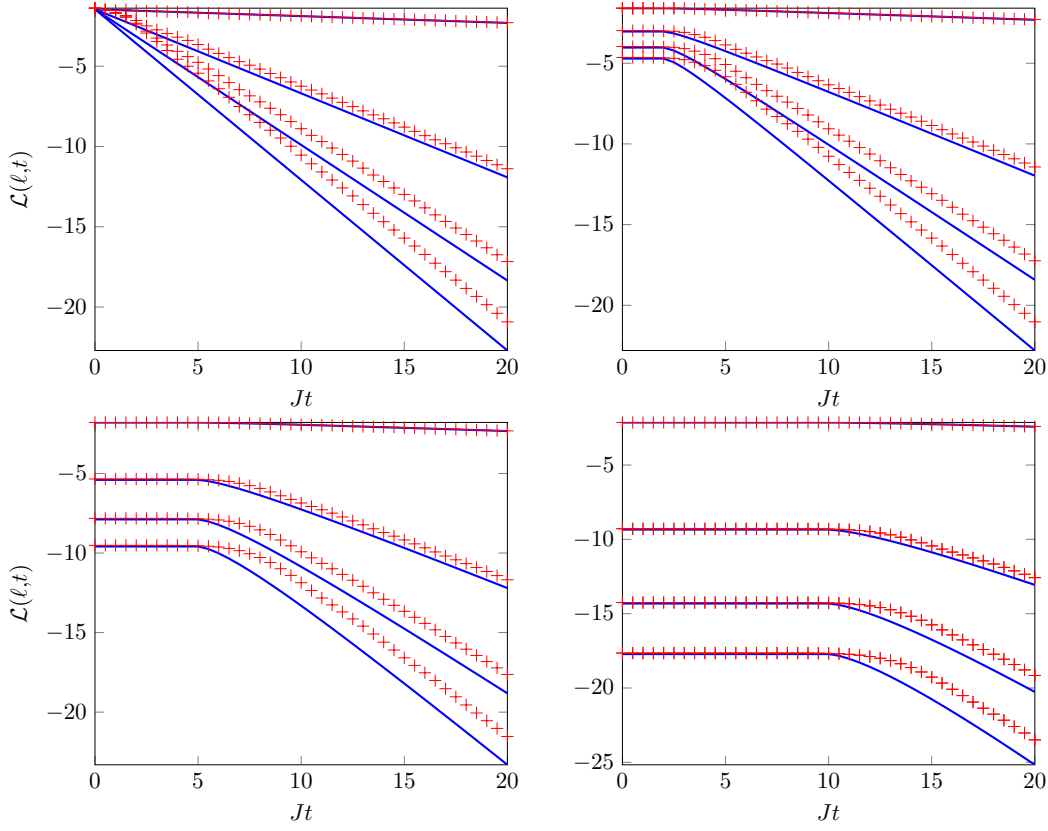
\begin{figure}[H]
\begin{tikzpicture}[scale=0.8]
\begin{axis}[
    enlargelimits=false,
    xlabel = $Jt$,
    ylabel = ${\cal L}(\ell{,} t)$,
    xmax=20,
     y tick label style={
        /pgf/number format/.cd,
            fixed,
            fixed zerofill,
            precision=0,
        /tikz/.cd
    }
]
\addplot[
    only marks,
    mark=+,
    mark size=2.9pt,
    color=red]
table[
           x expr=\thisrowno{3}, 
           y expr=ln(\thisrowno{4}/4 )
         ]{tempord_b2_h0v5_l0.dat};

\addplot[
    line width=1pt,
    color=blue]
table[
           x expr=\thisrowno{0}, 
           y expr=ln(\thisrowno{1} )
         ]{tempord_b2_h0v5_l0_th.dat};

\addplot[
    only marks,
    mark=+,
    mark size=2.9pt,
    color=red]
table[
           x expr=\thisrowno{3}, 
           y expr=ln(\thisrowno{4}/4 )
         ]{tempord_b0v5_h0v5_l0.dat};
\addplot[
    line width=1pt,
    color=blue]
table[
           x expr=\thisrowno{0}, 
           y expr=ln(\thisrowno{1} )
         ]{tempord_b0v5_h0v5_l0_th.dat};

\addplot[
    only marks,
    mark=+,
    mark size=2.9pt,
    color=red]
table[
           x expr=\thisrowno{3}, 
           y expr=ln(\thisrowno{4} /4)
         ]{tempord_b1s3v5_h0v5_l0.dat};
\addplot[
    line width=1pt,
    color=blue]
table[
           x expr=\thisrowno{0}, 
           y expr=ln(\thisrowno{1} )
         ]{tempord_b1s3v5_h0v5_l0_th.dat};
\addplot[
    only marks,
    mark=+,
    mark size=2.9pt,
    color=red]
table[
           x expr=\thisrowno{3}, 
           y expr=ln(\thisrowno{4}/4 )
         ]{tempord_b1s5_h0v5_l0.dat};
\addplot[
    line width=1pt,
    color=blue]
table[
           x expr=\thisrowno{0}, 
           y expr=ln(\thisrowno{1} )
         ]{tempord_b1s5_h0v5_l0_th.dat};

\end{axis}
\end{tikzpicture}
\quad
\begin{tikzpicture}[scale=0.8]
\begin{axis}[
    enlargelimits=false,
    xlabel = $Jt$,
    xmax=20,
     y tick label style={
        /pgf/number format/.cd,
            fixed,
            fixed zerofill,
            precision=0,
        /tikz/.cd
    }
]
\addplot[
    only marks,
    mark=+,
    mark size=2.9pt,
    color=red]
table[
           x expr=\thisrowno{3}, 
           y expr=ln(\thisrowno{4} /4)
         ]{tempord_b2_h0v5_l2.dat};

\addplot[
    line width=1pt,
    color=blue]
table[
           x expr=\thisrowno{0}, 
           y expr=ln(\thisrowno{1} )
         ]{tempord_b2_h0v5_l2_th.dat};

\addplot[
    only marks,
    mark=+,
    mark size=2.9pt,
    color=red]
table[
           x expr=\thisrowno{3}, 
           y expr=ln(\thisrowno{4}/4 )
         ]{tempord_b0v5_h0v5_l2.dat};
\addplot[
    line width=1pt,
    color=blue]
table[
           x expr=\thisrowno{0}, 
           y expr=ln(\thisrowno{1} )
         ]{tempord_b0v5_h0v5_l2_th.dat};

\addplot[
    only marks,
    mark=+,
    mark size=2.9pt,
    color=red]
table[
           x expr=\thisrowno{3}, 
           y expr=ln(\thisrowno{4} /4)
         ]{tempord_b1s3v5_h0v5_l2.dat};
\addplot[
    line width=1pt,
    color=blue]
table[
           x expr=\thisrowno{0}, 
           y expr=ln(\thisrowno{1} )
         ]{tempord_b1s3v5_h0v5_l2_th.dat};
\addplot[
    only marks,
    mark=+,
    mark size=2.9pt,
    color=red]
table[
           x expr=\thisrowno{3}, 
           y expr=ln(\thisrowno{4}/4 )
         ]{tempord_b1s5_h0v5_l2.dat};
\addplot[
    line width=1pt,
    color=blue]
table[
           x expr=\thisrowno{0}, 
           y expr=ln(\thisrowno{1} )
         ]{tempord_b1s5_h0v5_l2_th.dat};

\end{axis}
\end{tikzpicture}
\\
\begin{tikzpicture}[scale=0.8]
\begin{axis}[
    enlargelimits=false,
    xlabel = $Jt$,
    ylabel=${\cal L}(\ell{,}t)$,
    xmax=20,
     y tick label style={
        /pgf/number format/.cd,
            fixed,
            fixed zerofill,
            precision=0,
        /tikz/.cd
    }
]
\addplot[
    only marks,
    mark=+,
    mark size=2.9pt,
    color=red]
table[
           x expr=\thisrowno{3}, 
           y expr=ln(\thisrowno{4} /4)
         ]{tempord_b2_h0v5_l5.dat};

\addplot[
    line width=1pt,
    color=blue]
table[
           x expr=\thisrowno{0}, 
           y expr=ln(\thisrowno{1} )
         ]{tempord_b2_h0v5_l5_th.dat};

\addplot[
    only marks,
    mark=+,
    mark size=2.9pt,
    color=red]
table[
           x expr=\thisrowno{3}, 
           y expr=ln(\thisrowno{4} /4)
         ]{tempord_b0v5_h0v5_l5.dat};
\addplot[
    line width=1pt,
    color=blue]
table[
           x expr=\thisrowno{0}, 
           y expr=ln(\thisrowno{1} )
         ]{tempord_b0v5_h0v5_l5_th.dat};

\addplot[
    only marks,
    mark=+,
    mark size=2.9pt,
    color=red]
table[
           x expr=\thisrowno{3}, 
           y expr=ln(\thisrowno{4}/4 )
         ]{tempord_b1s3v5_h0v5_l5.dat};
\addplot[
    line width=1pt,
    color=blue]
table[
           x expr=\thisrowno{0}, 
           y expr=ln(\thisrowno{1} )
         ]{tempord_b1s3v5_h0v5_l5_th.dat};
\addplot[
    only marks,
    mark=+,
    mark size=2.9pt,
    color=red]
table[
           x expr=\thisrowno{3}, 
           y expr=ln(\thisrowno{4} /4)
         ]{tempord_b1s5_h0v5_l5.dat};
\addplot[
    line width=1pt,
    color=blue]
table[
           x expr=\thisrowno{0}, 
           y expr=ln(\thisrowno{1} )
         ]{tempord_b1s5_h0v5_l5_th.dat};

\end{axis}
\end{tikzpicture}
\quad
\begin{tikzpicture}[scale=0.8]
\begin{axis}[
    enlargelimits=false,
    xlabel = $Jt$,
    xmax=20,
     y tick label style={
        /pgf/number format/.cd,
            fixed,
            fixed zerofill,
            precision=0,
        /tikz/.cd
    }
]
\addplot[
    only marks,
    mark=+,
    mark size=2.9pt,
    color=red]
table[
           x expr=\thisrowno{3}, 
           y expr=ln(\thisrowno{4} /4)
         ]{tempord_b2_h0v5_l10.dat};

\addplot[
    line width=1pt,
    color=blue]
table[
           x expr=\thisrowno{0}, 
           y expr=ln(\thisrowno{1} )
         ]{tempord_b2_h0v5_l10_th.dat};

\addplot[
    only marks,
    mark=+,
    mark size=2.9pt,
    color=red]
table[
           x expr=\thisrowno{3}, 
           y expr=ln(\thisrowno{4}/4 )
         ]{tempord_b0v5_h0v5_l10.dat};
\addplot[
    line width=1pt,
    color=blue]
table[
           x expr=\thisrowno{0}, 
           y expr=ln(\thisrowno{1} )
         ]{tempord_b0v5_h0v5_l10_th.dat};

\addplot[
    only marks,
    mark=+,
    mark size=2.9pt,
    color=red]
table[
           x expr=\thisrowno{3}, 
           y expr=ln(\thisrowno{4}/4 )
         ]{tempord_b1s3v5_h0v5_l10.dat};
\addplot[
    line width=1pt,
    color=blue]
table[
           x expr=\thisrowno{0}, 
           y expr=ln(\thisrowno{1} )
         ]{tempord_b1s3v5_h0v5_l10_th.dat};
\addplot[
    only marks,
    mark=+,
    mark size=2.9pt,
    color=red]
table[
           x expr=\thisrowno{3}, 
           y expr=ln(\thisrowno{4}/4 )
         ]{tempord_b1s5_h0v5_l10.dat};
\addplot[
    line width=1pt,
    color=blue]
table[
           x expr=\thisrowno{0}, 
           y expr=ln(\thisrowno{1} )
         ]{tempord_b1s5_h0v5_l10_th.dat};

\end{axis}
\end{tikzpicture}
\caption{${\cal L}(\ell,t)$ as a function of $t$ for $h=0.5$ and
  $\beta=2,0.5,0.286,0.2$ from top to bottom inside each panel, with
  $\ell=0,2,5,10$ from top left to bottom right. Numerical results
  for a $L=200$ site open chain are
  shown by red crosses and equation \eqref{chitemperatureordered} by a
  straight blue line. The slopes of the numerical results are correctly
reproduced by \eqref{chitemperatureordered} in the time-like region,
i.e. for $v_{\rm max} t>\ell$.}  
\label {timeordered}
\end {figure}

We now turn to the time-like region. In Fig.~\ref{timeordered} we
compare numerical results for ${\cal   L}(\ell,t)$ as a function of
$t$ for different values of $\ell$ to
\eqref{chitemperatureordered}. We recall that in the time-like region
equation \eqref{chitemperatureordered} only gives the leading time
behaviour, i.e. the exponent of the exponential decay. Hence only
  the slope of our analytic result for ${\cal L}(\ell,t)$ has to match
  the numerics. This is indeed seen to be the case in
  Fig.~\ref{timeordered}. In the space-like regime, i.e. at
  sufficiently short times, \eqref{chitemperatureordered} is again
  seen to be in very good agreement with the numerical results.

\begin{figure}[H]
\begin{center}
\begin{tikzpicture}[scale=0.8]
\begin{axis}[
    enlargelimits=false,
    xlabel = $\log Jt$,
    ylabel = ${\cal L}(\ell{,} t)-\log \chi^{xx}(\bell{,}t)$,
    xmax=3,
     y tick label style={
        /pgf/number format/.cd,
            fixed,
            fixed zerofill,
            precision=1,
        /tikz/.cd
    }
]
%\mergetables{tempord_b1s5_h0v5_l0, tempord_b1s5_h0v5_l0_th}{\datatable}
\pgfplotstableread{tempord_b1s5_h0v5_l0.dat}{\tablea}
\pgfplotstableread{tempord_b1s5_h0v5_l0_th.dat}{\tableb}
\pgfplotstableread{tempord_b1s3v5_h0v5_l0.dat}{\tableaa}
\pgfplotstableread{tempord_b1s3v5_h0v5_l0_th.dat}{\tableba}
\pgfplotstableread{tempord_b0v5_h0v5_l0.dat}{\tableab}
\pgfplotstableread{tempord_b0v5_h0v5_l0_th.dat}{\tablebb}
\pgfplotstableset{
     create on use/x/.style={create col/copy column from table={\tablea}{3}}, % Copy x values from first (or 2nd) table
     create on use/y1/.style={create col/copy column from table={\tablea}{4}}, % Copy y values from first table
     create on use/y2/.style={create col/copy column from table={\tableb}{1}}, % Copy y values from second table
     create on use/abs/.style={create col/expr={ln(\thisrow{x})}},     % Sum y values
     create on use/sum/.style={create col/expr={ln(\thisrow{y1}/4/\thisrow{y2})}},     % Sum y values
     create on use/xa/.style={create col/copy column from table={\tableaa}{3}}, % Copy x values from first (or 2nd) table
     create on use/y1a/.style={create col/copy column from table={\tableaa}{4}}, % Copy y values from first table
     create on use/y2a/.style={create col/copy column from table={\tableba}{1}}, % Copy y values from second table
     create on use/absa/.style={create col/expr={ln(\thisrow{xa})}},     % Sum y values
     create on use/suma/.style={create col/expr={ln(\thisrow{y1a}/4/\thisrow{y2a})}},     % Sum y values
     create on use/xb/.style={create col/copy column from table={\tableab}{3}}, % Copy x values from first (or 2nd) table
     create on use/y1b/.style={create col/copy column from table={\tableab}{4}}, % Copy y values from first table
     create on use/y2b/.style={create col/copy column from table={\tablebb}{1}}, % Copy y values from second table
     create on use/absb/.style={create col/expr={ln(\thisrow{xb})}},     % Sum y values
     create on use/sumb/.style={create col/expr={ln(\thisrow{y1b}/4/\thisrow{y2b})}}     % Sum y values
}
\pgfplotstablenew[columns={x,y1,y2,sum}]{\pgfplotstablegetrowsof{\tablea}}\tablec
\pgfplotstablenew[columns={xa,y1a,y2a,suma}]{\pgfplotstablegetrowsof{\tableaa}}\tableca
\pgfplotstablenew[columns={xb,y1b,y2b,sumb}]{\pgfplotstablegetrowsof{\tableab}}\tablecb

\addplot [only marks,
    mark=+,
    mark size=2.9pt,
    color=red] table [x=abs,y=sum] {\tablec};
\addplot [only marks,
    mark=+,
    mark size=2.9pt,
    color=red] table [x=absa,y=suma] {\tableca};
\addplot [only marks,
    mark=+,
    mark size=2.9pt,
    color=red] table [x=absb,y=sumb] {\tablecb};
    
 \addplot [
    domain=-1:3, 
    samples=100, 
    color=olive,
    line width=1pt
    ]
    {0.77+x*0.34};
 \addplot [
    domain=-1:3, 
    samples=100, 
    color=olive,
    line width=1pt
    ]
    {0.48+x*0.24};
 \addplot [
    domain=-1:3, 
    samples=100, 
    color=olive,
    line width=1pt
    ]
    {0.23+x*0.1};
         
\end{axis}
\end{tikzpicture}
\begin{tikzpicture}[scale=0.8]
\begin{axis}[
    enlargelimits=false,
    xlabel = $\log Jt$,
    ylabel = ${\cal L}(\ell{,} t)-\log \chi^{xx}(\bell{,}t)$,
    xmax=3,
    xmin=1,
     y tick label style={
        /pgf/number format/.cd,
            fixed,
            fixed zerofill,
            precision=1,
        /tikz/.cd
    }
]
%\mergetables{tempord_b1s5_h0v5_l0, tempord_b1s5_h0v5_l0_th}{\datatable}
\pgfplotstableread{tempord_b1s5_h0v5_l5.dat}{\tablea}
\pgfplotstableread{tempord_b1s5_h0v5_l5_th.dat}{\tableb}
\pgfplotstableread{tempord_b1s3v5_h0v5_l5.dat}{\tableaa}
\pgfplotstableread{tempord_b1s3v5_h0v5_l5_th.dat}{\tableba}
\pgfplotstableread{tempord_b0v5_h0v5_l5.dat}{\tableab}
\pgfplotstableread{tempord_b0v5_h0v5_l5_th.dat}{\tablebb}
\pgfplotstableset{
     create on use/x/.style={create col/copy column from table={\tablea}{3}}, % Copy x values from first (or 2nd) table
     create on use/y1/.style={create col/copy column from table={\tablea}{4}}, % Copy y values from first table
     create on use/y2/.style={create col/copy column from table={\tableb}{1}}, % Copy y values from second table
     create on use/abs/.style={create col/expr={ln(\thisrow{x})}},     % Sum y values
     create on use/sum/.style={create col/expr={ln(\thisrow{y1}/4/\thisrow{y2})}},     % Sum y values
     create on use/xa/.style={create col/copy column from table={\tableaa}{3}}, % Copy x values from first (or 2nd) table
     create on use/y1a/.style={create col/copy column from table={\tableaa}{4}}, % Copy y values from first table
     create on use/y2a/.style={create col/copy column from table={\tableba}{1}}, % Copy y values from second table
     create on use/absa/.style={create col/expr={ln(\thisrow{xa})}},     % Sum y values
     create on use/suma/.style={create col/expr={ln(\thisrow{y1a}/4/\thisrow{y2a})}},     % Sum y values
     create on use/xb/.style={create col/copy column from table={\tableab}{3}}, % Copy x values from first (or 2nd) table
     create on use/y1b/.style={create col/copy column from table={\tableab}{4}}, % Copy y values from first table
     create on use/y2b/.style={create col/copy column from table={\tablebb}{1}}, % Copy y values from second table
     create on use/absb/.style={create col/expr={ln(\thisrow{xb})}},     % Sum y values
     create on use/sumb/.style={create col/expr={ln(\thisrow{y1b}/4/\thisrow{y2b})}}     % Sum y values
}
\pgfplotstablenew[columns={x,y1,y2,sum}]{\pgfplotstablegetrowsof{\tablea}}\tablec
\pgfplotstablenew[columns={xa,y1a,y2a,suma}]{\pgfplotstablegetrowsof{\tableaa}}\tableca
\pgfplotstablenew[columns={xb,y1b,y2b,sumb}]{\pgfplotstablegetrowsof{\tableab}}\tablecb

\addplot [only marks,
    mark=+,
    mark size=2.9pt,
    color=red] table [x=abs,y=sum] {\tablec};
\addplot [only marks,
    mark=+,
    mark size=2.9pt,
    color=red] table [x=absa,y=suma] {\tableca};
\addplot [only marks,
    mark=+,
    mark size=2.9pt,
    color=red] table [x=absb,y=sumb] {\tablecb};
    
 \addplot [
    domain=2:3, 
    samples=100, 
    color=olive,
    line width=1pt
    ]
    {1.275+(x-2)*0.5};
 \addplot [
    domain=2:3, 
    samples=100, 
    color=olive,
    line width=1pt
    ]
    {0.85+(x-2)*0.33};
 \addplot [
    domain=2:3, 
    samples=100, 
    color=olive,
    line width=1pt
    ]
    {0.38+(x-2)*0.15};
         
\end{axis}
\end{tikzpicture}
\end{center}
\caption{${\cal L}(\ell{,} t)-\log \chi^{xx}(\bell,t)$ with
  $\chi^{xx}(\bell,t)$ given by \eqref{chitemperatureordered} as a function
  of $\log Jt$ for $h=0.5$ and $\beta=0.5,0.286,0.2$ from bottom to top
  inside each panel, with $\bell=0$ (left) and $\bell=5$ (right) in red. In
  green is indicated a linear fit in the time-like region $v_{\rm
    max}t>\bell$.}   
\label {timeordered22}
\end {figure}
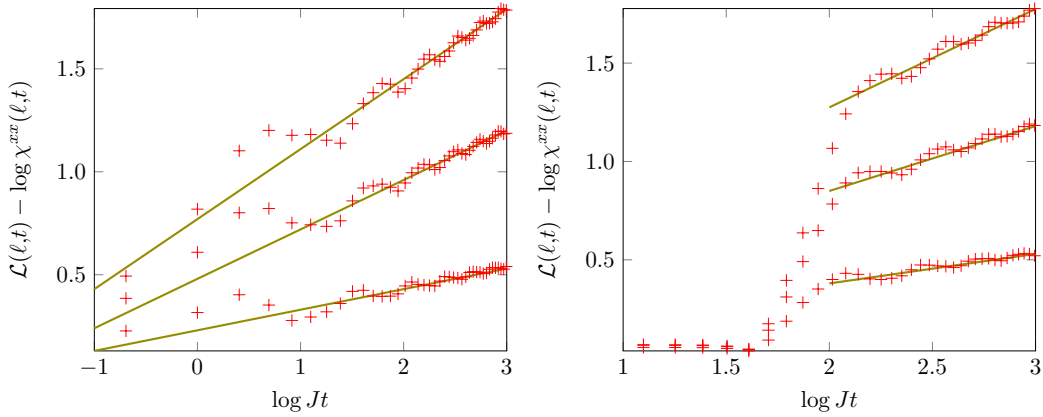

In order to have a more quantitative check on the exponential
factor in \eqref{chitemperatureordered} it is useful to consider the
difference 
\be
{\cal   L}(\ell,t)-\log \chi^{xx}(\ell,t)\ ,
\label{difference}
\ee
where $\chi^{xx}(\ell,t)$ is given by \eqref{chitemperatureordered}.
If our exponential factor is exact, the difference should decay at
late times more slowly than exponential, i.e. supposedly as a power
law. In Figure \ref{timeordered22} we plot \fr{difference} as a
function of $\log Jt$ and indeed observe a linear behaviour. This
confirms that the exponential factor in \eqref{chitemperatureordered}
is exact and moreover establishes that the subleading
asymptotics is a power law in time.

\subsection {\texorpdfstring{Disordered phase $h>1$}{Lg} \label{diso}}

\subsubsection {An approximate mapping to the ordered case}
In the disordered phase the form factors in \eqref{armtap} still
differ from the finite temperature ordered case by the fact that there
are no intermediate states with the same number of momenta as in the
representative state. From Section \ref{different} we know that the 
terms $M=N+1$ and $M=N-1$ are equally important at late times, but the
subleading terms are dominant in $\rho_\beta$ for $M=N+1$. Since we are
interested in the leading time behaviour only, with the order $\rho_\beta^0$
subleading corrections, we will restrict our analysis to the case $M=N+1$.

In this case the partial fraction decomposition \eqref{pfdhp1} of
$\vert {}_{\rm NS}\langle q_1,...,q_{N}\vert\sigma_l^x\vert
p_1,...,p_M\rangle_{\rm R}\vert ^2$ necessarily involves one $\nu_j=0$. To obtain
the $p_j$ dependence of the corresponding $A$'s in \eqref{pfdhp1}, it
suffices to decompose the form factor $\vert{}_{\rm NS} \langle
q_1,...,q_{N}\vert\sigma_l^x\vert p_1,...,p_M\rangle_{\rm R}\vert ^2$ by
starting to write the partial fraction decomposition with respect to
$p_j$ and retaining only the $\nu_j=0$ part (which corresponds to
$P_0(X)$ in \eqref{pfdgen}), and decomposing with respect to
the other momenta in the usual fashion. This part is $
\frac{\prod_{i=1}^N\varepsilon_{p_jq_i}^2}{\prod_{i=1}^{N+1}\varepsilon_{p_jp_i}}$,
that is $\frac{1}{\varepsilon(p_j)}$ times
$\frac{\prod_{i=1}^N\varepsilon_{p_jq_i}^2}{\prod_{i=1\neq
    j}^{N+1}\varepsilon_{p_jp_i}}$. This latter factor satisfies the
hypotheses of \eqref{approx} which allows us to replace it by $1$
as long as we are interested only in the leading time behaviour. It
follows that under this approximation the $\nu_j=0$ terms in
the partial fraction decomposition \eqref{pfdhp1} of $\vert{}_{\rm NS} \langle
q_1,...,q_{N}\vert\sigma_\bell^x\vert p_1,...,p_M\rangle_{\rm R}\vert ^2$
contribute to \eqref{armtap} as 
\begin{equation}
\begin{aligned}
&\frac{2J\sqrt{h}\xi}{(N+1)!L}\sum_{p_j\in\rm{R}}\frac{e^{-it\overline{\varepsilon}(p_j)}}{\varepsilon(p_{j})}\sum_{p_i\in\rm{R},i\neq j}\vert {}_{\rm{R}}\langle p_1,...,p_{j-1}p_{j+1},...,p_{N+1}\vert\sigma_\bell^x\vert q_1,...,q_{N}\rangle_{\rm{NS}}\vert ^2 \\
&\qquad\qquad\qquad\times e^{it[E\left( \left\{ q\right\} \right) -E( \left\{ p\right\} -\{p_j\} )]+ i\bell[P\left( \left\{ p\right\} -\{p_j\}\right) -P\left( \left\{ q\right\} \right) ]}\ .
\end{aligned}
\end{equation}
Taking into account all the possible $j=1,...,N+1$ for which one can have $\nu_j=0$ in \eqref{pfdhp1}, we obtain
\begin{equation}
\begin{aligned}
\chi^{xx}(\ell,t)\approx
&\left(\frac{2J\sqrt{h}\xi}{L}\sum_{p\in\rm{R}}\frac{e^{-it\overline{\varepsilon}(p)}}{\varepsilon(p)}\right)\\ 
&\times\frac {1} {N!}\sum_{\substack{p_1,...,p_{N}\\\in\rm{R}}}\vert {}_{\rm{R}}\langle p_1,...,p_{N}\vert\sigma_\bell^x\vert q_1,...,q_{N}\rangle_{\rm{NS}}\vert ^2 e^{it(E\left( \left\{ q\right\} \right) -E\left( \left\{ p\right\} \right))+ i\bell(P\left( \left\{ p\right\}\right) -P\left( \left\{ q\right\} \right) )}\ ,
\end{aligned}
\label{chixxdiso}
\end{equation}
where $\{p\}=\{p_1,...,p_N\}$ has now the same number of momenta as in
the representative state. The second factor in \fr{chixxdiso} is
precisely of the same form as the one we considered in the ordered
phase.

\subsubsection {Two-point dynamical correlation functions in the
  disordered phase} 
Putting everything together we obtain the following result for the
leading late time behaviour of the two-point function
\begin{equation}
\label {chitemperaturedisordered}
  \chi^{xx}(\ell,t)\approx C(\ell,t) \exp\left(\frac{1}{2\pi}\int_{-\pi}^\pi
  |t\varepsilon'(x)+\ell| \log(1-4\pi\rho(x))dx\right)\ , 
\end{equation}
where
%$C(\ell,t)=2J\sqrt{h}\xi\int_{-\pi}^\pi dx
%\frac{e^{-it\overline{\varepsilon}(x)}}{2\pi\varepsilon(x)}$.
% at leading order in $\rho$. 
(\emph{cf.} section \ref{different})
\begin{equation}
\begin{aligned}
&C(\ell,t)=
\begin{cases}
2J\sqrt{h}\xi\int_{-\pi}^\pi\left(\frac{e^{-it\overline{\varepsilon}(x)}}{2\pi}+2\rho(x)\cos(t\overline{\varepsilon}(x))\right)\frac{dx}{\varepsilon(x)}&
\text{if $v_{\rm max}t<\ell$ at }{\cal O}(\rho_\beta)\\ 
2J\sqrt{h}\xi\int_{-\pi}^\pi\frac{e^{-it\overline{\varepsilon}(x)}}{2\pi\varepsilon(x)}dx
&\text{if $v_{\rm max}t>\ell$ at }{\cal O}(\rho_\beta^0)\ ,
\end{cases}
\end{aligned}
\label{cellt}
\end{equation}
$\ell,t\geq 0$, and $v_{\rm max}$ is defined in \eqref{vmax}.

\subsubsection {Numerical checks}

We have checked the accuracy of \eqref {chitemperaturedisordered} by
comparing it to numerical calculations following Ref.~\cite
{derzkho}. In the following we show results for
\be
{\cal R}(\ell,t)={\rm Re}\big(\langle\sigma^x_{\ell+1}(t)\sigma^x_1(0)\rangle\big)\ .
\ee
In the space-like region $v_{\rm max}t<\ell$ we furthermore check our
result for the prefactor $C(\ell,t)$ \fr{cellt} by computing
\be
\Lambda(\ell,\beta)=\frac{{\cal R}(\ell,0)
e^{-\ell\int_{-\pi}^\pi\frac{dx}{2\pi}
\log(1-4\pi\rho(x))}-C_0(\ell,0)}{C(\ell,0)-C_0(\ell,0)}\ ,
\ee
where
$C_0(\ell,t)=\frac{J\sqrt{h}\xi}{\pi}\int_{-\pi}^\pi
\frac{dx}{\varepsilon(x)}e^{-it\overline{\varepsilon}(x)}$
is the 
${\cal O}(\rho_\beta^0)$ contribution to $C(\ell,t)$. 
If and only if the prefactor in
\eqref{cellt} is correct at order ${\cal O}(\rho_\beta)$, $\Lambda(\ell,\beta)\to 1$ when $\beta\to 0$. 

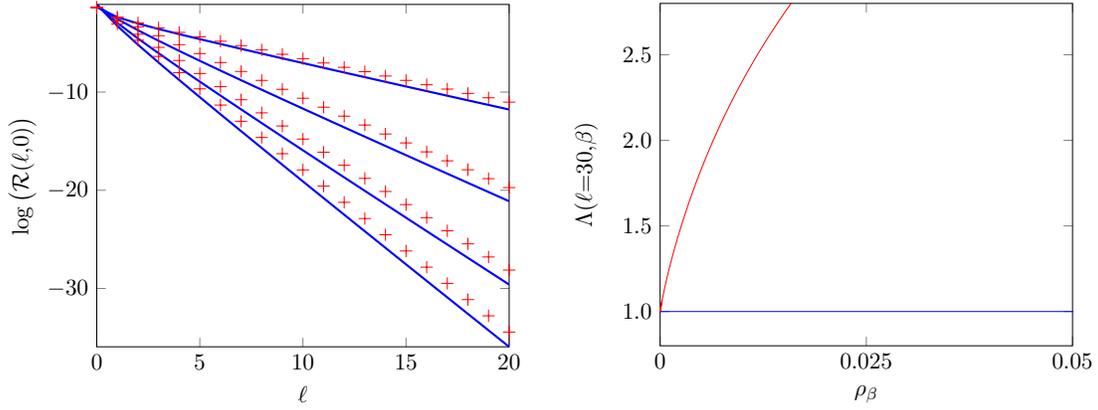
\begin{figure}[H]
\begin{center}
\begin{tikzpicture}[scale=0.80]
\begin{axis}[
    enlargelimits=false,
    xlabel = $\ell$,
    ylabel = $\log\big({\cal R}(\ell{,}0)\big)$,
     y tick label style={
        /pgf/number format/.cd,
            fixed,
            fixed zerofill,
            precision=0,
        /tikz/.cd
    }
]
\addplot[
    only marks,
    mark=+,
    mark size=2.9pt,
    color=red]
table[
           x expr=\thisrowno{2}, 
           y expr=ln(\thisrowno{4} /4)
         ]{tempdiso_b2_h1v5_l.dat};

\addplot[
    line width=1pt,
    color=blue]
table[
           x expr=\thisrowno{0}, 
           y expr=ln(\thisrowno{1} )
         ]{tempdiso_b2_h1v5_l_th.dat};

\addplot[
    only marks,
    mark=+,
    mark size=2.9pt,
    color=red]
table[
           x expr=\thisrowno{2}, 
           y expr=ln(\thisrowno{4}/4 )
         ]{tempdiso_b0v5_h1v5_l.dat};

\addplot[
    line width=1pt,
    color=blue]
table[
           x expr=\thisrowno{0}, 
           y expr=ln(\thisrowno{1} )
         ]{tempdiso_b0v5_h1v5_l_th.dat};

\addplot[
    only marks,
    mark=+,
    mark size=2.9pt,
    color=red]
table[
           x expr=\thisrowno{2}, 
           y expr=ln(\thisrowno{4} /4)
         ]{tempdiso_b1s3v5_h1v5_l.dat};

\addplot[
    line width=1pt,
    color=blue]
table[
           x expr=\thisrowno{0}, 
           y expr=ln(\thisrowno{1} )
         ]{tempdiso_b1s3v5_h1v5_l_th.dat};
\addplot[
    only marks,
    mark=+,
    mark size=2.9pt,
    color=red]
table[
           x expr=\thisrowno{2}, 
           y expr=ln(\thisrowno{4}/4)
         ]{tempdiso_b1s5_h1v5_l.dat};

\addplot[
    line width=1pt,
    color=blue]
table[
           x expr=\thisrowno{0}, 
           y expr=ln(\thisrowno{1} )
         ]{tempdiso_b1s5_h1v5_l_th.dat};

\end{axis}
\end{tikzpicture}
\quad
\begin{tikzpicture}[scale=0.80]
\begin{axis}[
    enlargelimits=false,
    xlabel =$\rho_\beta$,
    ylabel =$\Lambda(\ell{=}30{,}\beta)$,
    ymin=0.8,
    ymax =2.8,
    xmin =0 , 
    xmax=0.05,
    scaled x ticks = false,
    xtick={0,0.025,0.05},
    xticklabels={0,0.025,0.05},
    y tick label style={
        /pgf/number format/.cd,
            fixed,
            fixed zerofill,
            precision=1,
        /tikz/.cd
    }
]
\addplot[red] table[
x index=1,
y index=2,
mark=none]
{disord_prefact_h1v5_okrho2.dat};
 \addplot [
    domain=0:1, 
    samples=100, 
    color=blue,
    ]
          {1};
\end{axis}
\end{tikzpicture}
\end{center}
\caption {Left: $\log\big({\cal R}(\ell,0)\big)$ as a function of $\ell$ for
$h=1.5$ and $\beta=2,0.5,0.285,0.2$ from top to bottom. Numerical
  results for a $L=200$ site open chain are shown as red crosses and equation
  \eqref{chitemperaturedisordered} as straight blue lines. Right: 
numerical results for $\Lambda(\ell=30,\beta)$ as a function of
$\rho_\beta$ for $h=3/2$ (red line). The expected result when $\rho_\beta=0$ is shown in blue. }
\label {spacedisordered000}
\end {figure}

In Fig.~\ref{spacedisordered000} we compare our analytic expression
\eqref{chitemperaturedisordered} in the space-like region to numerical
results for $R(\ell,t)$. 
Our analytic expression is seen to be in good agreement with the
numerical results and the remaining discrepancy is due to ${\cal
  O}(\rho_\beta^2)$ corrections to the prefactor.

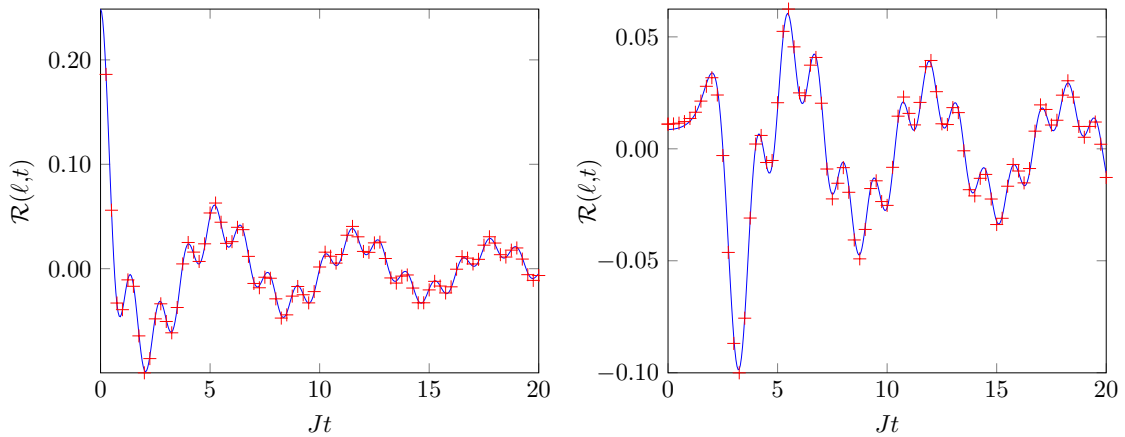
\begin{figure}[H]
\begin{center}
\begin{tikzpicture}[scale=0.85]
\begin{axis}[
    enlargelimits=false,
    xlabel = $Jt$,
    ylabel = ${\cal R}(\ell{,}t)$,
     y tick label style={
        /pgf/number format/.cd,
            fixed,
            fixed zerofill,
            precision=2,
        /tikz/.cd
    }
]
\addplot[
    only marks,
    mark=+,
    mark size=2.9pt,
    color=red]
table[
           x expr=\thisrowno{3}, 
           y expr=\thisrowno{4} /4
         ]{tempdiso_b3_h1v5_l0.dat};

\addplot[
    color=blue]
table[
           x expr=\thisrowno{0}, 
           y expr=\thisrowno{1} 
         ]{tempdiso_b3_h1v5_l0_th.dat};

\end{axis}
\end{tikzpicture}
\begin{tikzpicture}[scale=0.85]
\begin{axis}[
    enlargelimits=false,
    xlabel = $Jt$,
    ylabel = ${\cal R}(\ell{,}t)$,
     y tick label style={
        /pgf/number format/.cd,
            fixed,
            fixed zerofill,
            precision=2,
        /tikz/.cd
    }
]
\addplot[
    only marks,
    mark=+,
    mark size=2.9pt,
    color=red]
table[
           x expr=\thisrowno{3}, 
           y expr=\thisrowno{4} /4
         ]{tempdiso_b3_h1v5_l5.dat};

\addplot[
    color=blue]
table[
           x expr=\thisrowno{0}, 
           y expr=\thisrowno{1} 
         ]{tempdiso_b3_h1v5_l5_th.dat};

\end{axis}
\end{tikzpicture}
\caption {${\cal R}(\ell,t)$ as a function of $t$ with $h=1.5$ and
$\beta=3$ for $\ell=0$ (left) and $\ell=5$ (right). Numerical results
for a $L=200$ site open chain are shown as red crosses and equation \eqref{chitemperaturedisordered} as a
solid blue line. } 
\label {spacedisordered111}
\end{center}
\end{figure}

In Fig.~\ref{spacedisordered111} we present results for ${\cal
  R}(\ell,t)$ in the time-like region $v_{\rm max}t>\ell$ at low
temperatures. We see that the analytical result is in excellent
agreement with the numerics.

\begin{figure}[H]
\begin{center}
\begin{tikzpicture}[scale=0.8]
\begin{axis}[
    enlargelimits=false,
    xlabel = $Jt$,
    ylabel = $\log|{\cal R}(\ell{,}t)|$,
     y tick label style={
        /pgf/number format/.cd,
            fixed,
            fixed zerofill,
            precision=0,
        /tikz/.cd
    }
]
\addplot[
    only marks,
    mark=+,
    mark size=2.9pt,
    color=red]
table[
           x expr=\thisrowno{3}, 
           y expr=ln(abs(\thisrowno{4} /4))
         ]{tempdiso_b2_h1v5_l0.dat};
\addplot[
    only marks,
    mark=+,
    mark size=2.9pt,
    color=red]
table[
           x expr=\thisrowno{3}, 
           y expr=ln(abs(\thisrowno{4} /4))
         ]{tempdiso_b0v5_h1v5_l0.dat};
\addplot[
    only marks,
    mark=+,
    mark size=2.9pt,
    color=red]
table[
           x expr=\thisrowno{3}, 
           y expr=ln(abs(\thisrowno{4} /4))
         ]{tempdiso_b1s3v5_h1v5_l0.dat};
\addplot[
    only marks,
    mark=+,
    mark size=2.9pt,
    color=red]
table[
           x expr=\thisrowno{3}, 
           y expr=ln(abs(\thisrowno{4} /4))
         ]{tempdiso_b1s5_h1v5_l0.dat};
\addplot[
    only marks,
    mark=+,
    mark size=2.9pt,
    color=red]
table[
           x expr=\thisrowno{3}, 
           y expr=ln(abs(\thisrowno{4} /4))
         ]{tempdiso_b2_h1v5_l0_longtime.dat};
\addplot[
    line width=1pt,
    color=blue]
table[
           x expr=\thisrowno{0}, 
           y expr=ln(abs(\thisrowno{1} ))
         ]{tempdiso_b2_h1v5_l0_th.dat};
\addplot[
    line width=1pt,
    color=blue]
table[
           x expr=\thisrowno{0}, 
           y expr=ln(abs(\thisrowno{1} ))
         ]{tempdiso_b0v5_h1v5_l0_th.dat};
\addplot[
    line width=1pt,
    color=blue]
table[
           x expr=\thisrowno{0}, 
           y expr=ln(abs(\thisrowno{1} ))
         ]{tempdiso_b1s3v5_h1v5_l0_th.dat};
\addplot[
    line width=1pt,
    color=blue]
table[
           x expr=\thisrowno{0}, 
           y expr=ln(abs(\thisrowno{1} ))
         ]{tempdiso_b1s5_h1v5_l0_th.dat};

\end{axis}
\end{tikzpicture}
\quad
\begin{tikzpicture}[scale=0.8]
\begin{axis}[
    enlargelimits=false,
    xlabel = $Jt$,
    ylabel = $\log|{\cal R}(\ell{,}t)|$,
     y tick label style={
        /pgf/number format/.cd,
            fixed,
            fixed zerofill,
            precision=0,
        /tikz/.cd
    }
]

\addplot[
    only marks,
    mark=+,
    mark size=2.9pt,
    color=red]
table[
           x expr=\thisrowno{3}, 
           y expr=ln(abs(\thisrowno{4} /4))
         ]{tempdiso_b2_h1v5_l10.dat};
\addplot[
    line width=1pt,
    color=blue]
table[
           x expr=\thisrowno{0}, 
           y expr=ln(abs(\thisrowno{1} ))
         ]{tempdiso_b2_h1v5_l10_th.dat};
\addplot[
    line width=1pt,
    color=blue]
table[
           x expr=\thisrowno{0}, 
           y expr=ln(abs(\thisrowno{1} ))
         ]{tempdiso_b0v5_h1v5_l10_th.dat};
\addplot[
    line width=1pt,
    color=blue]
table[
           x expr=\thisrowno{0}, 
           y expr=ln(abs(\thisrowno{1} ))
         ]{tempdiso_b1s3v5_h1v5_l10_th.dat};
\addplot[
    line width=1pt,
    color=blue]
table[
           x expr=\thisrowno{0}, 
           y expr=ln(abs(\thisrowno{1} ))
         ]{tempdiso_b1s5_h1v5_l10_th.dat};
\end{axis}
\end{tikzpicture}
\end{center}
\caption {$\log |{\cal R}(\ell,t)|$ as a function of $t$ for $h=1.5$
and $\beta=2,0.5,0.286,0.2$ from top to bottom inside each panel,
with $\ell=0$ (left) and $\ell=10$ (right). Numerical results for
  a $L=200$ site open chain are shown as red crosses and equation
\eqref{chitemperaturedisordered} as a solid blue line. } 
\label{spacedisordered111b}
\end {figure}
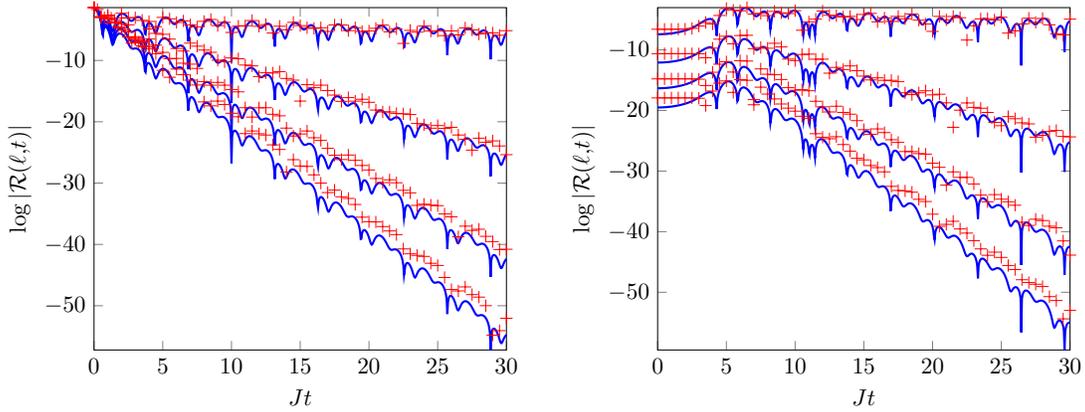

In order to check the accuracy of our result for the exponential decay
of the two-point function for intermediate and high temperatures we
compare \eqref{chitemperaturedisordered} to numerical results
$\log|{\cal R}(\ell,t)|$ in Fig.~\ref{spacedisordered111b}. As our
result for the prefactor $C(\ell,t)$ only holds at low temperatures we
expect the numerical results to differ from our analytical prediction
by an essentially constant offset. This is indeed what we observe.

%
%\begin{figure}[H]
%\begin {center}
%\begin{tikzpicture}
%\begin{axis}[
%width=\textwidth, 
%height=0.5\textwidth, 
%    enlargelimits=false,
%    xlabel = $Jt$,
%    ymin=-0.06,
%    ymax =0.06,
%    xmin =0, 
%     y tick label style={
%        /pgf/number format/.cd,
%            fixed,
%            fixed zerofill,
%            precision=1,
%        /tikz/.cd
%    }
%]
%\addplot+[
%    only marks,
%    mark=+,
%    mark size=2.9pt,
%    color=red]
%table{temp_b3_h1v5.dat};
%
%\addplot[
%    color=blue]
%table{temp_b3_h1v5_th.dat};
%
%\end{axis}
%\end{tikzpicture}
%
%\begin{tikzpicture}
%\begin{axis}[
%width=\textwidth, 
%height=0.5\textwidth, 
%    enlargelimits=false,
%    xlabel = $Jt$,
%    ymin=-0.06,
%    ymax =0.06,
%    xmin =0, 
%     y tick label style={
%        /pgf/number format/.cd,
%            fixed,
%            fixed zerofill,
%            precision=1,
%        /tikz/.cd
%    }
%]
%\addplot+[
%    only marks,
%    mark=+,
%    mark size=2.9pt,
%    color=red]
%table{temp_b3_h1v5_l20.dat};
%
%\addplot[
%    color=blue]
%table{temp_b3_h1v5_l20_th.dat};
%
%\end{axis}
%\end{tikzpicture}
%
%\caption {$\left\langle \sigma ^{x}_0\left( t\right) \sigma ^{x}_0\left( 0\right)\right\rangle $ (top) and $\left\langle \sigma ^{x}_{20}\left( t\right) \sigma ^{x}_0\left( 0\right)\right\rangle $  (bottom) as a function of $t$, for $\beta=3$ and $h=1.5$, numerical measures (red) and formula \eqref {temperaturedynamics} (blue).}
%\label {tempera}
%\end {center}
%\end{figure}

%%%%%%%%%%%%%%%%%%%%%%%%%%%%%%%%%%%
\section{Summary and Discussion}
%%%%%%%%%%%%%%%%%%%%%%%%%%%%%%%%%%%
In this work we have considered two problems in the transverse field
Ising model: (i) The time dependence of the order parameter after a
quantum quench; (ii) The dynamical order parameter two point function in
equilibrium at finite temperatures. Using the quench action approach
for (i) and a micro-canonical formulation combined with typicality
ideas for (ii) these problems can both be formulated in terms of spectral
representations using Hamiltonian eigenstates, i.e. sums over form factors. 

These highly intricate sums over a macroscopic quantity of momenta of
a form factor with many singularities represent however a considerable
technical challenge. We showed that in the case of semi-local
operators such as $\sigma^x_j$ in the TFIM, this difficulty can be
addressed by decomposing the form factor into partial fractions, which
permits the sums over momenta to be decoupled and the late time
behaviour to be determined.  These partial fractions can be organized in
terms of the degree and the position of their poles, which naturally
leads to an expansion in the density of particles of the representative
state. The leading behaviour at late times can be then computed at all
orders in the density through a determinant representation of these
poles, which leads invariably to an exponential decay. 

%for massive
%models\footnote{At criticality, i.e. for $h=1$ in the TFIM, at least in the zero temperature regime there should be a power-law decay of correlation functions. At $h=1$ there are other poles in the form factor arising from the $\varepsilon$ terms, which together with the discontinuity of the density at zero temperature is probably responsible for this power-law decay. Indeed for $h\neq 1$ the decay is exponential even if the density is discontinuous at some point.}.

Our analysis provides a precise characterization of the
excitations over the representative state that contributes to the late
time behaviour of correlation functions of semi-local operators. 
We find that \textit{simultaneous} particle-hole excitations of
\textit{all} particles in the representative state contribute to the
correlation function, and we altogether have to sum over ${\cal
  O}\big((\epsilon L)^N\big)$ excited states, where $\epsilon$ is a
fixed number that can be taken as small as desired and $N/L$ is the
density. In particular, this implies that the appealing picture of 
an expansion in terms of a finite number of particle-hole excitations
over the full momentum space \textit{fails} for semi-local operators
at finite temperatures. The form factor sum is dominated by
mesoscopic excitations (in the sense given in Section \ref{govern}) around each single particle, which is an exponential number of states, but subentropic in the sense that includes only states whose macroscopic state is the representative state itself.

We have compared our analytic results to numerical computations in
both the finite temperature and quench contexts. In the absence of
saddle points, thus where the sums \eqref{sumsval} are valid at all momenta, we find
remarkable agreement at all times. In cases where saddle points occur
our numerical results indicate the presence of a multiplicative  
power-law behaviour as a subleading correction to the exponential
decay. We believe that this will emerge from saddle point effects of
these classes of excitations, due to the fact that \eqref{sumsval} is not valid anymore close to the saddle point. Higher-order corrections in time will
also arise from contributions with $\nu_i=0$ in \eqref{decomp}; in this case there is no singularity and the full momentum space should be summed over to take them into account. We believe that partial fraction
decompositions for form factors of semi-local operators are not only
suited for extracting the late time asymptotics, but should also be
useful for determining intermediate and short time behaviours.

In interacting models, the singularity structure of the form factors
of semi-local operators is not fundamentally changed and a partial
fraction decomposition will provide a useful organizing principle as
well. A notable difference is that the Bethe equations link the
different particles so that the momenta sums can never fully decouple.
This will be the subject of a subsequent paper. 

For local operators however, the story is radically different. The
singularity structure of the form factors \eqref{localsing} is
completely dissimilar and the states that dominate the late time asymptotics are selected by different principles. Excitations over the full momentum
space and not only near the singularities will play equally
important roles. This will be discussed elsewhere.

%%%%%%%%%%%%%%%%%%%%%%%%%%%%%%%%%%%
\paragraph{Acknowledgements}
%%%%%%%%%%%%%%%%%%%%%%%%%%%%%%%%%%%

We are grateful to Jacopo de Nardis, Frank G\"ohmann and Karol
Kozlowski for very helpful discussions and comments. This work was
supported by the EPSRC under grant EP/S020527/1 (FHLE and EG) and by
the ERC under Starting Grant No. 805252 LoCoMacro (MF).  

\appendix
\section{Riemann sums of singular functions \label{appen}}
Sums of form factors lead to Riemann sums of functions with a quadratic singularity of the form
\begin{equation}
\frac {1} {L}\sum _{k=-L+1}^{L}f\left( \tfrac {k} {L}\right)\qquad\text{with }f(x) \underset{x\to 0}{\sim}\frac {1} {x^{2}}\ .
\end{equation}
Since the integral of $f(x)$ is divergent, the limit $L\to\infty$
cannot be directly taken as for regular functions and needs special
treatment. The purpose of this appendix is to explain techniques to
compute them. We note that an alternative way of treating such sums is
to employ contour integral techniques \cite{kozlowski4}.

\subsection{One-dimensional sums \label{appen1}}
Oscillatory Riemann sums of functions with a quadratic singularity cannot be estimated with the usual results on stationary phase approximation to obtain their large time behaviour. The principle is then to add and remove an elementary function with the same singularity, but whose Riemann sum can be computed directly, so that the remaining function has no singularity and thus has an oscillatory Riemann sum that vanishes at large times. 

\subsubsection{Generic example}
For concreteness, let us illustrate this procedure with the Riemann sum
\begin{equation}
  S_{L}\left( \theta\right) =\frac {1} {L}\sum _{k=-L+1\neq 0}^{L}f\left( \tfrac {k} {L}\right) e^{i\frac {k} {L}\theta}\ ,
\end{equation}
for $ f (x)$ a function such that $ f(x) =\frac {1} {x^{2}}+\mathcal{O}(x^0) $ for $ x\to 0$, $ f $ being regular otherwise. Then
\begin{equation}
  S_{L}\left( \theta\right) =\frac {1} {L}\sum _{k=-L+1\neq 0}^{L}\left( \tfrac {L} {k}\right) ^{2}e^{i\frac {k} {L}\theta}+\frac {1} {L}\sum _{k=-L+1}^{L}\bar {f}\left( \tfrac {k} {L}\right) e^{i\frac {k} {L}\theta}\ ,
\end{equation}
with $\bar{f}(x)=f(x)-1/x^2$ a regular function. The second term on the right-hand side can be turned into an
integral, while we rewrite the first as follows
\begin{equation}
\frac {1} {L}\sum _{k=-L+1\neq 0}^{L}\left( \tfrac {L} {k}\right)
^{2}e^{i\frac {k} {L}\theta}= L\sum _{k=-\infty,\neq 0}^{+\infty
}\tfrac {1} {k ^{2}}e^{i\frac {k} {L}\theta}-\frac {1} {L}\sum
_{k=L+1}^{+\infty }\left( \tfrac {L} {k}\right) ^{2}e^{i\frac {k}
  {L}\theta}-\frac {1} {L}\sum _{k=-\infty}^{-L}\left( \tfrac {L}
      {k}\right) ^{2}e^{i\frac {k} {L}\theta}\ . 
\end{equation}
The second and third Riemann sums can now be turned into integrals
  without any problems, which gives
%  are the Riemann sum of $ 1 / x^2$ on $\left] -\infty,-1 \right]\cup \left[ 1,+\infty \righ%t[ $ where it is regular. Thus
\begin{equation}
  S_{L}\left( \theta\right) =L\sum _{k=-\infty,\neq 0}^{+\infty }\frac {e^{i\frac {k} {L}\theta}} {k ^{2}} +\int _{-1}^{1}\bar {f}(x) e^{i\theta x}dx-\int _{1}^\infty\frac {e^{i\theta x}} {x^{2}}dx-\int _{-\infty}^{-1}\frac {e^{i\theta x}} {x^{2}}dx+\mathcal{O}\left( L^{-1}\right) \ .
\end{equation}
The three integrals are oscillatory integrals of bounded functions and hence
vanish for large $\theta$. We conclude that
\begin{equation}
S_{L}\left( \theta\right) =L\sum _{k=-\infty,\neq 0}^{+\infty }\frac {e^{i\frac {k} {L}\theta}} {k ^{2}} +\mathcal{O}\left( L^{0}\theta^{-1 / 2}\right) +\mathcal{O}\left( L^{-1}\right) \ .
\end{equation}

\subsubsection{Elementary oscillatory sums with singularities}
The above analysis shows that the leading asymptotics of Riemann sums with simple or
double poles involves the following sums 
\begin{equation}
\begin{aligned}
\sum_{n\in\mathbb{Z}}\frac{e^{iw(n+\alpha)}}{(n+\alpha)}&=\frac{\pi}{\sin \pi\alpha}e^{i\pi\alpha\sign(w)}\,,\qquad\text{for }-\pi<w\leq\pi\\
\sum_{n\in\mathbb{Z}}\frac{e^{iw(n+\alpha)}}{(n+\alpha)^2}&=\left(\frac{\pi}{\sin \pi\alpha}\right)^2 +\frac{i\pi}{\sin\pi\alpha}w e^{i\pi\alpha\sign(w)}\,,\qquad\text{for }-\pi<w\leq\pi\ .
\end{aligned}
\end{equation}
These are readily obtained by computing the Fourier series
coefficients of the right-hand sides multiplied by $e^{-iw\alpha}$ and
seen as a $2\pi$-periodic function of $w$. In particular we have

\begin{equation}
\label{sumuseful}
\begin{aligned}
&\sum_{n\in\mathbb{Z}}\frac{e^{iw(n+1/2)}}{(n+1/2)^m}=
\begin{cases}
\pi^2\big(1-\frac{|w|}{\pi}\big) &\text{ if } m=2\ ,\\
i\pi\ {\rm sgn}(w)&\text{ if } m=1\ .
\end{cases}\\
&\sum_{n\in\mathbb{Z}\backslash \{0\}}\frac{e^{iwn}}{n^m}=
\begin{cases}
\pi^2\big(\tfrac{1}{3}-\frac{|w|}{\pi}+\frac{w^2}{2\pi^2}\big) &\text{ if } m=2\ ,\\
i(\pi-|w|)\ {\rm sgn}(w)&\text{ if } m=1\ .
\end{cases}
\end{aligned}
\end{equation}

\subsubsection{\texorpdfstring{Sums arising in finite-temperature dynamics}{Lg}} 
\label{ssec:eqns40}
Following the steps outlined above we obtain for $q\in\rm{NS}$
\begin{equation}
\begin{aligned}
\sum _{p\in\rm{R}}\frac {e^{-it \overline{\varepsilon } (p) }} {L\sin^2  \left( \frac {p-q} {2}\right) }=&L\sum _{k=-\infty}^{+\infty }\frac {e^{-it \overline{\varepsilon }(q+2\pi (k+1/2)/L)}} {\pi^2(k+1/2) ^{2}}
-\int_{\pi}^\infty \frac{4e^{-it \overline{\varepsilon }(x)}}{(x-q)^2}\frac{dx}{2\pi}-\int_{-\infty}^{-\pi} \frac{4e^{-it \overline{\varepsilon }(x)}}{(x-q)^2}\frac{dx}{2\pi}\\
&+\int_{-\pi}^\pi e^{-it \overline{\varepsilon
    }(x)}\left(\frac{1}{\sin^2
    \tfrac{x-q}{2}}-\frac{1}{(x-q)^2/4}\right)\frac{dx}{2\pi}
  	+\mathcal{O}\left( L^{-1}\right)\ .
\end{aligned}
\label{40a}
\end{equation}

At leading order in time, one can Taylor expand the $
\overline{\varepsilon }$ in the remaining sum to fall back on
an elementary oscillatory sum. The three integrals involve oscillatory
bounded functions and hence decay to zero with time. 

Similarly we obtain for $q\in\rm{NS}$
\begin{equation}
\begin{aligned}
\sum _{p\in\rm{R}}\frac {e^{-it \overline{\varepsilon } \left(
    p\right) }} {L\sin  \left( \frac {p-q} {2}\right)
}&=\sum_{k=-\infty}^{+\infty }\frac {e^{-it \overline{\varepsilon
    }(q+2\pi (k+1/2)/L)}} {\pi(k+1/2) } -\int_{\pi}^\infty
\frac{2e^{-it \overline{\varepsilon
    }(x)}}{(x-q)}\frac{dx}{2\pi}-\int_{-\infty}^{-\pi} \frac{2e^{-it
    \overline{\varepsilon }(x)}}{(x-q)}\frac{dx}{2\pi}\\ 
&+\int_{-\pi}^\pi e^{-it \overline{\varepsilon }(x)}\left(\frac{1}{\sin \tfrac{x-q}{2}}-\frac{1}{(x-q)/2}\right)\frac{dx}{2\pi}+\mathcal{O}\left( L^{-1}\right)\ ,
\end{aligned}
\label{40b}
\end{equation}
where the integrals are again integrals of oscillatory bounded
functions and vanish at late times.

Finally we use \eqref{sumuseful} to obtain the asymptotic values of
the sums in \fr{40a} and \fr{40b} for $q\in\rm{NS}$ and  $\overline{\varepsilon}'(q)\neq 0$
\begin{equation}
  \begin {aligned}
  	\sum _{p\in\rm{R}}\frac {e^{-it\overline{\varepsilon }(p) }} {L\sin \left( \frac {p-q} {2}\right) }&=-i \sign (t\overline{\varepsilon } ^{'}(q) )e^{-it \overline{\varepsilon } (q) }+\mathcal{O}\left( L^{0}t^{-1 / 2}\right) +\mathcal{O}\left( L^{-1}\right), \\
  	\sum _{p\in\rm{R}}\frac {e^{-it \overline{\varepsilon } (p) }} {L^2\sin^2  \left( \frac {p-q} {2}\right) }&=\left(1-\frac {2\left| t \overline{\varepsilon } '(q) \right| } {L}\right)e^{-it \overline{\varepsilon } (q) }+\mathcal{O}\left( L^{-1}t^{-1 / 2}\right) +\mathcal{O}\left( L^{-2}\right) \ .
  \end {aligned}
\end{equation}

%%%%%%%%%%%%%%%%%%%%%%%%%%%%%%%%%%%%%%%%%%%%%%%%%%%%%%%%%%%%%%%%
\subsubsection{\texorpdfstring{Sums arising in quantum quench dynamics \eqref{use2}}{Lg}}
%%%%%%%%%%%%%%%%%%%%%%%%%%%%%%%%%%%%%%%%%%%%%%%%%%%%%%%%%%%%%%%%
We now turn to momentum sums of the form
\be
\Sigma^{(n)}(q,q',t)=\frac{4}{L^n}\sum_{0<p \in \rm{R}}\frac{\sin p \sin
  q' f(p)}{(\cos q-\cos
  p)^nf(q')}e^{2it(\varepsilon(p)-\varepsilon(q))}\ ,\quad n=1,2\ ,
\ee
where $f(q)$ is defined in \fr{f}. Using that $\frac{\sin x f(x)}{\cos
  q-\cos x}-\frac{f(q)}{x-q}$ is a bounded function of $x$ we can
proceed along the same lines as in Section~\ref{ssec:eqns40} to
conclude that for $q\in\rm{NS}$
\begin{equation}
\Sigma^{(1)}(q,q',t)=\frac{4}{L}\sum_{0<p\in\rm{R}}\frac{\sin
  q'}{p-q}e^{2it(\varepsilon(p)-\varepsilon(q))}\frac{f(q)}{f(q')}+\mathcal{O}(L^0
t^{-1/2})\ .
\end{equation}
At leading order in time we may then Taylor expand $\varepsilon(p)$
around $p=q$, and write $p=q+\frac{2\pi}{L}(n+1/2)$ to fall back on
one of the oscillatory sums in \eqref{sumuseful}. In this way we
obtain our final result
\be
\Sigma^{(1)}(q,q',t)=2i\sign(t\varepsilon'(q))\sin q'\frac{f(q)}{f(q')}+\mathcal{O}(L^{0}t^{-1/2}).
\ee
The analysis of the $\Sigma^{(2)}(q,q',t)$ proceeds in complete
analogy: we use that $\frac{\sin x f(x)}{(\cos x-\cos
  q)^2}-\frac{f(q)}{\sin q(x-q)^2}-\frac{f'(q)}{\sin q (x-q)}$ is
a bounded function of $x$ to conclude that
\begin{equation}
\begin{aligned}
\Sigma^{(2)}(q,q',t)&=\frac{4}{L^2}\sum_{p>0, \in
  \rm{R}}\frac{e^{2it(\varepsilon(p)-\varepsilon(q))}}{(p-q)^2}
+\frac{4}{L^2}\sum_{p>0,\in\rm{R}}\frac{f'(q)}{f(q)}\frac{e^{2it(\varepsilon(p)-\varepsilon(q))}}{p-q}
+\mathcal{O}(L^{-1} t^{-1/2})\ .
\end{aligned}
\end{equation}
Taylor expanding $\varepsilon(p)$ around $p=q$ and using
\eqref{sumuseful} we finally arrive at
\begin{equation}
\begin{aligned}
\Sigma^{(2)}(q,q',t)&=1-\frac{4  |t\varepsilon'(q)|}{L}+\frac{2i\sign(t\varepsilon'(q))}{L}\frac{f'(q)}{f(q)}+\mathcal{O}(L^{-1}t^{-1/2})\ .
\end{aligned}
\end{equation}

%%%%%%%%%%%%%%%%%%%%%%%%%%%%%%%%%%%%%%
\subsection{Two-dimensional sums}
%%%%%%%%%%%%%%%%%%%%%%%%%%%%%%%%%%%%%%

In this subsection we calculate the two-dimensional sums arising in Sections
\ref{rho2fixt} and \ref{qu}. Apart from being two-dimensional, the sums treated in this section differ from the previous section by the fact that they are performed over the particles of the representative state, and not over arbitrary, regularly spaced momenta in the Ramond sector.

\subsubsection{Sums arising in finite-temperature dynamics}
We consider
\begin{equation}
\begin{aligned}
\label{type1}
\Omega_1(t)&= \frac{1}{L^2}\sum _{i\neq j} \frac {e^{it( \overline{\varepsilon}(q_j)- \overline{\varepsilon}(q_i))}} {\sin ^{2}\left( \frac {q_{i}-q_{j}} {2}\right) }\sign( \overline{\varepsilon}'(q_j) \overline{\varepsilon}'(q_i))\,.
\end{aligned}
\end{equation}
This sum is divergent when $L\to\infty$. It grows as $\propto L$ and the proportionality constant depends on the realization of the root density $\rho$ in finite-size through the $q_i$'s, and as a consequence cannot be written in terms of the root density. 
%\footnote{The reader is encouraged to explicitly work out the case $\rho(\lambda)=\frac{1}{2\pi n}$ with $n$ an integer, that can be realized in finite-size with for example either $q_i=\frac{2\pi}{L}(ni+1/2)$, or $q_{2i}=\frac{2\pi}{L}(2ni+1/2),q_{2i+1}=\frac{2\pi}{L}(2ni+1+1/2)$. Although the density is the same, the prefactors of the $L$ divergence in the sum considered differ.}. 
However, this prefactor does not depend on $t$, and the difference $\Omega_1(t)-\Omega_1(0)$ which appears in the main text is not divergent in the thermodynamic limit. 

We have by symmetrizing the sum over $i,j$
\begin{equation}
\begin{aligned}
\Omega_1(t)-\Omega_1(0)&= \frac{1}{2L^2}\sum _{i\neq j} \frac {e^{it( \overline{\varepsilon}(q_j)- \overline{\varepsilon}(q_i))}+e^{it( \overline{\varepsilon}(q_i)- \overline{\varepsilon}(q_j))}-2} {\sin ^{2}\left( \frac {q_{i}-q_{j}} {2}\right) }\sign( \overline{\varepsilon}'(q_j) \overline{\varepsilon}'(q_i))\,.
\end{aligned}
\end{equation}
The summand does not have poles anymore, so that the sum can be turned into an integral in the $L\to\infty$ limit
\begin{equation}\label{comingback}
\begin{aligned}
&\Omega_1(t)-\Omega_1(0)= \\
&\frac{1}{2}\int_{-\pi}^\pi\int_{-\pi}^\pi\frac{e^{it( \overline{\varepsilon}(v)- \overline{\varepsilon}(u))}+e^{it( \overline{\varepsilon}(u)- \overline{\varepsilon}(v))}-2}{\sin ^{2}\left( \frac {u-v} {2}\right)}\sign( \overline{\varepsilon}'(u) \overline{\varepsilon}'(v))\rho(u)\rho(v)dudv+{\cal O}(L^{-1})\,.
\end{aligned}
\end{equation}
We now have to determine the large $t$ behaviour of this expression. We first write
\begin{equation}
\Omega_1(t)-\Omega_1(0)=-2\int_{-\pi}^\pi\int_{-\pi}^\pi\frac{\sin^2[t( \overline{\varepsilon}(v)- \overline{\varepsilon}(u))/2]}{\sin ^{2}\left( \frac {u-v} {2}\right)}\sign( \overline{\varepsilon}'(u) \overline{\varepsilon}'(v))\rho(u)\rho(v)dudv\,,
\end{equation}
and perform a change of variable $v=u+\eta/t$
\begin{equation}
\begin{aligned}
&\Omega_1(t)-\Omega_1(0)=\\
&-2\int_{-\pi}^\pi du \int_{(-\pi-u)|t|}^{(\pi-u)|t|}d\eta\frac{\sin^2[t( \overline{\varepsilon}(u+\tfrac{\eta}{t})- \overline{\varepsilon}(u))/2]}{|t|\sin ^{2}\left( \frac {\eta} {2t}\right)}\sign( \overline{\varepsilon}'(u) \overline{\varepsilon}'(u+\tfrac{\eta}{t}))\rho(u)\rho(u+\tfrac{\eta}{t})\,.
\end{aligned}
\end{equation}
At leading order in $t$, it yields
\begin{equation}
\begin{aligned}
&\Omega_1(t)-\Omega_1(0)=-8|t|\int_{-\pi}^\pi du \int_{-\infty}^{\infty}d\eta\frac{\sin^2[\overline{\varepsilon}'(u) \eta/2]}{\eta^2}\rho(u)^2+o(|t|)\,.
\end{aligned}
\end{equation}
Using $\int_{-\infty}^\infty \frac{\sin^2 x}{x^2}dx=\pi$, we obtain
\begin{equation}\label{leaddecay}
\begin{aligned}
&\Omega_1(t)-\Omega_1(0)=-4\pi\int_{-\pi}^\pi |t\overline{\varepsilon}'(u)|\rho(u)^2+o(|t|)\,.
\end{aligned}
\end{equation}
Let us determine the sub-leading term $o(|t|)$ in the space-like regime, i.e. when $\sign( \overline{\varepsilon}'(u))$ is constant, and when the root density $\rho$ is continuous. In this case, coming back to \eqref{comingback}, we integrate by part to obtain
\begin{equation}
\begin{aligned}
\Omega_1(t)-\Omega_1(0)&=2t\int_{-\pi}^\pi\int_{-\pi}^\pi\overline{\varepsilon}'(v)\frac{\sin[t( \overline{\varepsilon}(v)- \overline{\varepsilon}(u))]}{\tan\left( \frac {u-v} {2}\right)}\rho(u)\rho(v)dudv\\
&+2\int_{-\pi}^\pi\int_{-\pi}^\pi\frac{1}{\tan\left( \frac {u-v} {2}\right)}\rho(u)\rho'(v)dudv\\
&-\int_{-\pi}^\pi\int_{-\pi}^\pi\frac{e^{it( \overline{\varepsilon}(v)- \overline{\varepsilon}(u))}+e^{it( \overline{\varepsilon}(u)- \overline{\varepsilon}(v))}}{\tan\left( \frac {u-v} {2}\right)}\rho(u)\rho'(v)dudv\,,
\end{aligned}
\end{equation}
where the last two double integrals are understood in principal value. 

Let us focus first on the last double integral, that we separate into two integration regions, one with $|u-v|>\epsilon$ and one with $|u-v|<\epsilon$ for a small fixed $\epsilon>0$. In the first region, the term is an oscillatory integral of a bounded function, hence decays to zero with time. In the  second region, the $v$ integral at fixed $u$ may be approximated by
\begin{equation}
\approx 2\rho'(u)\int_{-f_-(u)}^{f_+(u)} \frac{e^{it x\overline{\varepsilon}'(u)}+e^{-it x\overline{\varepsilon}'(u)}}{x}dx\,,
\end{equation}
with $0<f_\pm(u)<\epsilon$, where $f_\pm(u)$ are some limits that depend on $u$. Assuming without loss of generality $f_-(u)<f_+(u)$, this integral is $\int_{f_-(u) t}^{f_+(u)t}\tfrac{e^{iy\overline{\varepsilon}'(u)}}{y}dy-\int_{-f_+(u) t}^{-f_-(u)t}\tfrac{e^{iy\overline{\varepsilon}'(u)}}{y}dy$, which decays to zero when $t\to\infty$. Hence 
\begin{equation}
\begin{aligned}
\Omega_1(t)-\Omega_1(0)&=2t\int_{-\pi}^\pi\int_{-\pi}^\pi\overline{\varepsilon}'(v)\frac{\sin[t( \overline{\varepsilon}(v)- \overline{\varepsilon}(u))]}{\tan\left( \frac {u-v} {2}\right)}\rho(u)\rho(v)dudv\\
&+2\int_{-\pi}^\pi\int_{-\pi}^\pi\frac{1}{\tan\left( \frac {u-v} {2}\right)}\rho(u)\rho'(v)dudv\\
&+o(t^0)\,.
\end{aligned}
\end{equation}
We now focus on the first double integral. First, since in the space-like regime $\overline{\varepsilon}'(u)$ never vanishes, $\overline{\varepsilon}(u)$ is one-to-one from $[-\pi,\pi]$ to $[\varepsilon_{\min},\varepsilon_{\max}]$. Hence one can perform a change of variable $x=\overline{\varepsilon}(u),y=\overline{\varepsilon}(v)$ to obtain
\begin{equation}\label{toadd}
\begin{aligned}
&\int_{-\pi}^\pi\int_{-\pi}^\pi\overline{\varepsilon}'(v)\frac{\sin[t( \overline{\varepsilon}(v)- \overline{\varepsilon}(u))]}{\tan\left( \frac {u-v} {2}\right)}\rho(u)\rho(v)dudv=\\
&\int_{\varepsilon_{\min}}^{\varepsilon_{\max}}\int_{\varepsilon_{\min}}^{\varepsilon_{\max}}\frac{1}{\overline{\varepsilon}'(\overline{\varepsilon}^{-1}(x))}\frac{\sin[t(y-x)]}{\tan\left( \frac {\overline{\varepsilon}^{-1}(x)-\overline{\varepsilon}^{-1}(y)} {2}\right)}\rho(\overline{\varepsilon}^{-1}(x))\rho(\overline{\varepsilon}^{-1}(y))dxdy\,.
\end{aligned}
\end{equation}
We now make the following observation. For any regular function $f(x,y)$, the integral
\begin{equation}
\int_{\varepsilon_{\min}}^{\varepsilon_{\max}}\int_{\varepsilon_{\min}}^{\varepsilon_{\max}}\sin[t(y-x)]f(x,y)dxdy
\end{equation}
is $o(t^{-1})$. Indeed, by integrating by part the $y$ integral we make appear a $1/t$, and the remaining integrals are oscillatory integrals, hence decay to zero with time. By adding and subtracting the appropriate term to the right-hand side of \eqref{toadd} so as to cancel the pole, we conclude from this 
\begin{equation}
\begin{aligned}
&\int_{\varepsilon_{\min}}^{\varepsilon_{\max}}\int_{\varepsilon_{\min}}^{\varepsilon_{\max}}\frac{1}{\overline{\varepsilon}'(\overline{\varepsilon}^{-1}(x))}\frac{\sin[t(y-x)]}{\tan\left( \frac {\overline{\varepsilon}^{-1}(x)-\overline{\varepsilon}^{-1}(y)} {2}\right)}\rho(\overline{\varepsilon}^{-1}(x))\rho(\overline{\varepsilon}^{-1}(y))dxdy\\
&\qquad\qquad\qquad=2\int_{\varepsilon_{\min}}^{\varepsilon_{\max}}\int_{\varepsilon_{\min}}^{\varepsilon_{\max}}\frac{\sin[t(y-x)]}{x-y}\rho(\overline{\varepsilon}^{-1}(x))^2dxdy+o(t^{-1})\,.
\end{aligned}
\end{equation}
We now split the $x$ integral into several pieces $[x_n,x_{x+1}]$ with $x_{n+1}-x_n$ small enough so that $\rho$ can be approximated to be constant on these pieces. Then we have
\begin{equation}
\int_{x_n}^{x_{n+1}}dx\int_{\varepsilon_{\min}}^{\varepsilon_{\max}}dy \frac{\sin[t(y-x)]}{x-y}=-\frac{1}{t}\int_{|t|(\varepsilon_{\max}-x_{n+1})}^{|t|(\varepsilon_{\max}-x_{n})} \si(u)du+\frac{1}{t}\int_{|t|(\varepsilon_{\min}-x_{n+1})}^{|t|(\varepsilon_{\min}-x_{n})} \si(u)du\,,
\end{equation}
with $\si(u)=\int_0^u \frac{\sin x}{x}dx$ the sinus integral. Using the expansion at large $u>0$
\begin{equation}
\si(u)=\frac{\pi}{2}-\frac{\cos u}{u}+{\cal O}(u^{-2})\,,
\end{equation}
one obtains
\begin{equation}
\int_{x_n}^{x_{n+1}}dx\int_{\varepsilon_{\min}}^{\varepsilon_{\max}}dy \frac{\sin[t(y-x)]}{x-y}=-\pi \sign(t)(x_{n+1}-x_n)+{\cal O}(t^{-2})\,.
\end{equation}
Summing over the windows $[x_n,x_{n+1}]$ we obtain
\begin{equation}
\begin{aligned}
\int_{\varepsilon_{\min}}^{\varepsilon_{\max}}\int_{\varepsilon_{\min}}^{\varepsilon_{\max}}\frac{\sin[t(y-x)]}{x-y}\rho(\overline{\varepsilon}^{-1}(x))^2dxdy=-\pi\sign(t)\int_{\varepsilon_{\min}}^{\varepsilon_{\max}}\rho(\overline{\varepsilon}^{-1}(x))^2dx+{\cal O}(t^{-2})\,,
\end{aligned}
\end{equation}
which yields in the space-like regime
\begin{equation}
\begin{aligned}
&\Omega_1(t)-\Omega_1(0)=-4\pi\int_{-\pi}^\pi |t\overline{\varepsilon}'(u)|\rho(u)^2+2\int_{-\pi}^\pi\int_{-\pi}^\pi\frac{1}{\tan\left( \frac {u-v} {2}\right)}\rho(u)\rho'(v)dudv+o(t^0)\,.
\end{aligned}
\end{equation}
\subsubsection{\texorpdfstring{Sums arising in quantum quench dynamics}{Lg}}
%%%%%%%%%%%%%%%%%%%%%%%%%%%%%%%%%%%%%%%%%%%%%%%%%%%%%%%%%%%%%%%%
We consider
\begin{equation}\label{qqq}
\begin{aligned}
\Omega_2(t)&= \frac{8}{L^2}\sum _{i\neq j} \frac {\sin q_i \sin q_j} {(\cos q_i-\cos q_j)^2}e^{2it( \overline{\varepsilon}(q_i)- \overline{\varepsilon}(q_j))}\\
\tilde{\Omega}_2(t)&= \frac{8}{L^2}\sum _{i\neq j} \frac {\sin q_i \sin q_j} {(\cos q_i-\cos q_j)^2}\frac{f(q_i)}{f(q_j)}e^{2it( \overline{\varepsilon}(q_i)- \overline{\varepsilon}(q_j))}\,.
\end{aligned}
\end{equation}
Again, $\Omega_2(t)$ and $\tilde{\Omega}_2(t)$ diverge as $L$ in the thermodynamic limit. But the coefficient of the divergence does not depend on $t$ and is the same for $\Omega_2(t)$ and $\tilde{\Omega}_2(t)$. The differences $\tilde{\Omega}_2(t)-\Omega_2(0)$ are well-defined in the thermodynamic limit. Using the same approach as in the previous section, one obtains
\begin{equation}
\begin{aligned}
\tilde{\Omega}_2(t)-\Omega_2(0)&=-16\pi\int_0^\pi
\rho(x)^2 |t\overline{\varepsilon}'(x)|dx-8\int_0^\pi\int_0^\pi dx dy\  \frac {\sin y} {\cos
  y-\cos x} \rho'(x) \rho( y)\\
  &+o(t^0)+\mathcal{O}(L^{-1})\ .
\end{aligned}
\end{equation}


\begin{thebibliography}{99}

%%%%%%%%%%% finite temperature dynamics %%%%%%%%%%%%%%%%

%%%%%%% finiteT %%%%%%%%%%
\bibitem{LSM}
E. Lieb, T. Schultz and D. Mattis, \emph{Two soluble models of an
  antiferromagnetic chain}, \href{http://dx.doi.org/j.aop.10.1016}
{\bf 16}, 407 (1961).

\bibitem{Perk80}
J.H.H. Perk, \emph{Equations of motion for the transverse correlations
of the one-dimensional XY-model at finite temperature},
Phys. Lett. A \href{https://doi.org/10.1016/0375-9601(80)90298-4}{{\bf 79},1 (1980).}

\bibitem{Perk84}
J.H.H. Perk, H.W. Capel, G.R.W. Quispel, and F.W. Nijhoff,
\emph{Finite-temperature correlations for the Ising chain in a transverse field},
Physica \href{https://doi.org/10.1016/0378-4371(84)90102-X}{{\bf A 123}, 1 (1984).}

\bibitem{Perk09}
J.H.H. Perk and H. Au-Yang,
\emph{New Results for the Correlation Functions of the Ising Model and
  the Transverse Ising Chain}, J. Stat. Phys. \href{https://doi.org/10.1007/s10955-009-9758-5}{{\bf 135}, 599 (2009).}
     
\bibitem{MPS83a}
B.M. McCoy, J.H.H. Perk, and R.E. Shrock,
\emph{Time-dependent correlation functions of the transverse Ising
  chain at the critical magnetic field},  Nucl. Phys. \href{https://doi.org/10.1016/0550-3213(83)90132-3}{{\bf B 220}, 35 (1983).}

\bibitem{MPS83b}
B.M. McCoy, J.H.H. Perk, and R.E. Shrock,
\emph{Correlation functions of the transverse Ising chain at the
critical field for large temporal and spatial separations},
Nucl. Phys. \href{https://doi.org/10.1016/0550-3213(83)90041-X}{{\bf B 220}, 269 (1983).}

\bibitem{IIKS93}
A. R. Its, A. G. Izergin, V. E. Korepin, and N. A. Slavnov,
\emph{Temperature correlations of quantum spins},
Phys. Rev. Lett. \href{https://doi.org/10.1103/PhysRevLett.70.1704}{{\bf
    70}, 1704 (1993).}

\bibitem{Colomo93}
F. Colomo, A. G. Izergin, V. E. Korepin and V. Tognetti,
\emph{Temperature correlation functions in the XX0 Heisenberg chain. I},
Theor. Math. Phys.\href{https://doi.org/10.1007/BF01016992}{{\bf 94}, 11 (1993).}

\bibitem{vladbook}
V.E. Korepin, A.G. Izergin and N.M. Bogoliubov, {\em {Quantum Inverse
  Scattering Method, Correlation Functions and Algebraic Bethe Ansatz}}
  (Cambridge University Press, 1993).

\bibitem{leclair1996} 
A. Leclair, F. Lesage, S. Sachdev and H. Saleur, \emph{Finite
  temperature correlations in the one-dimensional quantum Ising
  model}, Nucl. Phys. B {\bf 482}, 579 (1996). 

%%%%%%%%%%% semiclassics %%%%%%%%%%%%%%%%%%%

\bibitem{young}
S. Sachdev, A. P. Young, \emph{Low Temperature Relaxational Dynamics
  of the Ising Chain in a Transverse Field}, Phys. Rev. Lett. \href{https://doi.org/10.1103/PhysRevLett.78.2220}{{\bf 78}, 2220 (1997).}

\bibitem{damle98}
K. Damle and S. Sachdev, \emph{Spin dynamics and transport in gapped
  one-dimensional Heisenberg antiferromagnets at nonzero
  temperatures}, Phys. Rev. B \href{https://doi.org/10.1103/PhysRevB.57.8307}{{\bf 57}, 8307 (1998).}

\bibitem{damle05}
K. Damle, S. Sachdev, \emph{Universal Relaxational Dynamics of Gapped
  One-Dimensional Models in the Quantum Sine-Gordon Universality
  Class}, Phys. Rev. Lett. \href{https://doi.org/10.1103/PhysRevLett.95.187201}{{\bf
    95}, 187201 (2005).}

\bibitem{zarand09}
A. Rapp, G. Zarand, \emph{Universal diffusive decay of correlations in
  gapped one-dimensional systems}, Eur. Phys. Jour. \href{https://doi.org/10.1140/epjb/e2008-00465-5}{{\bf B67}, 7 (2009).}

\bibitem{muss}  
A. LeClair, G. Mussardo, \emph{Finite temperature correlation
  functions in integrable QFT}, Nucl. Phys. B \href{https://doi.org/10.1016/S0550-3213(99)00665-3}{ {\bf 552}, 624 (1999).}

\bibitem{Saleur00}
H. Saleur, \emph{A comment on finite temperature correlations in
  integrable QFT}, Nucl. Phys. \href{https://doi.org/10.1016/S0550-3213(99)00665-3}{{\bf
    B567} 602 (2000).} 

\bibitem{CF02}
O.A. Castro Alvaredo and A. Fring, \emph{Finite temperature
  correlation functions from form factors}, Nucl. Phys. \href{https://doi.org/10.1016/S0550-3213(02)00409-1}{{\bf B636} 611 (2002).}

\bibitem{rmk} 
R.M. Konik, \emph{Haldane-gapped spin chains : Exact low-temperature
  expansion of correlation functions}, Phys. Rev. B
\href{https://doi.org/10.1103/PhysRevB.68.104435}{{\bf 68}, 104435 (2003).}

\bibitem{AKT} 
B.L. Altshuler, R.M. Konik and A.M. Tsvelik, \emph{Low temperature
  correlation functions in integrable models: Derivation of the large
  distance and time asymptotics from the form factor expansion},
Nucl. Phys. \href{https://doi.org/10.1016/j.nuclphysb.2006.01.022}{{\bf
    B739}, 311 (2006).} 

\bibitem{Reyes06}
S.A. Reyes, A. Tsvelik, Phys. Rev. B {\bf 73}, 220405(R) (2006).

\bibitem{PT08} 
B. Pozsgay and G. Takacs, \emph{Form factors in finite volume I:
form factor bootstrap and truncated conformal space},
Nucl. Phys. B. \href{https://doi.org/10.1016/j.nuclphysb.2007.06.027}{{\bf 788} 167 (2008). }

\bibitem{takacs} 
B. Pozsgay and G. Takacs, \emph{Form factors in finite volume II:
  Disconnected terms and finite temperature correlators},
Nucl. Phys. B. \href{https://doi.org/10.1016/j.nuclphysb.2007.07.008}{{\bf 788} 209 (2008). }

\bibitem{EK08}
F.H.L. Essler and R.M. Konik, \emph{Finite-temperature lineshapes in
  gapped quantum spin chains}, Phys. Rev. B
\href{https://doi.org/10.1103/PhysRevB.78.100403}{{\bf 78}, 100403(R) (2008).}

\bibitem{EK:finiteT}
F.H.L. Essler and R.M. Konik, \emph{Finite Temperature Dynamical
  Correlations in Massive Integrable Quantum Field Theories},
J. Stat. Mech. (2009) \href{http://dx.doi.org/10.1088/1742-5468/2009/09/P09018}{P09018}.

\bibitem{PT10}
B. Pozsgay and G Takacs, \emph{Form factor expansion for thermal
  correlators}, J. Stat. Mech. \href{https://doi.org/10.1088/1742-5468/2010/11/P11012}{(2010) P11012.}

\bibitem{Steinberg}
J. Steinberg, N.P. Armitage, F.H.L. Essler and S. Sachdev, Phys. Rev. B 99, 035156 (2019).

\bibitem{Doyon05}
B. Doyon, \emph{Finite-temperature form factors in the free Majorana
  theory},
J. Stat. Mech. \href{https://doi.org/10.1088/1742-5468/2005/11/P11006}{P11006
  (2005).} 

\bibitem{doyongamsa}
B. Doyon and A. Gamsa, \emph{Integral equations and long-time asymptotics for
finite-temperature Ising chain correlation functions}, J. Stat. Mech. \href{https://doi.org/10.1088/1742-5468/2008/03/P03012}{ P03012 (2008).}

\bibitem{PC14}
M. Panfil and J.-S. Caux, \emph{Finite-temperature correlations in the
  Lieb-Liniger one-dimensional Bose gas},
Phys. Rev. A\href{http://dx.doi.org/10.1103/PhysRevA.89.033605}{89,
  033605 (2014).}

\bibitem{kozlowski1}
K. K. Kozlowski, \emph{On the thermodynamic limit of form factor expansions of dynamical correlation functions in the massless regime of the {XXZ} spin {1/2} chain},  J. Math. Phys. \href{\doi10.1063/1.5021892}{{\bf 59}, 091408, (2018).}

\bibitem{kozlowski2}
K. K. Kozlowski, \emph{On singularities of dynamic response functions
in the massless regime of the XXZ spin-1/2 chain}, preprint arXiv:1811.06076.

\bibitem{kozlowski3}
K. K. Kozlowski, \emph{Long-distance and large-time asymptotic behaviour of dynamic correlation functions in the massless regime of the {XXZ} spin-{1/2} chain},  J. Math. Phys. \href{\doi10.1063/1.5094332}{{\bf 60}, 073303, (2019).}

\bibitem{kozlowskimaillet}
K. K. Kozlowski and J.-M. Maillet, \emph{Microscopic approach to a class of {1D} quantum critical models},  J. Phys. A  \href{\doi10.1088/1751-8113/48/48/484004}{{\bf 48}, 484004, (2015).}

\bibitem{DNP18}
J. de Nardis and M. Panfil, \emph{Particle-hole pairs and
density-density correlations in the Lieb-Liniger model},
J. Stat. Mech. \href{https://doi.org/10.1088/1742-5468/aab012}
{(2018) 033102.}

\bibitem{CCP19}
A. Cortes Cubero and M. Panfil, \emph{Thermodynamic bootstrap program
for integrable QFTs: form factors and correlation functions at
  finite energy density.}, J. High
Energ. Phys. \href{https://doi.org/10.1007/JHEP01(2019)104}{104 (2019).} 

\bibitem{Klumper93}
A. Kl\"umper, \emph{Thermodynamics of the anisotropic spin-1/2
  Heisenberg chain and related quantum chains}, Z. Phys. B \href{https://doi.org/10.1007/BF01316831}{{\bf 91}, 507 (1993).}

\bibitem{DDV95}
C. Destri and H.J. de Vega, \emph{Unified Approach to Thermodynamic
  Bethe Ansatz and Finite Size Corrections for Lattice Models and
  Field Theories}, Nucl. Phys. B \href{https://doi.org/10.1016/0550-3213(94)00547-R}{{\bf 438}, 314 (1995).}

\bibitem{Damerau07}
J. Damerau, F. G\"ohmann, N.P. Hasenclever and A. Kl\"umper,
\emph{Density matrices for finite segments of Heisenberg chains of
  arbitrary length}, J. Phys. A  \href{https://doi.org/10.1088/1751-8113/40/17/002}{{\bf 40} 4439 (2007).}

\bibitem{Boos08}
H.E. Boos, J. Damerau, F. G\"ohmann, A. Kl\"umper, J. Suzuki, and
A. Wei{\ss}e, \emph{Short-distance thermal correlations in the XXZ
  chain}, J. Stat. Mech. \href{https://doi.org/10.1088/1742-5468/2008/08/P08010}{(2008) P08010.}

\bibitem{Trippe10}
C. Trippe, F. G\"ohmann, and A. Kl\"umper, \emph{Short-distance
thermal correlations in the massive XXZ chain}, Eur. Phys. J. B
\href{10.1140/epjb/e2009-00417-7}{{\bf 73}, 253 (2010).} 

\bibitem{Dugave13}
M. Dugave, F. G\"ohmann and K.K. Kozlowski, \emph{Thermal form factors
  of the XXZ chain and the large-distance asymptotics of its
  temperature dependent correlation functions},
J. Stat. Mech. \href{https://doi.org/10.1088/1742-5468/2013/07/P07010}{(2013)
  P07010.}

\bibitem{Dugave14}
M. Dugave, F. G\"ohmann and K.K. Kozlowski, \emph{Low-temperature
  large-distance asymptotics of the transversal two-point functions of
  the XXZ chain},
J. Stat. Mech. \href{https://doi.org/10.1088/1742-5468/2014/04/P04012}{(2014)
  P04012.}

\bibitem{GKKKS17}
F. G\"ohmann, M. Karbach, A. Kl\"umper, K.K. Kozlowski and J. Suzuki, \emph{Thermal form-factor approach to dynamical correlation functions of integrable lattice models},
J. Stat. Mech. \href{https://doi.org/10.1088/1742-5468/aa9678}{ (2017) P113106.}

\bibitem{GKS18}
F. G\"ohmann, K.K. Kozlowski and J. Suzuki, \emph{High-temperature
  analysis of the transverse dynamical two-point correlation function
  of the XX quantum-spin chain},
J. Math. Phys.\href{https://doi.org/10.1063/1.5111039}{61, 013301
  (2020).}

\bibitem{GKSS19}
F. G\"ohmann, K.K. Kozlowski, J. Sirker and J. Suzuki, \emph{
Equilibrium dynamics of the XX chain}, Phys. Rev. B\href{https://doi.org/10.1103/PhysRevB.100.155428}{100, 155428 (2019).}

\bibitem{GKS19}
F. G\"ohmann, K.K. Kozlowski and J. Suzuki, \emph{Late-time
  large-distance asymptotics of the transverse correlation functions
  of the XX chain in the space-like regime}, 	arXiv:1908.11555. 


%%%%%%%%%%%%%% Ising %%%%%%%%%%%%%%%%%%%%%%%

%\bibitem{leclairbernard94}
%A. Leclair and D. Bernard, \emph{Differential equations for
%  sine-Gordon correlation functions at the free fermion point},
%Nucl. Phys. \href{https://doi.org/10.1016/0550-3213(94)90020-5}{{\bf B426}, 534 (1994)}; Erratum-%ibid. B {\bf 498}, 619 (1997).

%%%%%%%%%%%%%% GHD %%%%%%%%%%%%%%%%%%%%

\bibitem{CADY16}
O. A. Castro-Alvaredo, B. Doyon, T. Yoshimura, \emph{Emergent hydrodynamics
 in integrable quantum systems out of equilibrium}, Phys. Rev. X \href{https://doi.org/10.1103/PhysRevX.6.041065}{{\bf 6}, 041065 (2016).}

\bibitem{BCDF16}
B. Bertini, M. Collura, J. De Nardis, M. Fagotti, \emph{Transport in
out-of-equilibrium xxz chains: Exact profiles of charges and
currents}, Phys. Rev. Lett. \href{https://doi.org/10.1103/PhysRevLett.117.207201}{{\bf 117}, 207201 (2016).}

\bibitem{Doyon18}
B. Doyon, \emph{Exact large-scale correlations in integrable systems
  out of equilibrium}, SciPost Phys. \href{https://doi:
  10.21468/SciPostPhys.5.5.054}{{\bf 5}, 054 (2018).}

\bibitem{Sarang19}
S. Gopalakrishnan, R. Vasseur and B. Ware, \emph{Anomalous relaxation
  and the high-temperature structure factor of XXZ spin chains}, 
PNAS \href{ https://doi.org/10.1073/pnas.1906914116}{{\bf 116}, 16250 (2019).}



%%%%%%%%%%% quench dynamics %%%%%%%%%%%%%%%%

%%%% Ising %%%%%%%

\bibitem{BMD70}
E. Barouch, B. McCoy, and M. Dresden, \emph{Statistical Mechanics of
  the XY Model. I}, Phys. Rev. A
\href{http://dx.doi.org/10.1103/PhysRevA.2.1075}{\bf 2}, 1075 (1970).

\bibitem{BM71}
E. Barouch and B. McCoy, \emph{Statistical Mechanics of
  the XY Model. III}, Phys. Rev. A
\href{https://doi.org/10.1103/PhysRevA.3.2137}{{\bf 3}, 2137 (1971).}

\bibitem{SPS:04} K. Sengupta, S. Powell, and S. Sachdev, \emph{Quench
dynamics across quantum critical points}, Phys. Rev. A
  \href{http://dx.doi.org/10.1103/PhysRevA.69.053616}{\bf 69}, 053616 (2004).

\bibitem{RSMS09}
D. Rossini, A. Silva, G. Mussardo, and G.E. Santoro, \emph{Effective
  Thermal Dynamics Following a Quantum Quench in a Spin Chain},
Phys. Rev. Lett. \href{http://dx.doi.org/10.1103/PhysRevLett.102.127204}{\bf 
  102}, 127204 (2009).

\bibitem{RSMS10}
D. Rossini, S. Suzuki, G. Mussardo,
  G.E. Santoro, and A. Silva, \emph{Long time dynamics following a
    quench in an integrable quantum spin chain: Local versus nonlocal
    operators and effective thermal behavior}, Phys. Rev. B
  \href{http://dx.doi.org/10.1103/PhysRevB.82.144302}{\bf 82}, 144302
  (2010).

%%%%%%%%%%%%%%%%%%%%%%%%%%

\bibitem{CEF1} 
P. Calabrese, F.H.L. Essler, and M. Fagotti, \emph{Quantum Quench in the Transverse-Field Ising Chain}, Phys. Rev. Lett. \href{http://dx.doi.org/10.1103/PhysRevLett.106.227203}{\bf
  106}, 227203 (2011).

\bibitem{CEF2} 
P. Calabrese, F.H.L. Essler, and M. Fagotti, \emph{Quantum quench in the transverse field Ising chain: I. Time evolution of order parameter correlators}. J. Stat. Mech. (2012) \href{http://dx.doi.org/10.1088/1742-5468/2012/07/P07016}{P07016}.

\bibitem{CEF3} 
P. Calabrese, F.H.L. Essler, and M. Fagotti, \emph{Quantum Quench in the Transverse Field Ising Chain II: Stationary State Properties}, J. Stat. Mech. (2012) \href{http://dx.doi.org/10.1088/1742-5468/2012/07/P07022}{P07022}. 

\bibitem{FE13}
M.~Fagotti and F.H.L.~Essler, \emph{Reduced density matrix after a quantum quench}, Phys. Rev. B \href{http://dx.doi.org/10.1103/PhysRevB.87.245107}{\bf 87}, 245107 (2013).

\bibitem{SE12}
D. Schuricht and F.H.L. Essler, \emph{Dynamics in the Ising field theory after a quantum quench}, J. Stat. Mech. (2012) \href{http://dx.doi.org/10.1088/1742-5468/2012/04/P04017}{P04017}.

\bibitem{EEF12}
F.H.L. Essler, S. Evangelisti and M. Fagotti, \emph{Dynamical
  Correlations After a Quantum Quench},
Phys. Rev. Lett. \href{http://dx.doi.org/10.1103/PhysRevLett.109.247206}{\bf
  109}, 247206 (2012). 

\bibitem{KCC14}
M. Kormos, M. Collura and P. Calabrese, \emph{Analytic results for a
  quantum quench from free to hard-core one-dimensional bosons},
Phys. Rev. A \href{http://dx.doi.org/10.1103/PhysRevA.89.013609}{\bf
  89}, 013609 (2014). 

\bibitem{quenchLL}
J. De Nardis and J.-S. Caux, \emph{Analytical expression for a
  post-quench time evolution of the one-body density matrix of
  one-dimensional hard-core bosons}, J. Stat. Mech. (2014)
\href{http://dx.doi.org/10.1088/1742-5468/2014/12/P12012}{P12012}. 

\bibitem{Bertini14}
B. Bertini, D. Schuricht, and F.H.L. Essler, \emph{Quantum quench in the sine-Gordon model}, J. Stat. Mech. (2014) \href{http://dx.doi.org/10.1088/1742-5468/2014/10/P10035}{P10035}. 

\bibitem{CCS17}
A. Cort\'es Cubero and D. Schuricht, \emph{Quantum quench in the
  attractive regime of the sine-Gordon model}, J. Stat. Mech. (2017)
\href{10.1088/1742-5468/aa8c2e}{103106}.

\bibitem{Horvath18}
D. X. Horvath, M. Kormos and G. Takacs, \emph{Overlap singularity and
  time evolution in integrable quantum field theory},  JHEP 08 (2018) \href{10.1007/JHEP08(2018)170}{170}.

%%%%%%%%%%% semiclassical quench %%%%%%%%%%%%%
  
\bibitem{IR11} F. Igl\'oi and H. Rieger, \emph{Quantum
Relaxation after a Quench in Systems with Boundaries},
Phys. Rev. Lett. \href{http://dx.doi.org/10.1103/PhysRevLett.106.035701}{\bf
106}, 035701 (2011). 

\bibitem{RI11} H. Rieger and F. Igl\'oi, \emph{Semiclassical theory
  for quantum quenches in finite transverse Ising chains},
  Phys. Rev. B \href{http://dx.doi.org/10.1103/PhysRevB.84.165117}{\bf
    84}, 165117 (2011).

\bibitem{BRI12} B. Blass, H. Rieger, and F. Igl\'oi, \emph{Quantum relaxation and finite-size effects in the XY chain in a transverse field after global quenches}, EPL \href{http://dx.doi.org/10.1209/0295-5075/99/30004}{\bf 99} 30004 (2012).

\bibitem{Evangelisti13} 
S. Evangelisti, \emph{Semi-classical theory for quantum quenches in the O(3) non-linear sigma model}, J. Stat. Mech. (2013) \href{http://dx.doi.org/10.1088/1742-5468/2013/04/P04003}{P04003}. 

\bibitem{Kormos16} 
M. Kormos and G. Zar\'and, \emph{Quantum quenches in the sine-Gordon
  model: a semiclassical approach}, Phys. Rev. E
\href{https://doi.org/10.1103/PhysRevE.93.062101}{{\bf 93}, 062101 (2016).}

%%%%%%%%%  quench action %%%%%%%%%%%%%%%%%%%
  
\bibitem{CE13} 
J.-S.~Caux and F.H.L.~Essler, \emph{Time Evolution of Local Observables After Quenching to an Integrable Model}, Phys. Rev. Lett. \href{http://dx.doi.org/10.1103/PhysRevLett.110.257203}{\bf 110}, 257203 (2013).

\bibitem{granetjacobsensaleur} 
E. Granet, J. L. Jacobsen and H. Saleur, \emph{Analytical results on the Heisenberg spin chain in a magnetic field}, J. Phys. A: Math. Theor. \href{http://dx.doi.org/10.1088/1751-8121/ab1f97}{\bf 52}, 255302 (2019).

\bibitem{ristivojevic} 
Z. Ristivojevic, \emph{Excitation spectrum of the Lieb-Liniger model}, Phys. Rev. Lett. \href{http://dx.doi.org/10.1103/PhysRevLett.113.015301}{\bf 113}, 015301 (2014).

\bibitem{Kitanine}
N. Kitanine, K.K. Kozlowski, J.M. Maillet, N.A. Slavnov and V. Terras,
\emph{Algebraic Bethe ansatz approach to the asymptotic behavior of
  correlation functions},
J. Stat. Mech. \href{\doi10.1088/1742-5468/2009/04/P04003}, P04003 (2009).

\bibitem{Wouters14} B. Wouters, J. De Nardis, M. Brockmann,
  D. Fioretto, M. Rigol, and J.-S. Caux, \emph{Quenching the
    Anisotropic Heisenberg Chain: Exact Solution and Generalized Gibbs
    Ensemble Predictions},
  Phys. Rev. Lett. \href{http://dx.doi.org/10.1103/PhysRevLett.113.117202}{\bf
    113}, 117202 (2014). 

\bibitem{Brockmann14}
  M. Brockmann, B. Wouters, D. Fioretto, J. De Nardis, R. Vlijm and J.-S. Caux, \emph{Quench action approach for releasing the N\'eel state into the spin-1/2 XXZ chain}, Stat. Mech. (2014) \href{http://dx.doi.org/10.1088/1742-5468/2014/12/P12009}{P12009}.


\bibitem{Pozsgay14} B. Pozsgay, M. Mesty\'an, M.A. Werner, M. Kormos,
  G. Zar\'and, and G. Tak\'acs, \emph{Correlations after Quantum
    Quenches in the XXZ Spin Chain: Failure of the Generalized Gibbs
    Ensemble},
  Phys. Rev. Lett. \href{http://dx.doi.org/10.1103/PhysRevLett.113.117203}{\bf
    113}, 117203 (2014).

\bibitem{Mestyan15}
M. Mesty\'an, B. Pozsgay, G. Tak\'acs, and M.A. Werner, \emph{Quenching the XXZ spin chain: quench action approach versus generalized Gibbs ensemble}, J. Stat. Mech. (2015) \href{http://dx.doi.org/10.1088/1742-5468/2015/04/P04001}{P04001}.

\bibitem{denardis14}
J. De Nardis, B. Wouters, M. Brockmann, and J.-S. Caux, \emph{Solution for an interaction quench in the Lieb-Liniger Bose gas}, Phys. Rev. A \href{http://dx.doi.org/10.1103/PhysRevA.89.033601}{\bf 89}, 033601 (2014).

\bibitem{piroli16}
L. Piroli, P. Calabrese, and F.H.L. Essler,
  \emph{Multiparticle Bound-State Formation following a Quantum Quench
    to the One-Dimensional Bose Gas with Attractive Interactions},
  Phys. Rev. Lett. \href{http://dx.doi.org/10.1103/PhysRevLett.116.070408}{\bf
    116}, 070408 (2016).
  
\bibitem{DNPC15} 
J. De Nardis, L. Piroli and J.-S. Caux, \emph{Relaxation dynamics of local observables in integrable systems}, Phys. A: Math. Theor. \href{http://dx.doi.org/10.1088/1751-8113/48/43/43FT01}{\bf 48} 43FT01 (2015).

%%%%%%%%%%%%%%%%% locality %%%%%%%%%%%%%%%%%%%%
\bibitem{lukyanov}
S. Lukyanov, \emph{Free Field Representation For Massive Integrable Models},
Commun. Math. Phys. \href{https://doi.org/10.1007/BF02099357}{{\bf 167}, 183 (1995).}

\bibitem{smirnov}
F.~A. Smirnov, {\em Form factors in completely integrable models of quantum
  field theory} (World Scientific, Singapore, 1992).
%%%%%%%%%%%%%%%%%%%%%%%%%%%%%%%%%%%%%%%%%%%%%%%

\bibitem{sachdevbook}
S. Sachdev, \emph{Quantum Phase Transitions}, Cambridge University Press,
2001.

\bibitem{Bugrij}
A. Bugrij, \emph{Correlation function of the two-dimensional Ising
  model on the finite lattice. I},
Theor. Math. Phys. \href{\doi10.1023/A:1010320126700}{\bf 127}, 528
(2001).

\bibitem{BL03}
A. Bugrij and O. Lisovyy, \emph{Spin matrix elements in 2D Ising model
  on the finite lattice},
Phys. Lett. \href{\doi10.1016/j.physleta.2003.10.039}{\bf A 319}, 390
(2003).

\bibitem{Gehlen}
G. von Gehlen, N. Iorgov, S. Pakuliak, V. Shadura and Y. Tykhyy,
\emph{Form-factors in the Baxter-Bazhanov-Stroganov model II: Ising
  model on the finite lattice},
J. Phys. \href{\doi10.1088/1751-8113/41/9/095003}{\bf A 41}, 095003
(2008).

\bibitem{iorgov11}
N. Iorgov, V. Shadura and Yu. Tykhyy, \emph{Spin operator matrix elements in
the quantum Ising chain: fermion approach}, J. Stat. Mech. (2011) \href{\doi10.1088/1742-5468/2011/02/P02028}{P02028}.

\bibitem{korepinslavnov}
V. E. Korepin and N. A. Slavnov, \emph{The time dependent correlation
  function of an impenetrable {B}ose gas as a {F}redholm minor. {I}},
Commun. Math. Phys. \href{\doi10.1007/BF02096781}{{\bf 129}, 103,
  (1990).} 


\bibitem{derzkho}
O.~Derzhko and T.~Krokhmalskii, \emph{Dynamic structure factor of the spin-1/2
  transverse {I}sing chain}, Phys. Rev. B, \href{https://doi.org/10.1103/PhysRevB.56.11659}{{\bf 56}, 11659, (1997).}

\bibitem{kozlowski4}
K.K. Kozlowski, \emph{Large-Distance and Long-Time Asymptotic Behavior
  of the Reduced Density Matrix in the Non-Linear Schr\"odinger Model},
Ann. Henri Poincar\'e \href{\doi10.1007/s00023-014-0327-3}{{\bf 16}, 437 (2015).}

\end{thebibliography}
\end{document}